%% file: BNVv3.tex
\documentclass[12pt,prd,aps,epsfig,floats,preprint,superscriptaddress,eqsecnum,nofootinbib]{revtex4-1}

\usepackage{color}
\usepackage{amsmath}
\usepackage{amsfonts}
\usepackage{graphicx}
\usepackage{subfigure}
\usepackage{etoolbox}
\usepackage{fontenc}
\usepackage{textcomp}
\usepackage{xspace}
\usepackage{siunitx}
\usepackage{graphicx,amsmath, nccmath, amssymb,epsfig,verbatim,mathrsfs,array,layout,latexsym}
\usepackage{multirow}
\usepackage{slashed}
\usepackage{braket}
\usepackage{booktabs}

\usepackage{tikz-feynman}
\usepackage{tikz}
\usetikzlibrary{arrows}
\usepackage[compat=1.1.0]{tikz-feynhand}
\usepackage{diagbox}

\tikzfeynmanset{
  fermion/.style={
    /tikz/postaction={
      /tikz/decoration={
        markings,
        mark=at position 0.5 with {
          \arrow{>[length=6pt, width=5pt]};
        },
      },
      /tikz/decorate=true,
    },
  },
}

\usepackage{hyperref}
\newcommand{\mg}{{\sc\small MadGraph5\_aMC@NLO}~}

\newcommand{\feyncalc}{\texttt{feyncalc}\,\cite{Mertig:1990an,Shtabovenko:2016sxi,Shtabovenko:2020gxv}\,}
\newcommand{\feynrules}{\texttt{feynrules}\,\cite{Alloul:2013bka}\,}

\DeclareSIUnit{\ab}{\text{ab}}

\newcommand{\emiss}{E_T^{\text{miss}}}
\newcommand{\highPt}{high-$p_T$}
\begin{document}

\preprint{CERN-TH-2026-013}

\title{Probing baryon number with missing energy}

\author{Gudrun~Hiller}
\email{gudrun.hiller@cern.ch}
\affiliation{Theoretical Physics Department, CERN, 1211 Geneva 23, Switzerland}
\affiliation{TU Dortmund University, Department of Physics, Otto-Hahn-Str.4, D-44221 Dortmund, Germany}
\author{Antonio Rodríguez-Sánchez}
\email{antonio.rsanchez@uclm.es}
\affiliation{Departamento de Física, Universidad de Castilla-La Mancha, Avenida de Carlos III,
s/n, 45004 Toledo, Spain}
\author{Daniel~Wendler}
\email{daniel.wendler@tu-dortmund.de}
\affiliation{TU Dortmund University, Department of Physics, Otto-Hahn-Str.4, D-44221 Dortmund, Germany}

\begin{abstract}
Quark portal interactions $qqqN$ with a light singlet fermion $N$ 
make baryon number testable through missing transverse energy (MET).
We find that present LHC data constrain scales up to 10 TeV (MET plus jet), 8  TeV (MET plus top) and 11 TeV (MET plus bjet). With increasing mass, or larger portal couplings, $N$ becomes less long-lived, and gives clean displaced vertex  
signatures, which encourage dedicated searches.
We also narrow down viable mesogenesis models with color triplet scalars to a mass range  $\sim y \,\cdot 3 \, \text{TeV}$ 
(with charm) and $\sim y \, \cdot 5 \,  \text{TeV}$ (charmless)
couplings $y$ to $b$ and lighter quarks, a window that can be scrutinized by HL-LHC.
Interactions also induce rare decays of type baryon (meson) to meson (baryon) plus invisible, 
which complement high-$p_T$ searches and can prove baryon number violation.
We explore charm decays $\Lambda_c \to (\pi,K) + \mathrm{invisible}$.
Their branching ratios are 
subject to sizable hadronic uncertainties and 
require  high luminosity flavor facilities such as a Tera-Z facility (FCC-ee, CEPC).
Branching ratios of top quarks into one or two $b$-jets plus $N$ can reach few$\times 10^{-6}$.

\end{abstract}

\maketitle

\tableofcontents

\section{Introduction}
\label{sec:introduction}

The lifetime of the proton is much longer than the age of the universe \cite{ParticleDataGroup:2024cfk}.
Yet, baryon number 
is not too well protected; it is only an accidental symmetry in the  standard model (SM), and broken in extensions such as GUTS, leptoquark models, and 
R-parity violating supersymmetry.  Non-observation of proton decay, or more
general baryon (meson) to meson (baryon) processes pose strong limits on baryon number violation (BNV). At the same time, baryon number, 
however small its violation, is intimately linked to  our existence. Improving and extending searches to wider areas is therefore highly motivated.

Probing BNV-operators at colliders looks like a hopeless enterprise, considering the astonishing stability of protons
with the associated BNV-scale generically $\sim 10^{16} \, \text{GeV}$ way out of reach.
The simplest way of realizing a collider-type scale is by imposing that the decay products carry baryon number and that this is only conserved when considering these otherwise feebly interacting particles. 
This way, one also avoids constraints from neutron oscillations \cite{Fornal:2018eol,Alonso-Alvarez:2021qfd,Alonso-Alvarez:2021oaj}, and leaves the lepton sector intact.

Here we consider the SM amended by a light SM-singlet fermion $N$ that carries baryon number
with mass above the proton's.
 While less simplified model-building can also
invoke flavorful Majorana fermions to avoid various constraints, we stress that the limits on quark portal interactions $qqq N$ and
$\bar q q \bar N N$
hold also for alternative viable models as long as $N$ is light and has apart from the contact terms at most feeble interaction with the SM.

Concretely, we perform a comprehensive 
search for the elusive  $N$-particle
in missing transverse energy (MET) plus jet, MET plus bjet, and MET plus top signals from a recast of LHC measurements 
\cite{ATLAS:2024vqf,ATLAS:2020yzc,CMS:2022usq,CMS:2024rkj}. 
See, e.g.~\cite{Allahverdi:2017edd} for earlier works.
The $pp$-data allow for all flavors to be probed as they add incoherently to the cross sections.
In addition, we propose searches in charm baryon decays and in decays of the top-quark, into $N$ plus SM quarks or hadrons.
The overarching feature is that missing energy observables are 
clean, as SM backgrounds are often negligible which makes them null tests, 
or under theoretical control.
At the same time, MET is inclusive, making searches versatile and cover interactions with neutrinos, sterile neutrinos, axions, 
and a dark sector.

Here, MET is linked to baryon number, carried away by a  singlet fermion $N$, which restores baryon number,
but leads to apparent BNV in the SM sector (For $N$ being Majorana, the BNV is real.)
This setting is a generalization of the one by \cite{Alonso-Alvarez:2021qfd,Alonso-Alvarez:2021oaj} which aims at mesogenesis and has restrictions from $B$-meson decays.
Here, we would like to have baryon number visible at colliders, and do not aim to address cosmology. On the other hand, our findings in terms of EFT apply to 
the mesogenesis models based on scalar color triplets.

Constructing an EFT is very much alike to SMEFT with the $N$ added as a light degree of freedom.
This allows us to study portals, Dim 6 operators that connect the SM to the new sectors, as $\bar qq \bar NN$ and $qqqN$.
The resulting EFT very much resembles the $\nu$-SMEFT ~\cite{Liao:2016qyd}, featuring light right-handed neutrinos.
(We consider only one $N$-particle but one could do more).
UV completions are also analogous.

Searches at colliders so far have been mostly limited to top production or decays in association with charged leptons,
e.g.,~\cite{Dong:2011rh,Bahl:2023xkw}, or to processes with EFT operators of high dimensions, e.g.~\cite{Durieux:2012gj}, 
or to scenarios with super-symmetric particles~\cite{Durieux:2013uqa,ATLAS:2017tmw}.

Rare decays offer alternative ways to look for  apparent BNV. Efforts include searches in $B$-decays \cite{BaBar:2023dtq,Belle:2021gmc}, 
and hyperons \cite{BESIII:2025sfl}.
Here we explore  opportunities from charm decays to invisibles.
They could be studied at 
future high luminosity flavor facilities such as a Tera-$Z$-factory, FCC-ee \cite{FCC:2025lpp}, CEPC \cite{Ai:2024nmn}, and a   
super-$\tau$-charm factory \cite{Achasov:2023gey}.

The plan of the paper is as follows:
In Sec.~\ref{sec:framework} we introduce the  EFT set-up and the operators we are going to focus on. 
Single UV-mediator models are also presented.
Constraints from low energy observables are discussed in Sec.~\ref{sec:low_energy_theory}.
The bulk of this work consists of a comprehensive analysis of collider searches and their complementarity depending on the lifetime of the dark fermion,
given in Sec.~\ref{sec:highPt_theory}.
Resulting constraints from the high-$p_T$ LHC data  and correlations are presented in Sec.~\ref{sec:res}. 
Complementarily we explore whether  similar effects can be seen in low-energy hadron decays. 
We study charm baryon decays in
Sec.~\ref{sec:charm_decays}, and work out the sensitivities at current and future charm facilities. Implications of the $qqqN$-operators for top-decays
are worked out in Sec.~\ref{sec:top}.
We conclude in Sec.~\ref{sec:con}. Summary tables with numerical values of the  collider limits are provided in App.~\ref{app:AddResults}. 
In App.~\ref{app:QCDF} we detail
the computation of the $\Lambda_c \to (\pi,K)+\bar N$ decay amplitudes. In App~\ref{app:UV} we list the lagrangians of possible UV completions, and their tree-level EFT.

\section{Model framework}
\label{sec:framework}

We add a massive vector-like fermion $N$, singlet under the SM gauge group, assumed to be 
heavier than the proton to avoid proton decay.
In Sec.~\ref{sec:eft} we discuss the EFT, and in Sec.~\ref{sec:uv} the corresponding single, heavy  mediators.
In Sec.~\ref{sec:mix} we consider constraints from meson mixing and under which instances they can be avoided.

\subsection{EFT for BNV \label{sec:eft}}
In this study, we explore the viable options one has to observe (apparent) BNV in colliders with $D=6$ operators, considering which UV completions one can conceive and how to avoid critical low-energy bounds, also aiming to motivate simplified textures for the collider fits.
There are two sets of operators, induced at the scale $\Lambda$,
\begin{equation} \label{eq:L-BNV_4F}
\mathcal{L}=\frac{C_{ijk}^{qqdN}}{\Lambda^2}\mathcal{O}^{qqdN}_{ijk}+ \frac{ C_{ijk}^{uddN}}{\Lambda^2} \mathcal{O}^{uddN}_{ijk} + h.c. \, 
\end{equation}
where,
\begin{align}
  \label{eqn:BNV_4F}
\mathcal{O}^{qqdN}_{ijk} &= \epsilon^{ab} \epsilon^{\alpha \beta \gamma} ( \bar{q}^{C}_{L,a,\alpha,i}  q_{L,b,\beta,j} ) ( \bar{d}_{R,\gamma,k}^{C} N_R ) \, ,\\
\mathcal{O}^{uddN}_{ijk} &= \epsilon^{\alpha \beta \gamma} ( \bar{u}^{C}_{R,\alpha,i} d_{R,\beta,j} ) ( \bar{d}_{R,\gamma,k}^{C} N_R ) \, .  \nonumber
\end{align}

$a$ and $b$ are $SU(2)_L^{W}$ indices, $\alpha, \beta , \gamma$ are $SU(3)_C$ color indices and $i,j,k$ are generation indices.

Due to the properties of the spinor bilinears, the operator $\mathcal{O}^{qqdN}_{ijk}$ is symmetric in $i,j$, and
\begin{equation} \label{eq:sym}
C^{qqdN}_{ijk}=C^{qqdN}_{jik} = C^{qqdN}_{\{ij\}k}\, , 
\end{equation}
where we use curly brackets $\{ \}$ to further indicate this within this work. 
In total both operators of Eqn.~\eqref{eqn:BNV_4F}  have $27 + 18 = 45 $ parameters, which we assume to be real-valued.

\subsection{UV mediators \label{sec:uv}}

Only three heavy mediators, which are color triplet bosons and also  termed leptoquarks (LQ) can generate operators (\ref{eqn:BNV_4F}) at tree-level~\cite{Beltran:2023ymm}. They are summarized in Tab.~\ref{tab:UV_mediators}.

\begin{table}[]
    \centering
    \begin{tabular}{c| c c c}
       Model &  Mediator name & SM Rep. & Operators generated  \\ \toprule
    1&     $\Psi$ ~\cite{Alonso-Alvarez:2021oaj,Alonso-Alvarez:2021qfd} ,$(\bar{S}_1)^{\dagger}$~\cite{Dorsner:2016wpm},$\omega_2$~\cite{deBlas:2017xtg,Beltran:2023ymm} & $(3,1, 2/3)$ & $ {\mathcal{O}}^{uddN}_{i[jk]}$ \\
    2&     $\Phi$~\cite{Alonso-Alvarez:2021oaj,Alonso-Alvarez:2021qfd}, $(S_1)^{\dagger}$~\cite{Dorsner:2016wpm},$\omega_1$ ~\cite{deBlas:2017xtg,Beltran:2023ymm} &  $(3,1,-1/3)$ & $ {\mathcal{O}}^{uddN}_{ijk}, {\mathcal{O}}^{qqdN}_{\{ij\}k}$  \\
     3&    $X_{\mu}$~\cite{Alonso-Alvarez:2021qfd},$ (\tilde{V}_2)^{\dagger}$~\cite{Dorsner:2016wpm},$Q_1$~\cite{deBlas:2017xtg,Beltran:2023ymm} & $(3,2,1/6)$ & ${\mathcal{O}}^{qqdN}_{\{ij\}k}$  \\
    \end{tabular}
    \caption{Heavy UV mediators, which generate the BNV operators introduced in Eqn.~(\ref{eqn:BNV_4F}). Listed are the names, used in different references, the SM representation and the BNV operators generated at tree level. The brackets $\{ \} ([])$ denote the (anti-)symmetric part of the operator. }
    \label{tab:UV_mediators}
\end{table}

Let us discuss the low energy Lagrangian $\mathcal{L}_1$ of leptoquark model 1, with Yukawa terms 
$-y^{dd}_{rs}\,\Psi\,(\bar d^{C}_{R,r} d_{R,s})-y^{Nu}_p \, \Psi^\dagger\,(\bar u^{C}_{R,p} N_R) +\mathrm{h.c.}$, where $N_R=P_R N$ with the right-handed projector $P_R$.
 The tree-level matching of the other models is given in App.~\ref{app:UV}.  Integrating out $\Psi$ at tree level, one finds
\begin{equation}
\label{eqn:LQ_model_1}
\begin{aligned}
\mathcal{L}_1
&=\Big(-2\frac{\epsilon_{rst}\tilde{y}_t^{dd}y^{Nu}_{p}}{M_{\Psi}^2} \quad
[\bar{u}_{R,p}^{C}d_{R,r}\,\bar{d}^{C}_{R,s}N_R] +\mathrm{h.c.}\Big)\\
&+\frac{\epsilon_{r's'p'}\epsilon_{rsp}\tilde{y}^{*,dd}_p \tilde{y}^{dd}_{p'} }{M_{\Psi}^2} \quad
[\bar{d}_{R}^r \gamma^{\mu}d_{R,r'} \, \bar{d}_{R}^s\gamma_{\mu}d_{R,s'}] \\
&+\frac{y_{p}^{Nu}y_{p'}^{*,Nu}}{2M_{\Psi}^2}  \quad
[\bar{N}_R \gamma^{\mu} N_R \, \bar{u}_{R,p'}\gamma_{\mu}u_{R,p}] \, .
\end{aligned}
\end{equation}
Here, $M_{\Psi}$ denotes the mediator mass, $y^{Nu}, \tilde y^{dd}$ are the Yukawa couplings and other small letters are flavor indices.
The purpose of using $\tilde y^{dd}$  with the epsilon-tensor, i.e.,  $y^{dd}_{rs}=\epsilon_{rst}\tilde{y}^{dd}_t$
instead of the Yukawa with the two flavor-indices of the down-quarks is to make the asymmetry in $rs$ manifest. It also
implies that there are only three independent couplings to the diquarks.
We observe that the following types of operators are induced, a feature which holds generically  for all UV models:

\begin{enumerate}
\item[{\it i)}] 4-quark operators,

\item[{\it ii)}] 3-quark operators with a single fermion $N$; these induce the apparent BNV (\ref{eqn:BNV_4F}),

\item[{\it iii)}] 2-quark vector operators involving a pair of singlets $\sim \bar N_R \gamma^\mu N_R$.
Constraints on the latter have been obtained from MET observables in \cite{Hiller:2024vtr}.
\end{enumerate}

The 2- and 3-quark operators induce missing energy.
The 4-quark operators contribute to dijet signals, and also meson mixing, depending
on the model. Notably, model 1  has no tree-level contribution to meson-mixing, due to the flavor-asymmetric diquark Yukawa \cite{Dorsner:2016wpm}. 
This can be understood from exemplarily writing down the 4-quark operator in (\ref{eqn:LQ_model_1}) for $p=p'=3$:
$2  (\tilde y^{dd}_3/M_\Psi )^2(\bar d_R \gamma_\mu d_R \bar s_R \gamma^\mu s_R- \bar d_R \gamma_\mu s_R \bar s_R \gamma^\mu d_R)$. Kaon mixing, on the other hand,
is induced by the operator $(\bar d_R \gamma_\mu s_R )^2+\mathrm{h.c.}$.
Model 2 also has no tree-level contribution because the diquark-couplings  simultaneously involve  up- and down-type quarks.
To understand the importance of one-loop contributions one can perform a naive dimensional analysis:
Ignoring differences in the loop-functions between the SM and BNV box-diagrams,
the consistency of meson-mixing in the SM with data requires that the BNV contributions do not exceed the SM one:
\begin{align}
    \frac{y^4}{M_\Psi^2} 
  \lesssim (V_{ij} V_{ik}^*)^2 G_F^2 m_W^2 \sim (V_{ij} V_{ik}^*)^2 \frac{0.9}{\text{TeV}^2} \, .
\end{align}
Here, $V_{ij}$ denote the CKM matrix elements, $G_F$ is Fermi's constant,
and $m_W$ is  the $W$-boson mass. This rough estimate yields for order one
BNV-couplings leptoquarks outside the direct LHC reach, 
$M_\Psi \gtrsim 3000, 124, 26 \, \text{TeV}$  
for kaon, $B_d$- and $B_s$-mixing, respectively, in agreement with \cite{Aebischer:2020dsw}
up to factors of order one.
However, allowing for only a subset of  flavor couplings switched on can evade these constraints.
This is further discussed in Sec.~\ref{sec:mix}.

In the EFT-analysis the main focus of this work is on the 3-quark operators. 
As indicated in Tab.~\ref{tab:UV_mediators} the mediator models induce them with symmetry-properties in the flavor indices.

\subsection{Meson mixing in UV-models \label{sec:mix}}

BSM scenarios that generate four-quark operators are stringently constrained by bounds on meson mixing. However, these bounds can be largely evaded in simple UV setups that generate the BNV operators studied in this work.

\begin{figure}
    \centering
\includegraphics[width=0.85\linewidth]{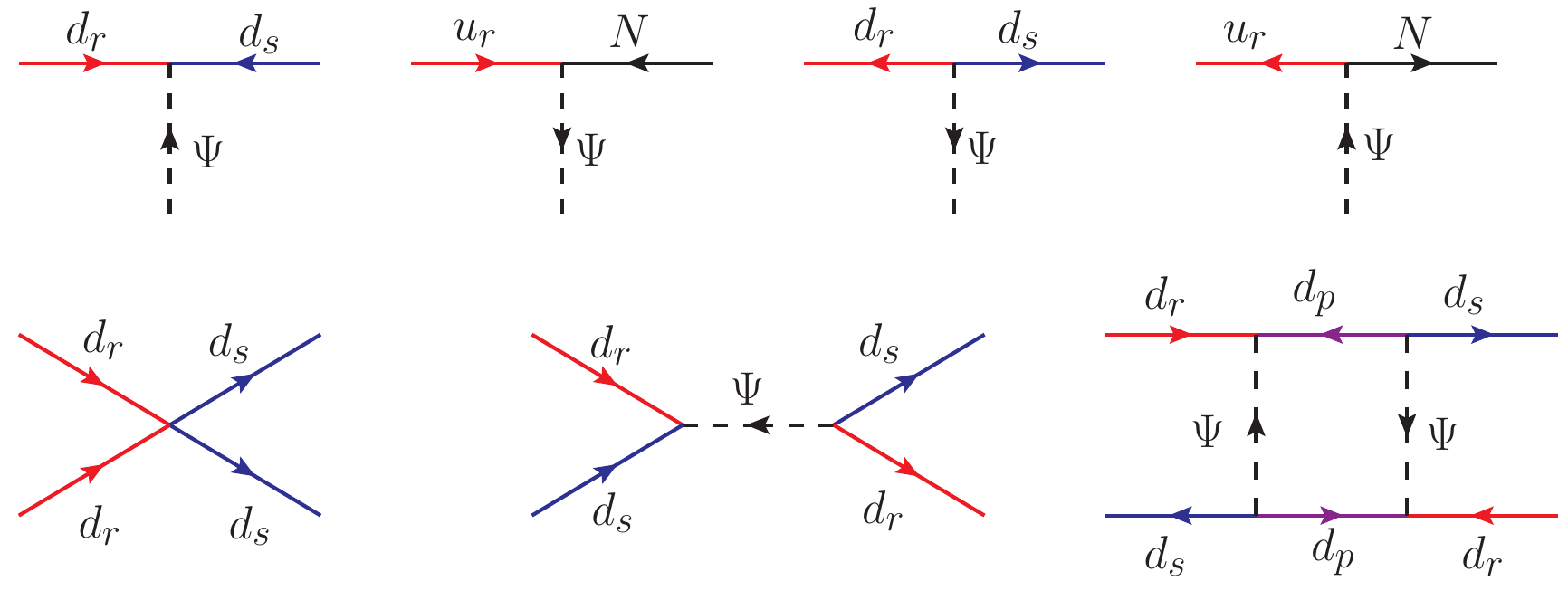}
    \caption{Top panel: vertices for interactions between $\Psi\sim(3,1,2/3)$ 
 and fermions. Bottom panel (left to right): topology relevant for meson mixing; four-quark operators generated at tree level by the model that do not induce meson mixing; and four-quark operators generated by the model that induce meson mixing at loop level, avoided if couplings are allowed only between two quark generations.}
    \label{fig:mixing}
\end{figure}

Let us illustrate this with the simplest model, $\Psi\sim(3,1,2/3)$. 
Its two interaction terms in the Lagrangian read, omitting color indices but keeping generation labels, $\propto y^{dd}_{rs}\,\Psi\,(\bar d^{C}_{R,r} d_{R,s})$ and $\propto y^{uN}_{r}\,\Psi^\dagger\,(\bar u^{C}_{R,r} N_R)$, plus the hermitian conjugate. The corresponding vertices are shown in Fig.~\ref{fig:mixing}. Notice that $y^{dd}_{rs}$ can be chosen antisymmetric (as the corresponding operator is), and then, crucially, there are no flavor-diagonal contributions. Only $d_{r}\overset{\Psi}{\to}\bar{d}_{s\neq r}$ and $\bar{u}_{r}\overset{\Psi}{\to}N$ (and equivalents under crossing and/or reversing all arrows) are allowed.
The topology relevant for $(\Delta F=2)$ meson mixing requires two incoming quarks of flavor $r$ and two outgoing quarks of a different flavor $s$, e.g.\ $d_{r}d_{r}\to d_s d_s$, or equivalents under crossing, as also illustrated in Fig.~\ref{fig:mixing}. However, starting from $d_{r}$, the allowed fermion chains in the down sector can only generate at tree level $d_{r}\overset{\Psi}{\to}\bar{d}_{s\neq r}$, and thus do not contribute to meson mixing. At loop level, such contributions can be generated through two chains,
$d_{r}\overset{\Psi}{\to}\bar{d}_{p\neq r}\overset{\Psi^{\dagger}}\to d_{s\neq p}$,
as shown in Fig.~\ref{fig:mixing}.

In general setups, the corresponding loop contributions 
are still more stringent than the bounds obtained in this work. However, they can be completely evaded, as argued in Ref.~\cite{Alonso-Alvarez:2021oaj}, by allowing only one $y^{dd}_{rs}$ coupling at a time, i.e.\ only $y^{dd}_{12}$, or only $y^{dd}_{13}$, or only $y^{dd}_{23}$. Analogously, in the up sector, charm meson mixing at loop level is avoided as long as $y^{uN}_{1}$ and $y^{uN}_{2}$ are not simultaneously active. In Sec.~\ref{sec:resultsUV} we present some of the bounds motivated by this type of scenario.

Similarly, $\Phi\sim(3,1,-1/3)$ can generate new meson-mixing topologies at loop level, in this case also involving $W$ exchange. But again these can be evaded in simple scenarios, for example by switching on only one $y^{ud}_{rs}$ and one $y^{Nd}_{p}$ at a time. As argued in Ref.~\cite{Alonso-Alvarez:2021oaj}, evading the meson-mixing bounds in the vector $X_{\mu}\sim(3,2,1/6)$ model is less straightforward, since it is already induced at tree level.

\section{Low-energy probes of BNV}
\label{sec:low_energy_theory}

We discuss constraints from low-energy processes.
In Sec.~\ref{sec:rare} we discuss limits from rare decays into invisibles using existing searches with $B$-mesons and hyperons.
We identify decays of charm hadrons suitable to probe BNV in Sec.~\ref{sec:charm}.
In Sec.~\ref{sec:neutrons} we work out limits from radiative neutron decay into the $N$ particle allowed in the very small window between the proton and the neutron mass.

\subsection{Constraints from $B$ and $\Xi$ decays \label{sec:rare}}

We discuss branching ratios for invisible decays of $B$-mesons and hyperons.
Predictions for $\Lambda_c$-decays are given in Sec.~\ref{sec:charm_decays}.

\begin{figure}
    \centering
    \includegraphics[width=0.49\linewidth]{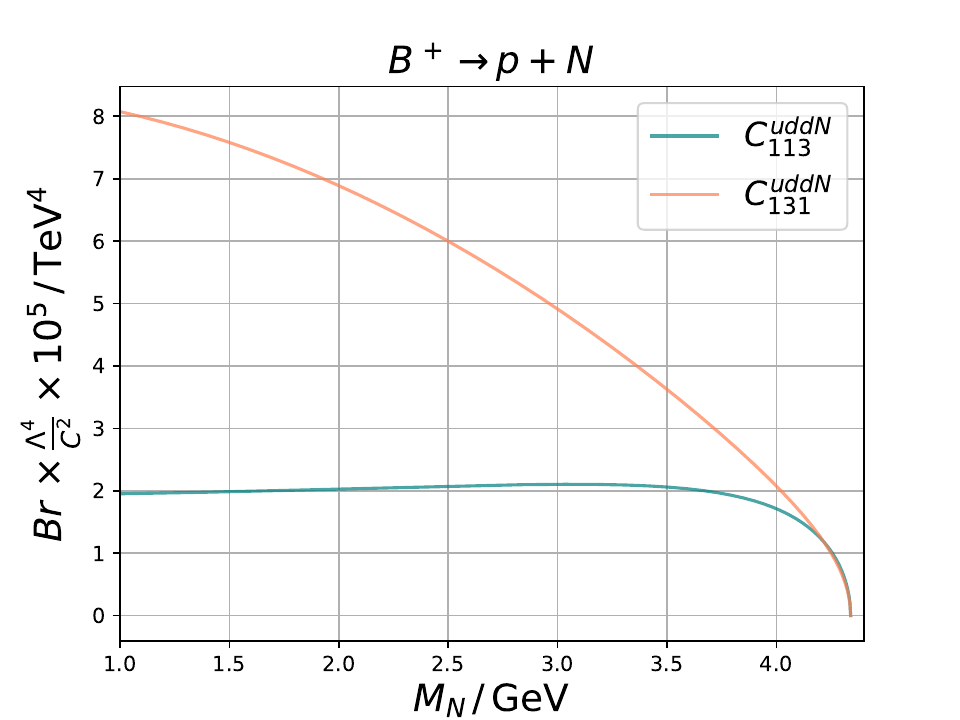}
    \includegraphics[width=0.49\linewidth]{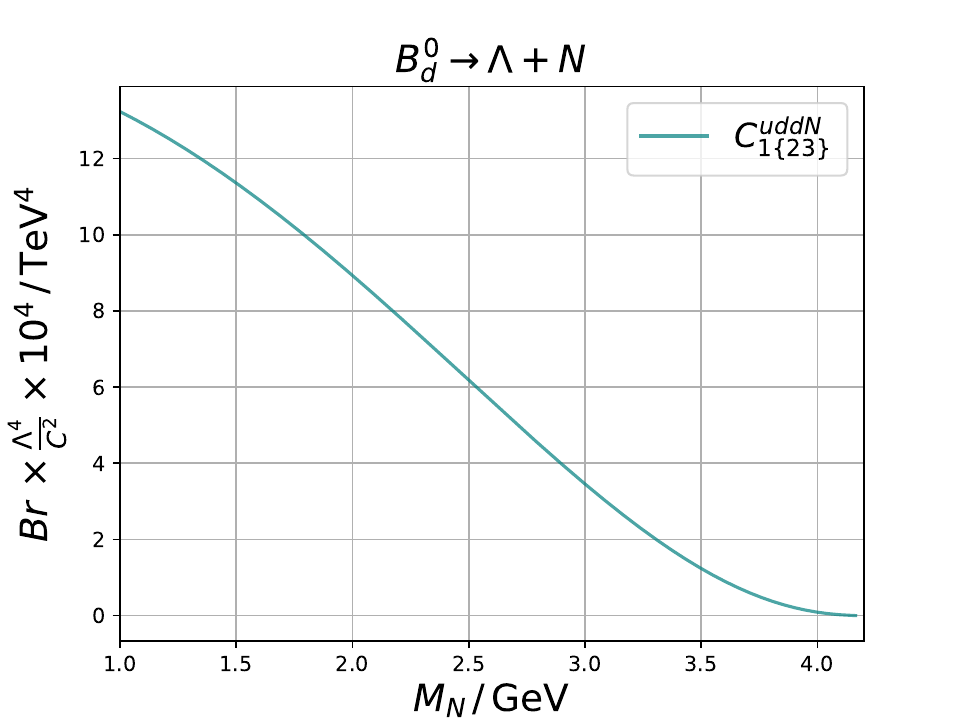}
    \includegraphics[width=0.49\linewidth]{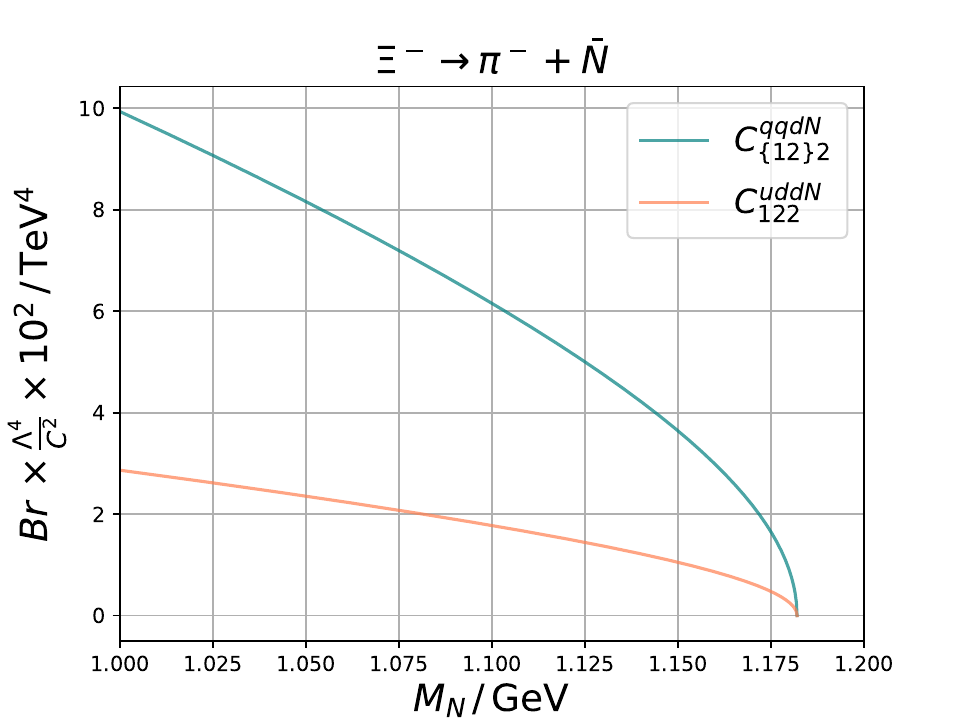}
        \caption{Branching ratios of exclusive $B$ meson (upper row) and hyperon (lower row)  decays into the singlet $N (\bar N)$  in units of $C^2/\Lambda^4 \text{ TeV}^4$.
    Predictions are based on Refs.~\cite{Elor:2022jxy,Alonso-Alvarez:2021oaj}. }
    \label{fig:brs}
\end{figure}

For $B$ and $\Xi$  decays we reproduce results available in the literature \cite{Alonso-Alvarez:2021oaj,Elor:2022jxy} to obtain predictions for branching ratios. 
In Tab.~\ref{tab:low-input} we list the decays, the theory input used and the available experimental upper limit. In Fig.~\ref{fig:brs} we show the branching ratios for $B^+ \to p + N$, $B^0 \to \Lambda + N $ and $\Xi^- \to \pi^- + \bar N $ and in Tab.~\ref{tab:current_bounds} we summarize the respective bounds. Note that Ref.~\cite{Elor:2022jxy} only considers right-handed quarks, which means predictions and bounds are only available on $C^{uddN}_{ijk}$. Furthermore $B^0 \to \Lambda + N $ form factors are identical for both operators 
considered in Ref.~\cite{Elor:2022jxy}, which implies that only the symmetric combination  $C^{uddN}_{1\{23\}}$ is probed.

To illustrate the order of magnitude in Fig.~\ref{fig:brs} we show central values only, in view of the significant theory uncertainties. In particular subleading terms can be important \cite{Khodjamirian:2022vta,Boushmelev:2023huu,Mohamed:2025zgx}.

\begin{table}[]
    \centering
    \begin{tabular}{c  c  c c c c c}
        Decay                          &WCs probed& Theory  & Exp. limit & $M_N$ max \\ \toprule                       
        $B^+ \to p +  N $              &$C^{uddN}_{131},C^{uddN}_{113}$ & \cite{Elor:2022jxy}& $\num{2e-6}$\cite{BaBar:2023dtq} & $4.34 $ GeV\\
        $B^0 \to \Lambda + N $         &$C^{uddN}_{1\{23\}}$ &\cite{Elor:2022jxy}& $\num{2e-5}$\cite{Belle:2021gmc} & $4.16 $ GeV\\
        $\Xi^- \to \pi^- + \bar N $    &$C^{qqdN}_{\{12\}2},C^{uddN}_{122}$ &\cite{Alonso-Alvarez:2021oaj}& $\num{4.2e-5}$\cite{BESIII:2025sfl} & $1.18 $ GeV\\
         $\Lambda_c \to \pi^++ \bar N $ &$C^{qqdN}_{\{21\}1},C^{uddN}_{211}$ & this work & 0.25$^a$ & $2.15 $ GeV \\
       $\Lambda_c \to K^+ + \bar N $  &$ C^{qqdN}_{2\{12\}},C^{uddN}_{212},C^{uddN}_{221}$ & this work & 0.25$^a$& $1.79$ GeV
    \end{tabular}
    \caption{Decays considered in the low-energy analysis, the WCs probed using the respective theory inputs 
    and the experimental upper limits  at 90 \%  \text{C.L.} on the branching ratios for $M_N = \SI{1}{\GeV}$. The last column shows the maximum value of  $M_N$ that is kinematically accessible. 
    Note that only specific flavor combinations of the WCs are probed, as indicated  by the brackets where $\{ \} ([])$ denote (anti-)symmetrization, in addition to the general property (\ref{eq:sym}). For charm decays see Secs.~\ref{sec:charm} and \ref{sec:charm_decays}. $^a$Lifetime constraint \cite{Hiller:2025zgr}.}
        \label{tab:low-input}
\end{table}

\begin{table}[]
    \centering
    \begin{tabular}{cc| cc | cc}
      \multicolumn{2}{c|}{$ B^+ \to p +  N $  \cite{BaBar:2023dtq}}  & \multicolumn{2}{c|}{$B^0 \to \Lambda + N $ \cite{Belle:2021gmc} } & \multicolumn{2}{c}{$\Xi^- \to \pi^- + \bar N $  \cite{BESIII:2025sfl} }  \\ \midrule 
      $\vert C^{uddN}_{131 } \vert \, / \, \Lambda^2 \leq $ &$\SI{0.70}{\TeV}^{-2} $  & $\vert C^{uddN}_{1\{23\}}\vert  \, / \, \Lambda^2 \leq$ &$ \SI{0.15}{\TeV}^{-2} $& $\vert C^{uddN}_{122} \vert \, / \, \Lambda^2 \leq$ & $\SI{0.026}{\TeV}^{-2}$\\
      $\vert C^{uddN}_{113} \vert \, / \, \Lambda^2  \leq$ & $\SI{1.52}{\TeV}^{-2} $ &\multicolumn{2}{c|}{$–$ } & $\vert C^{qqdN}_{\{12\}2}\vert \, / \, \Lambda^2 \leq $ & $\SI{0.014}{\TeV}^{-2}$   \\ 
    \end{tabular}
    \caption{Constraints on BNV four-fermion operators from data on $B$-meson and hyperon decays given in Tab.~\ref{tab:low-input} for $M_N = \SI{1}{GeV}$.}
    \label{tab:current_bounds}
\end{table}

\subsection{Target modes in charm \label{sec:charm}}

In Table \ref{tab:charmdecays} we give two-body charm hadron decays to $N$. 
Further final states with the same flavor content but different spin are implied. For instance, $\pi$ stands for $\pi, \rho, \pi \pi, ...$, and $K$ for $K, K^*, K \pi,...$,
although higher resonances and multi-bodies are heavier and reduce the mass range of $N$. Requiring $M_N \gtrsim 1$ GeV,
only charm baryon decays remain as targets with sufficient phase space for low energy searches, see also \cite{Fajfer:2020tqf}.

\begin{table}[]
    \centering
    \begin{tabular}{c| c| c }
    $cdd$ & $ csd, cds$  & $css$  \\
    \toprule 
    $\bar D^0 \to \Delta^0,n$  & $\bar D^0 \to \Lambda,\Sigma^0$ & $ \bar D^0 \to \Xi^0$\\
    $D^- \to \Delta^-$ & $D^- \to \Sigma^-$ & $D^- \to \Xi^-$\\
    $D_s^- \to \Sigma^-$ & $D_s^- \to \Xi^-$ & $D_s^- \to \Omega^-$ \\
    \hline
    $\Lambda_c \to \pi^+$ & $\Lambda_c \to K^+$ &  --\\
     $\Xi_c^0 \to K^0$  &   $\Xi_c^0 \to \pi^0,\eta^{(\prime)}, \phi$& $\Xi_c^0 \to \bar K^0 $\\
     --   &   $\Xi_c^+ \to \pi^+ $& $\Xi_c^+\to K^+ $\\
      --  & $\Omega_c \to K^0$ & $\Omega_c \to \eta^{(\prime)}, \phi$\\
    \end{tabular}
    \caption{Charm hadron decay modes, plus $N$ or $\bar N$ (omitted for brevity). For the mesons, we show $\bar c \to qq N$ decays, because the  baryon naming schemes are given in terms of quarks. For $M_N \gtrsim 1$GeV $D$-meson decays are kinematically forbidden, while charm baryon ones are possible.}
    \label{tab:charmdecays}
\end{table}

The flavor content can be realized with the Wilson coefficients as follows:
\begin{align}
cdd: & \quad C^{uddN}_{211} , \quad C^{qqdN}_{\{12\}1}  \\
cds: & \quad C^{uddN}_{212} , \quad C^{qqdN}_{\{12\}2}\\
csd: &  \quad C^{uddN}_{221} ,\quad C^{qqdN}_{\{22\}1}\\
css: & \quad C^{uddN}_{222} , \quad C^{qqdN}_{\{22\}2} 
    \end{align}

To look for BNV-decays is complementary to the MET-collider searches, as the latter provide  rather flavor-inclusive constraints, while exclusive hadron decays probe specific combinations.
Rare charm baryon decays are analyzed in Sec.~\ref{sec:charm_decays}.

\subsection{Neutrons \label{sec:neutrons}}

In case the mass of the particle \(N\) lies in the narrow window that kinematically forbids proton decay but still allows for 
neutron decay into \(N\), neutron decay observables constrain the corresponding exotic 
branching fraction to $\text{Br}_X<10^
{-4}$~\cite{McKeen:2020zni}.

To estimate \(\mathrm{BR}(n\to \bar{N}\gamma)\), we follow and generalize the derivation of Ref.~\cite{Fornal:2018eol,Berezhiani:2018eds}. We obtain

\begin{align}
\Gamma(n\to \bar{N}\gamma)&=
\frac{g_n^{2}\,\alpha_{\mathrm{em}}}{32 \Lambda^4}\left(1-\frac{M_{N}^{2}}{m_n^{2}}\right)^{3} m_n
\left[
\left|\frac{2\,\alpha\, C^{qqdN}_{111}}{m_n-M_N}\right|^{2}
+
\left|\frac{\beta\, C^{uddN}_{111}}{m_n-M_N}\right|^{2}
\right], \\ \nonumber
&\simeq \frac{g_n^{2}\,\alpha_{\mathrm{em}}}{4 \Lambda^4} \frac{m_n-M_N}{m_n^2}
\left[
\left|2\,\alpha\, C^{qqdN}_{111}\right|^{2}
+
\left|\beta\, C^{uddN}_{111}\right|^{2}
\right], 
\end{align}
where \(g_n\) is the neutron \(g\)-factor, \(g_n\simeq -3.826\), and \(\alpha,\beta\) are the relevant three-quark matrix elements.  In the second line we expanded in $m_n-M_N \ll m_n$. For a numerical estimate we take \(\beta\simeq -\alpha\simeq 0.014~\mathrm{GeV}^3\) from Ref.~\cite{Aoki:2017puj}, neglecting the running of the Wilson coefficients.

As a benchmark, for \(m_N=938~\mathrm{MeV}\) one finds
\begin{equation}
\frac{\text{Br}(n\to \bar{N}\gamma)}{10^{-4}}\;\sim\;
\left(\frac{600\, \mathrm{TeV}}{\Lambda}\right)^{4}\;
\Big(|2\, C^{qqdN}_{111}|^2+| C^{uddN}_{111}|^2\Big)_{\rm}\,.
\end{equation}
Therefore, in the small kinematic window \(937.900~\mathrm{MeV}\lesssim M_N \lesssim 938.783~\mathrm{MeV}\), neutron-decay bounds dominate over collider constraints. Note that, as \(M_N\) approaches the lower bound, even if free proton decays are forbidden, binding effects within nuclei can kinematically allow transitions that spoil the stability of some nuclei. Strictly speaking, only \(M_N \gtrsim 937.992\,\mathrm{MeV}\) guarantees nuclear stability, including that of \({}^{9}\mathrm{Be}\), which yields the largest mass gap. Complementary constraints above that mass may also be obtained from \({}^{1}\mathrm{H}\to \bar{N}\nu_{e}\), proceeding via electron capture. See Refs.~\cite{Fornal:2018eol,Berezhiani:2018udo} for further discussion.

\section{High-$p_T$  Analysis}
\label{sec:highPt_theory}

We summarize the requisite theory frameworks for the production (Sec.~\ref{sec:prod}) and inclusive decay of $N$ (Sec.~\ref{sec:N_decay}).
As the singlet $N$ has a nonzero mass, it can also decay, specifically  through the same operators that produces them, Eqn.~\eqref{eqn:BNV_4F}.
We therefore consider multiple signatures: MET + jets, in which $N$ escapes the detector and leads to missing energy and production of multiple (un)tagged jets, a displaced vertex, in which the $N$ decays inside the detector after traveling a finite length and a multijet signature, in which the $N$ decays at the primary vertex and only hadronic jets remain.
To describe all signatures we need the inclusive production cross section $pp \to N(\bar N) +  X$, where $X$ denotes any hadronic final state, as well as the decay width $N(\bar N ) \to X$, which both depends on the WCs in Eqn.~\eqref{eq:L-BNV_4F} and the mass $M_N$ of $N$. 
We  assume the WCs in Eqn. \eqref{eq:L-BNV_4F} to be real-valued, which implies that amplitudes in which quarks and antiquarks are swapped, are identical.
We frequently use generically $N$ for both $N$ and $\bar N$,
as corresponding transitions,
for instance $N \to \bar u \bar d \bar d $ and $\bar N \to  u  d  d $ are equivalent after charge conjugation. 
We comment on possible combinatorial factors if $N$ is
Majorana~~\footnote{If $N$ is a Majorana fermion constraints from $n -\bar n$ and $\Lambda-\bar \Lambda$ oscillations need to be considered.}.
The recast of LHC data is performed in Sec.~\ref{sec:recast}.

\subsection{Production cross section \label{sec:prod}}

\begin{figure}
    \input{feynman/xsec.tex}
    \caption{Chiral amplitudes for the process $u_i d_j \to \bar N \bar d_k$ highlighting the different contractions of the indices.
    In the high-energy limit interference is only possible between diagrams with identical chiralities. 
    This allows for interference between $\mathcal{O}^{uddN}_{ijk}$ and $\mathcal{O}^{uddN}_{ikj}$, while for $\mathcal{O}^{qqdN}$ all interference terms vanish.
    Antiquark channels are related by charge conjugation and down-quark fusion channels are related by crossing.  
    The indicated factors of 2 for $C^{qqdN}_{\{ij\}k}$ arise from  expanding the doublets in $\mathcal{O}^{qqdN}_{ijk}$ into components, see  Eqn.~\ref{eqn:BNV_4F}.
     }
    \label{fig:BNV_diagramms}
\end{figure}
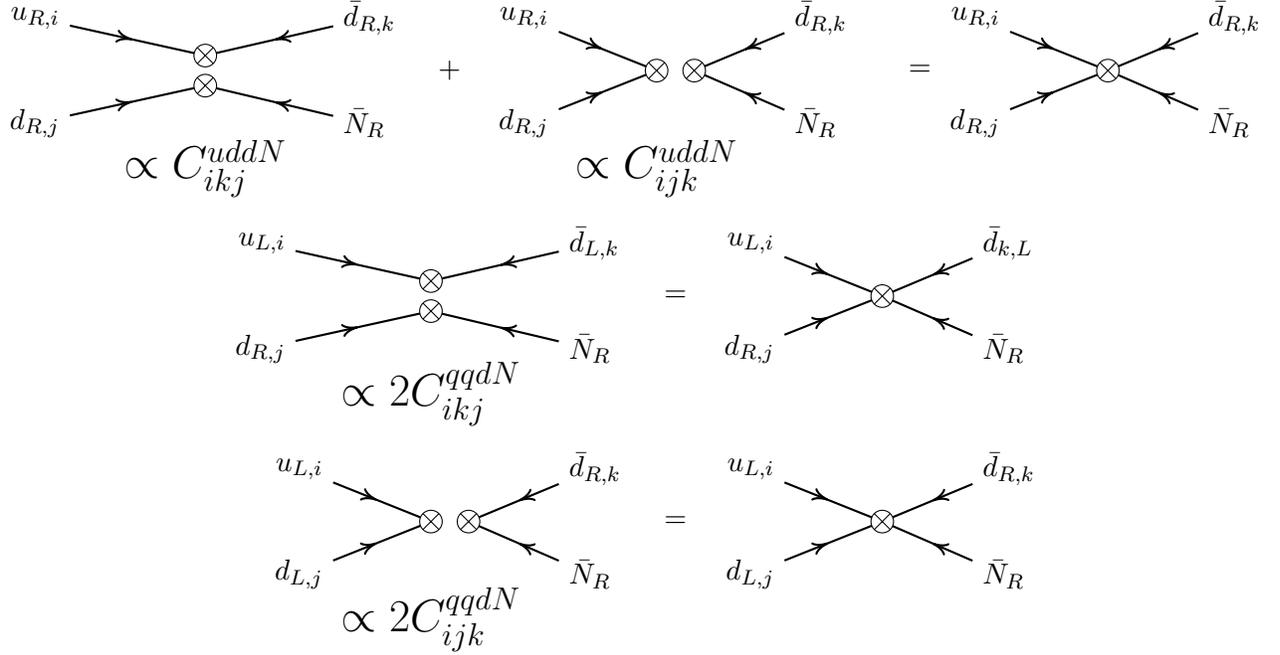

Multiple partonic processes contribute to the inclusive production of the singlet $ p p \to N (\bar N) X$, where $X$ denotes any hadronic final state.
Contributing initial states are 
$dd'$, $\bar d \bar d' $, $u d'$ and $ \bar u \bar d'$
, where $d$ ($u$) stands for a down (up)-type quark of any flavor except the top and we allow for different flavors by a prime. 
Initial states involving two up-quarks only contribute to BNV interactions involving charged leptons, which have been examined in Ref.~\cite{Dong:2011rh} for top quarks.
We calculate the cross section using \feyncalc with the Feynman rules generated by \feynrules, based on a self implemented UFO model \cite{Degrande:2011ua}.
In Fig.~\ref{fig:BNV_diagramms} we show the diagrams contributing to the matrix elements, with explicit chiralities of the final and initial states.
For the analytical consideration of the inclusive cross section, we neglect all fermion masses except for the mass $M_N$ of $N$, since the energy scales we are interested in are well separated from the quark masses.
The unpolarized cross section for $ u_i(p_1) d_j(p_2) \to \bar N(k_1) \bar d_k(k_2)$ reads 
\begin{equation}
    \label{eqn:xsec_ud}
    \begin{aligned}
     \frac{\mathrm{d} \hat  \sigma_{u_i d_j \to \bar{N} \bar{d}_k} }{\mathrm{d} \hat t } = \frac{1 }{ 96 \pi \Lambda^4} \Bigg\{&  \frac{\left(\hat s + \hat t \right)\left( \hat s + \hat t - M_N^2\right)}{\hat s^2 } \left(  4  {C^{qqdN}_{\{ik\}j}}^2   +{C^{uddN}_{ikj}}^2 \right) \\
     &+ \left( \frac{\hat s - M_N^2 }{\hat s} \right) \left( 4 {C^{qqdN}_{\{ij\}k}}^2 +{C^{uddN}_{ijk}}^2  \right) \\
     &+ 2  \frac{\left( \hat s + \hat t - M_N^2 \right)}{\hat s} C^{uddN}_{ijk} C^{uddN}_{ikj} \Bigg \} \Theta( \hat s + \hat t -M_N^2)
    \end{aligned}
\end{equation}

and the crossing-related cross section for $ d_i(p_1) d_j(p_2) \to \bar  N(k_1) \bar u_k(k_2)$ reads 
\begin{equation}
    \label{eqn:xsec_dd}
    \begin{aligned}
     \frac{\mathrm{d} \hat  \sigma_{d_i d_j \to \bar{N} \bar{u}_k} }{\mathrm{d} \hat t } = \frac{1 }{ 96 \pi \Lambda^4} \Bigg\{&  \frac{\hat t \left(\hat t - M_N^2\right) }{\hat s^2 }  \left( 4 {C^{qqdN}_{\{kj\}i}}^2 + {C^{uddN}_{kji}}^2 \right) \\
      &+ \frac{ (\hat s + \hat t ) ( \hat s + \hat t - M_N^2)}{\hat s^2} \left( 4 {C^{qqdN}_{\{ki\}j}}^2 + {C^{uddN}_{kij}}^2 \right)\\
     &+ \frac{2 \hat t  \left( \hat s + \hat t - M_N^2 \right)}{\hat s^2 } C^{uddN}_{kij} C^{uddN}_{kji}   \Bigg \}\Theta\left( \hat s + \hat t -M_N^2\right)
    \end{aligned}
\end{equation}
where $\hat s = (p_1 + p_2)^2, \hat t = ( p_1 - k_1)^2$.
Since $\hat t <0$, the $ud$ ($dd$)-annihilation has a positive (negative) interference term for same-sign WCs ($C^{uddN}_{ijk} \times C^{uddN}_{ikj} >0$).
 Unlike $\mathcal{O}^{uddN}$,  interference effects involving $\mathcal{O}^{qqdN}$ vanish in the high-energy limit, as they would require two 
quark mass insertions, see Fig.~\ref{fig:BNV_diagramms}, and  are hence irrelevant for the production of $N$.
The partonic production cross section can be obtained by integrating Eqn.~\eqref{eqn:xsec_dd} and Eqn.~\eqref{eqn:xsec_ud} for $\hat t \in (M_N^2 - \hat s ,0 )$, which leads to 
\begin{equation}
    \begin{aligned}
    \hat  \sigma_{u_i d_j \to \bar N \bar d_k} =& \frac{ \left( \hat s - M_N^2\right)^2}{576 \pi \Lambda^4 \hat s^2}\Bigg\{  \left( M_N^2 + 2 \hat s \right)  \left( 4 {C^{qqdN}_{\{ik\}j}}^2 +  {C^{uddN}_{ikj}}^2 \right) +6 \hat s  \left(4{C^{qqdN}_{\{ij\}k}}^2  + {C^{uddN}_{ijk}}^2\right)\\
         & +6 \hat s C^{uddN}_{kij} C^{uddN}_{kji}  \Bigg \} \Theta\left( \hat s -M_N^2\right)\,, \\
        \hat  \sigma_{d_i d_j \to \bar N \bar u_k} =& \frac{ \left( \hat s - M_N^2\right)^2}{576 \pi \Lambda^4 \hat s^2}\Bigg\{  \left( M_N^2 + 2 \hat s \right)  \left( 4 {C^{qqdN}_{\{ki\}j}}^2 + 4 {C^{qqdN}_{\{kj\}i}}^2 + {C^{uddN}_{kij}}^2  + {C^{uddN}_{kji}}^2\right)\\
         & -2 \left( \hat s - M_N^2 \right) C^{uddN}_{kij} C^{uddN}_{kji}  \Bigg \}\Theta\left( \hat s -M_N^2\right) \,.
    \end{aligned}
\end{equation}
Neglecting $M_N \ll  \sqrt{\hat s}$ the energy-enhancement of the BNV-contributions in the cross sections becomes apparent, i.e. $\hat \sigma \propto \frac{\hat s}{\Lambda^4}$.
At LO, following standard nomenclature,   $ \vec{E}_T^{\text{miss}} = -\vec{P}_T$, where $\vec{P}_T = (k^x_2,k_2^y)$ is the two-dimensional vector transverse to the z-direction. 
Therefore, missing transverse energy is given by $\emiss =| \vec{E}_T^{\text{miss}}| = P_T$, where 
\begin{equation}
\label{eqn:PTrelation}
\begin{aligned}
    P_T^2 &\equiv |\vec{P}_T|^2  =\frac{\hat t \left( M_N^2 - \hat s -\hat t\right) }{\hat s} \,.
\end{aligned}
\end{equation}
The latter Eqn.~(\ref{eqn:PTrelation}) can be inverted to obtain 
\begin{equation}
    \begin{aligned}
        \hat t_{\pm} &= -\frac{1}{2}  \left( \hat s - M_N^2 \pm \sqrt{ \left( \hat s - M_N^2\right)^2 - 4 \hat s P_T^2}  \right) \, , \\
        \left|\frac{\mathrm{d} \hat t_{\pm}}{\mathrm{d} P_T} \right|  &= \frac{2 P_T \hat s }{\sqrt{ \left( \hat s - M_N^2\right)^2 - 4 \hat s P_T^2}}\,,
    \end{aligned}
\end{equation}
which leads to the total differential $P_T$ cross section 
\begin{equation}
    \label{eqn:BNV_hadronxsec}
    \begin{aligned}
    \frac{ \mathrm{d}\sigma_{ p p \to N} }{\mathrm{d} P_T} =& 2 \sum_{\underset{\hat t =  \hat t_{\pm}}{i j k} } \int_{\tau_0}^1 \frac{\mathrm{d} \tau}{\tau} \frac{2 P_T s \tau  }{\sqrt{ \left( s \tau - M_N^2\right)^2 - 4 s \tau P_T^2}}\\
    &\times  \Big(  \mathcal{L}^{ud}_{i j}(\tau,\mu_F^2) 
     \frac{\mathrm{d} \hat  \sigma_{u_i d_j \to \bar  N \bar d_k} }{\mathrm{d} \hat t }
    +  \mathcal{L}^{dd}_{i j}(\tau,\mu_F^2) \frac{\mathrm{d} \hat  \sigma_{d_i d_j \to \bar N \bar u_k} }{\mathrm{d} \hat t } \Big) 
    \end{aligned}
\end{equation}
with $\tau = \hat s \, / \, s$, $\tau_0 = \left( P_T + \sqrt{ P_T^2 + M_N^2}\right)^2 \, / \, s$.
 The sum in Eqn.~\eqref{eqn:BNV_hadronxsec} runs over both branches of $\hat t $ as well as over all generation indices $i,j,k = 1,2,3$
with the exception of initial state top-quarks. It also includes final state quark flavors $k$, as jets are in general not tagged.
Furthermore we define the parton luminosity functions (PLFs) for $ud$ and $dd$ channels, at factorization scale $\mu_F$, by 
\begin{equation}
    \begin{aligned}
    \mathcal{L}^{ud}_{ij}(\tau,\mu_F^2) = \tau \int_{\tau}^1 \frac{\mathrm{d} x}{x} \left(f_{u_i}(x,\mu_F) f_{d_j}(\tau/x ,\mu_F)    +f_{\bar u_i}(x,\mu_F) f_{\bar d_j}(\tau/x ,\mu_F) \right) \, , \\
     \mathcal{L}^{dd}_{ij}(\tau,\mu_F^2) = \tau \frac{1}{1 + \delta_{ij}}\int_{\tau}^1 \frac{\mathrm{d} x}{x} \left(f_{d_i}(x,\mu_F) f_{d_j}(\tau/x ,\mu_F)    +f_{\bar d_i}(x,\mu_F) f_{\bar d_j}(\tau/x ,\mu_F) \right) \,,         \label{eq:PLF}     
    \end{aligned}
\end{equation}
which 
include contributions from quark-quark and antiquark-antiquark channels, as the parton level cross sections are identical.
The kinematic factor in the integral of Eqn.~\eqref{eqn:BNV_hadronxsec} accounts for the jacobian between $\hat t$ and $P_T$ and the factor $2$ accounts for the contributions 
from both initial state protons. 
Depending on the production process $u(d) d \to \bar N \bar{d}(u) $ or $\bar{u}(\bar d ) \bar{d} \to N d(u)$ we produce either an $N$ or $\bar N $ in the dirac case.
 Final states are not reconstructed and therefore all four channels are counted independently and summed inclusively, 
 which is why no additional factor in the majorana case appears. 
\begin{figure}
    \centering
    \includegraphics[width = 0.48 \textwidth]{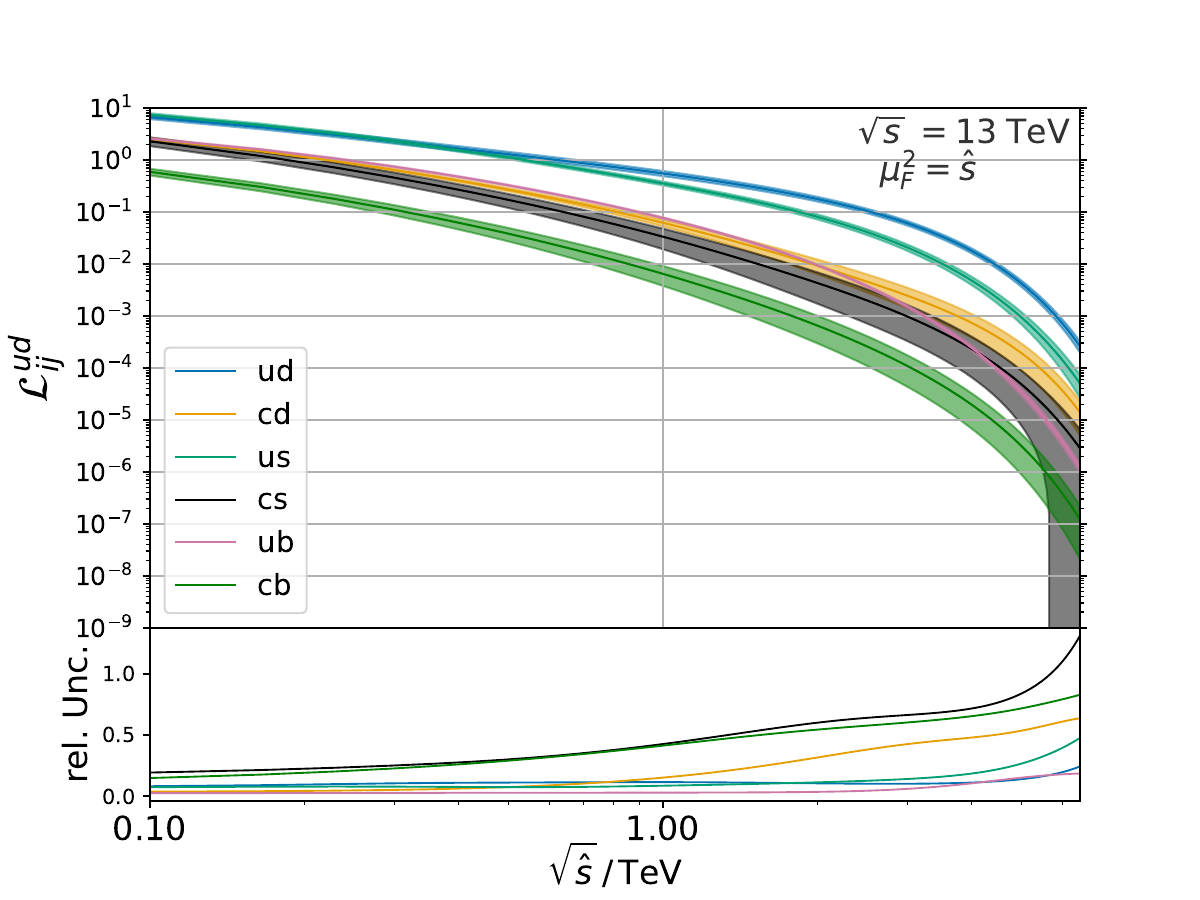}
    \includegraphics[width = 0.48 \textwidth]{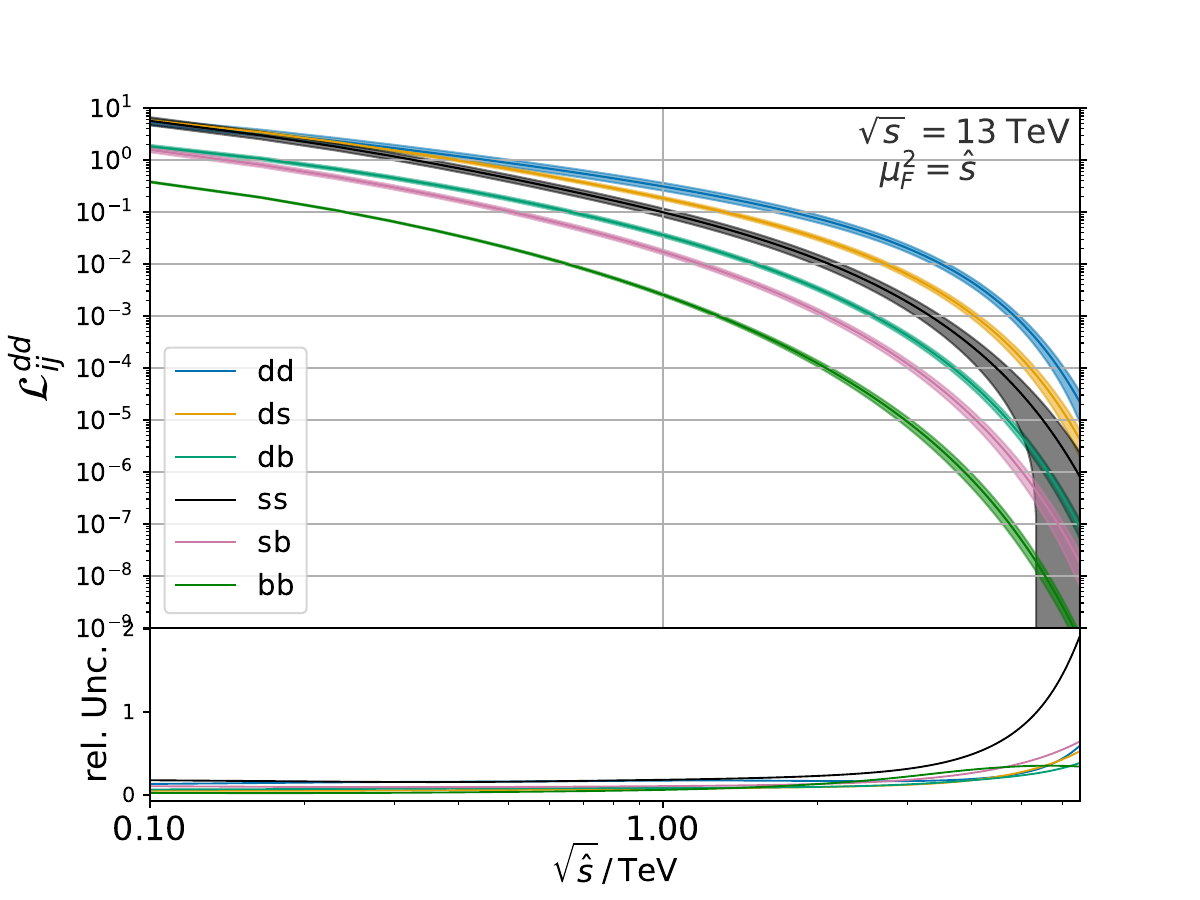}
    \label{fig:PLFs_BNV}
    \caption{The PLFs (\ref{eq:PLF}) that allow for BNV initial 
    quark-combinations. The results are given as summed over quark and antiquark combinations and the lower plot shows the relative uncertainties.
    We use the PDFset  \texttt{NNPDF40\_lo\_as\_01180} \cite{NNPDF:2021njg}  with the  factorization scale  $\mu_F = \sqrt{\hat s}$. Solid lines correspond to the central values and the envelope shows the $1 \sigma$-ranges for the PDF uncertainties. }
\end{figure}

\subsection{Inclusive decay of $N$ }
\label{sec:N_decay}

For the inclusive decay of $N$ to hadrons, we consider $N \to \bar u_i(k_1) \bar d_j(k_2) \bar d_k(k_3)$ and $\bar N \to  u_i(k_1)  d_j(k_2)  d_k(k_3)$, which are also related by charge conjugation. 
Therefore we only consider the partial differential width for $N \to \bar  u_i(k_1) \bar d_j(k_2) \bar d_k(k_3)$, which is given by 
\begin{equation}
    \label{eqn:diff_width}
    \begin{aligned}
        \mathrm{d}^2 \Gamma =& \frac{3 \mathrm{d}m_{1,2}^2 \mathrm{d}m_{2,3}^2}{256 \pi^3 \Lambda^4 M_N^3\left(\delta_{jk}+1 \right)} \\
        &\times \Bigg\{ \left(4 {C^{qqdN}_{\{ik\}j}}^2 + {C^{uddN}_{ikj}}^2\right) \left( M_N^2 - m_{1,2}^2 + m_j^2 - m_{2,3}^2 \right) \left( m_{1,2}^2 + m_{2,3}^2  -m_i^2 - m_k^2 \right) \\
            &\qquad-\left(4 {C^{qqdN}_{\{ij\}k}}^2 + {C^{uddN}_{ijk}}^2 \right) \left( M_N^2 - m_{1,2}^2 + m_k^{2} \right) \left( m_i^2 - m_{1,2}^2 + m_j^2 \right) \\
            &\qquad -2 C^{uddN}_{ijk} C^{uddN}_{ikj}(M_N^2 (m_i-m_{1,2}) (m_i+m_{1,2})+m_{12}^2 (m_{1,2}^2+m_{2,3}^2 -m_j^2 -m_i^2 - m_k^2 )+m_k^2 m_j^2)) \\
            &\qquad +   8 C^{qqdN}_{\{ij\}k} C^{qqdN}_{\{ik\}j}  m_j  m_k (M_N^2+m_i^2-m_{2,3}^2)  \\
            &\qquad -4 C^{qqdN}_{\{ij\}k} \left(2 C^{uddN}_{ijk}+C^{uddN}_{ikj} \right) m_i m_j (M_N^2-m_{1,2}^2+m_k^2)  \\
            &\qquad +4 C^{qqdN}_{\{ik\}j} \left(C^{uddN}_{ijk}+2 C^{uddN}_{ikj}\right)  m_i m_k (m_i^2-m_{1,2}^2-m_{2,3}^2+m_k^2)   \Bigg\}  
    \end{aligned}
\end{equation}
with $m_{1,2}^2 = (k_1 + k_2)^2, m_{2,3}^2 = (k_2 + k_3)^2$, $m_{i,j,k}$ the masses of the final state quarks and $\delta_{j,k}$ accounts for identical down quarks in the final state. 
Notably in Eqn.~\eqref{eqn:diff_width} we observe interference between different contractions of $C^{qqdN}$, as well as interference between $C^{uddN}$ and $C^{qqdN}$. This is consistent with Fig.~\ref{fig:BNV_diagramms}, as interference between $C^{qqdN}_{\{ij\}k}$ and $C^{qqdN}_{\{ik\}j}$ is always proportional to two down-quark masses needed to flip the chirality of the initial and final state quarks. 
Likewise, interference between $C^{qqdN}$ and $C^{uddN}$ requires one down-type and one up-type quark mass insertion.
Therefore all interference effects involving $\mathcal{O}^{qqdN}$ vanish in the high-energy limit, unlike interference between different insertions of $\mathcal{O}^{uddN}$.
Moreover, they are also subleading for the decay of $N$. For instance, for $M_N=1 \, \text{GeV}$, they come with suppression
factors of light masses, at most $m_s^2$ and $m_s m_u$, and until  $M_N <5 \, \text{GeV}$, at most $m_s^2, m_d m_b$ and $m_s m_c$.
In the limit of vanishing down quark masses ($m_j = m_k = 0$) we can integrate the width analytically and obtain, using $x = m_i \, / \, M_N$,
 and consistent with \cite{Dong:2011rh},
\begin{equation}
    \label{eqn:BNV_partial_width}
    \begin{aligned}
    \Gamma(N \to \bar u_i \bar d_j \bar d_k) =& \frac{ M_N^5}{1024 \pi^3 \Lambda^4\left(\delta_{jk}+1 \right) } \left( 4  \left({C^{qqdN}_{\{ij\}k}}^2 +  {C^{qqdN}_{\{ik\}j}}^2 \right) + \left( {C^{uddN}_{ijk}}^2 + {C^{uddN}_{ikj}}^2  +C^{uddN}_{ijk} C^{uddN}_{ikj} \right) \right) \\
                                & \times \left( 1 - x^8 + 8 \left( x^6 - x^2 \right) - 24 x^4 \log x \right) \\
                                \approx&   \frac{ M_N^5}{1024 \pi^3 \Lambda^4\left(\delta_{jk}+1 \right)} \left( 4  \left({C^{qqdN}_{\{ij\}k}}^2 +  {C^{qqdN}_{\{ik\}j}}^2 \right) + \left( {C^{uddN}_{ijk}}^2 + {C^{uddN}_{ikj}}^2  +C^{uddN}_{ijk} C^{uddN}_{ikj} \right) \right) \, , 
    \end{aligned}
\end{equation}
where the last line holds in the limit of all quark masses vanishing. 
In reality, near the kinematic threshold, the fermion will decay to pairs of a meson $\mathcal{M}$ and a baryon $\mathcal{B}$, i.e. $N \to \mathcal{B} + \mathcal{M}$ and not to free quarks.
To account for the difference in the masses on partonic and hadronic level, we introduce an additional constraint to the decay of $N$, where the decay is only possible if a hadronic channel exist, i.e. $M_N \geq M_{\mathcal{B}} + M_{\mathcal{M}}$. 
The explicit hadronic channel, in which $N$ decays into, does not matter, as we are only interested in the inclusive width and if a possible channel exists.
The total mean decay width is then given, by summing over all possible combinations consistent with kinematic constraints, by 
\begin{equation}
    \label{eqn:BNV_total_width}
      \Gamma_N = \xi_N \sum_{ijk} \Gamma(N \to \bar{u}_i \bar{d}_j \bar{d}_k) \Theta\left(M_N - (M_{\mathcal{B}} + M_{\mathcal{M}})\right)\,,
\end{equation}
with a factor $\xi_N = 1(2)$ for the dirac (majorana) case, accounting for the decay of $N$ to antiquarks and quarks in the majorana case. 
In the dirac case $\Gamma_{\bar{N}}$ equals $\Gamma_N$, with the corresponding final state antiquarks swapped to quarks. 
The Heaviside function accounts for the kinematics of the hadronic channels.

The decay of $N$ plays a significant role in the following analysis, as the experimental signature changes, depending on the lifetime of $N$.
To study this effect we consider the total width, given in Eqn.~\eqref{eqn:BNV_total_width} and the boost $\gamma_N \beta_N$ of the produced particle $N$, which contribute to the decay length \cite{Nemevsek:2023hwx} given by 
\begin{equation}
\label{eqn:decay_length}
    L_N(P_T,\eta) = \frac{\gamma_N \beta_N }{\Gamma_N} = \frac{P_T}{M_N \Gamma_N} \sqrt{1 +  \left(1 + \frac{M_N^2}{P_T^2} \right) \sinh^2 \eta} \,,
\end{equation}
which explicitly depends on the $P_T$ and $\eta$ of $N$. 
The differential number of events $n$ decaying per length interval $\mathrm{d}l$ are distributed according to  
\begin{equation}
    \label{eqn:decay_law}
    \frac{\mathrm{d}^3 n}{\mathrm{d}P_T \mathrm{d} \eta  \mathrm{d} l} = \mathcal{L} \frac{\mathrm{d}^2 \sigma }{\mathrm{d} P_T \mathrm{d} \eta} \frac{\exp( -l/L_N(P_T,\eta))}{L_N(P_T,\eta)} \, 
\end{equation} 
where $\mathcal{L}$ is the luminosity and $ \mathrm{d} \sigma$ is the total production cross section.
The total number of events within $l \in ( l_0,l_1)$, is then obtained by integrating Eqn.~\eqref{eqn:decay_law}
\begin{equation}
    \label{eqn:Nevents}
    \begin{aligned}
       n\left( l_0,l_1 \right) =& \int_{l_0}^{l_1} \mathrm{d} l  \int \int \mathrm{d} P_T \mathrm{d}  \eta  \left(\frac{\mathrm{d}^3 n}{\mathrm{d}P_T \mathrm{d} \eta  \mathrm{d} l} \right) \\
    =&  -\mathcal{L}  \int  \mathrm{d} P_T \mathrm{d} \eta  \left( \frac{\mathrm{d}^2  \sigma }{\mathrm{d} P_T \mathrm{d} \eta }\right) \left( \exp( -l_1/L_N(P_T,\eta))  -  \exp( -l_0/L_N(P_T,\eta)) \right)  \, . 
    \end{aligned}
\end{equation}
Based on Eqn.~\eqref{eqn:Nevents} we can define different regions of interest by considering different values of $l_0,l_1$.
Using the geometry of general purpose detectors at the LHC, we consider $n^{\text{MET}} $, events which escape the detector, $n^{\text{DV}} $, events with a DV within the inner detector, $n^{\text{Prompt}}$, events decaying at the primary vertex, and $n^0$ the total number of events.
The respective values of $l_0,l_1$ are summarized in Tab.~\ref{tab:ATLAS_regions}.

\begin{table}[h]
    \centering
    \begin{tabular}{c c c c c}
        &$ n^{\text{MET}}  $& $n^{\text{DV}} $ & $n^{\text{Prompt}}$ &$n^0$ \\ \toprule
  $l_0$ &$\SI{5}{\m }      $& $\SI{0.1}{\mm }$ & $0      $ &$0 $  \\
  $l_1$ &$\infty           $&$\SI{1}{\m}     $ &  $\SI{0.1}{\mm }     $ & $\infty$
    \end{tabular}
    \caption{The definition of the number of events, with a decay length  $l_0 \leq L_N \leq l_1$, based on the geometry of general purpose LHC detectors. }
    \label{tab:ATLAS_regions}
\end{table}

\begin{figure}
    \centering
    \includegraphics[width = 0.7\textwidth ]{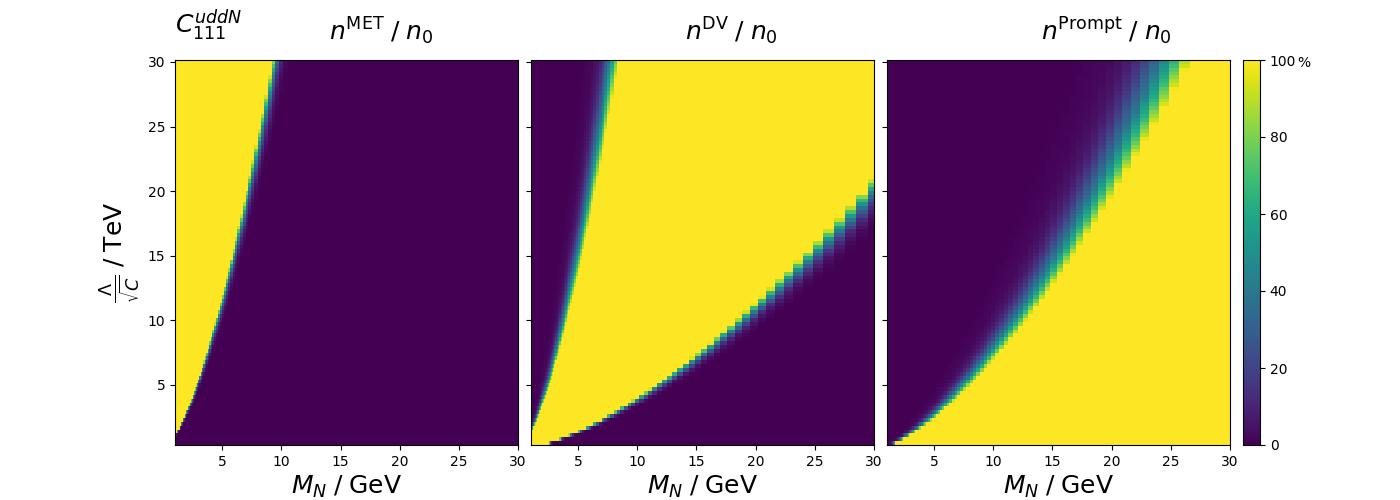}
    \includegraphics[width = 0.7\textwidth ]{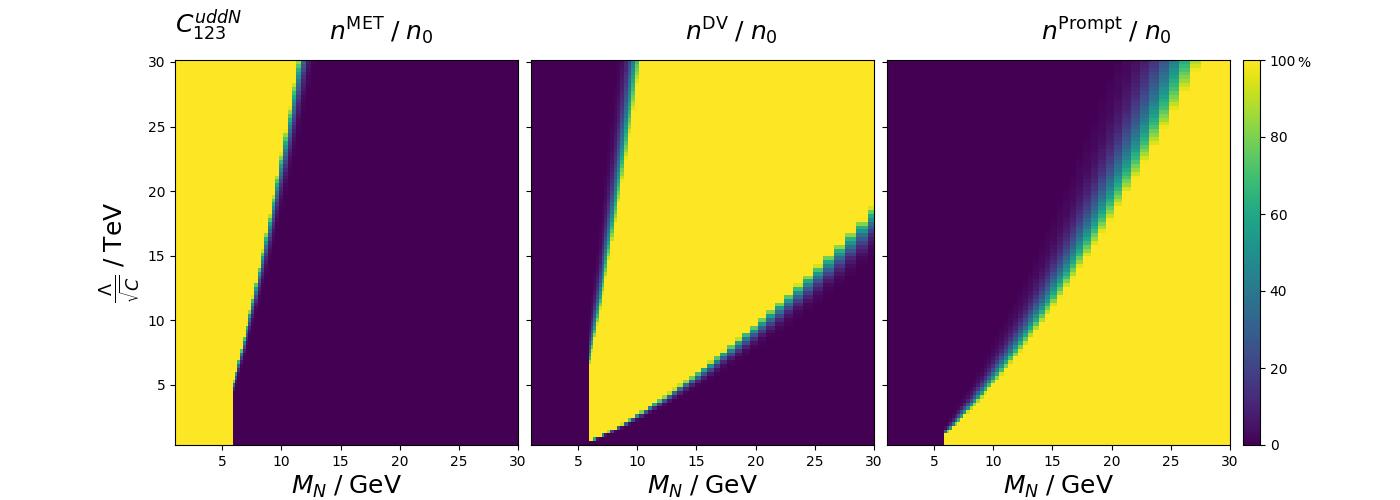}
    \includegraphics[width = 0.7\textwidth ]{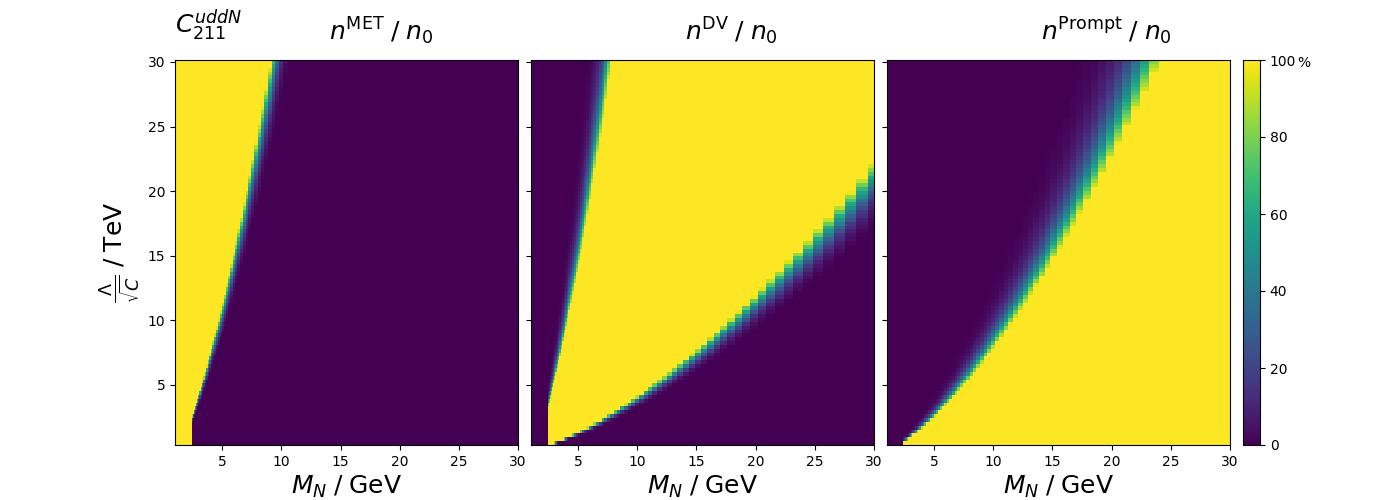}
    \includegraphics[width = 0.7\textwidth ]{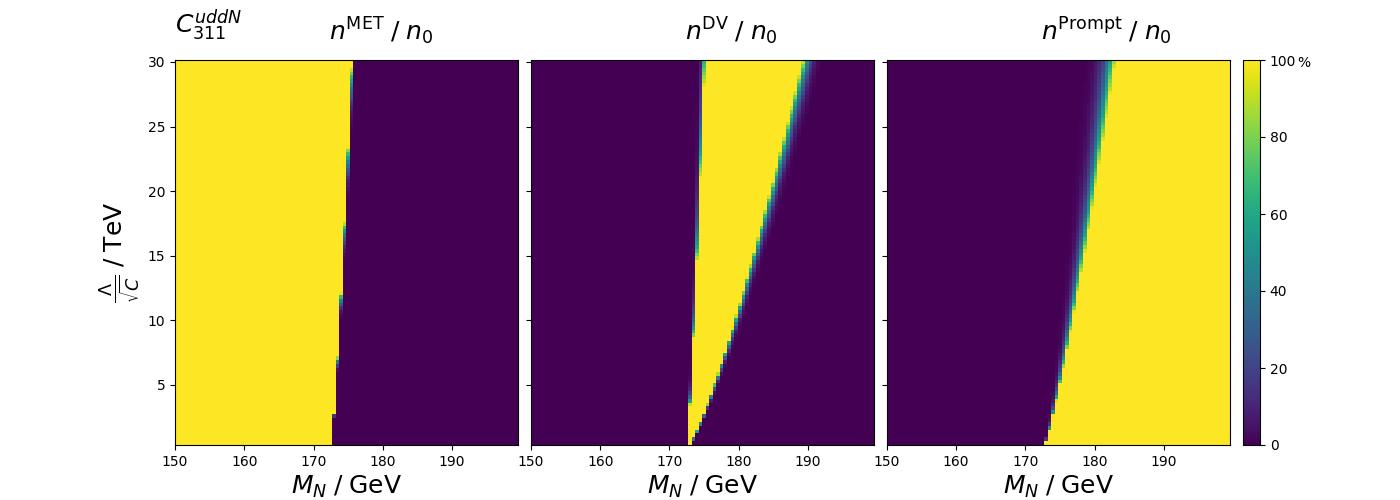}
    \caption{Number of events in the $ (M_N,\Lambda \, / \, \sqrt{C})$-plane with signature of MET (left), a DV (middle) and a prompt decay (right) of the singlet  $N$, relative to the total number of produced particles $n_0$ for flavors $C_{111}^{uddN}$,  $C_{123}^{uddN}$,
     $C_{211}^{uddN}$, and, with a top,  $C_{311}^{uddN}$ (from top to bottom).
    The ranges of $M_N$ are shown close to the respective kinematic cutoffs given by the decay products, see Eqn.~\eqref{eqn:BNV_total_width}.}
     \label{fig:decay_plots}
\end{figure}

We explicitly estimate Eqn.~\eqref{eqn:Nevents} by generating events using \mg  and applying a cut on particle level, that considers all particles with a decay length $ l_0 \leq L_N \leq l_1$, depending on the region defined in Tab.~\ref{tab:ATLAS_regions}.
In Fig. \ref{fig:decay_plots} we show the different $n^{I} \, / \, n_0$, normalized to the total number of events, in the $ (M_N,\Lambda \, / \, \sqrt{C})$-plane for decays to light quarks ($C^{uddN}_{111}$), one involving $u$-, $s$ and $b$-quark ($C^{uddN}_{123}$), one involving $c$-quarks and two $d$-quarks ( $C^{uddN}_{211}$) and one involving one $t$-quark and two light quarks ($C^{uddN}_{311}$).
This highlights the meaningful different cases depending on the masses of the final state quarks. Results for $C^{qqdN}_{\{ij\}k}$ are analogous, as can be seen in Eqn.~\eqref{eqn:BNV_partial_width}.

\subsection{Recast of \highPt~ observables \label{sec:recast}}%
To derive constraints on WCs we recast multiple searches by the ATLAS and CMS collaboration, which are summarized in Tab.~\ref{tab:BNV_obs}.
In addition to the MET observables we employ multijet searches. The latter are relevant as the $N$, depending on its mass, can decay  through the same
4-fermion operators that produces it,  which leads to jet signatures if the decay length is sufficiently short.
\begin{table}[h]
    \centering
    \begin{tabular}{c c c c }
      Process & Observable & $\mathcal{L}_{\text{int}}$ & Ref.  \\
      \toprule
       MET + jets  & $\frac{\text{d}\sigma}{\text{d}P_{T}^Z}$ & 140 fb$^{-1}$ & \cite{ATLAS:2024vqf} \\
     MET + top & $\frac{\text{d}\sigma}{\text{d}E_{T}^{\text{miss}} }$ &  139 fb$^{-1}$ & \cite{ATLAS:2020yzc} \\
      MET + bjet & $\frac{\text{d}\sigma}{\text{d}P_{T}^{\text{miss}} }$ &  138 fb$^{-1}$ & \cite{CMS:2024rkj} \\
     Multijets & $\frac{\text{d}\sigma}{\text{d}m_{4j}} $ &  138 fb$^{-1}$ & \cite{CMS:2022usq} \\
    \end{tabular}
    \caption{Data sets and observables used in the analysis, together with the corresponding integrated luminosity $\mathcal{L}_{\text{int}}$. All data is taken at $\sqrt{s} = 13$ TeV.
}
    \label{tab:BNV_obs}
  \end{table} 
\subsubsection{MET and Multijet observables}
We implement the Lagrangian~\eqref{eqn:BNV_4F} using Feynrules as an UFO model, which is then read in by \texttt{MadGraph5\_aMC@NLO} \cite{Alwall:2011uj}.
To validate our model we compare numerical results with Ref.~\cite{Dong:2011rh}, where cross sections for BNV processes including top quarks and charged leptons are analyzed.
We analyze all results obtained within the \texttt{Pythia8} framework, where we cluster jets with the anti-$k_T$ algorithm with a jet radius $\Delta R = 0.4$ using the program \texttt{fastJet} \cite{Cacciari:2011ma}
and factorization and renormalization scales are chosen as $\mu_R = \mu_F = \frac{1}{2}\sum_i P_{T,i}$, where the sum is over all final states.  

\begin{figure}
    \centering
    \includegraphics[width=0.48\linewidth]{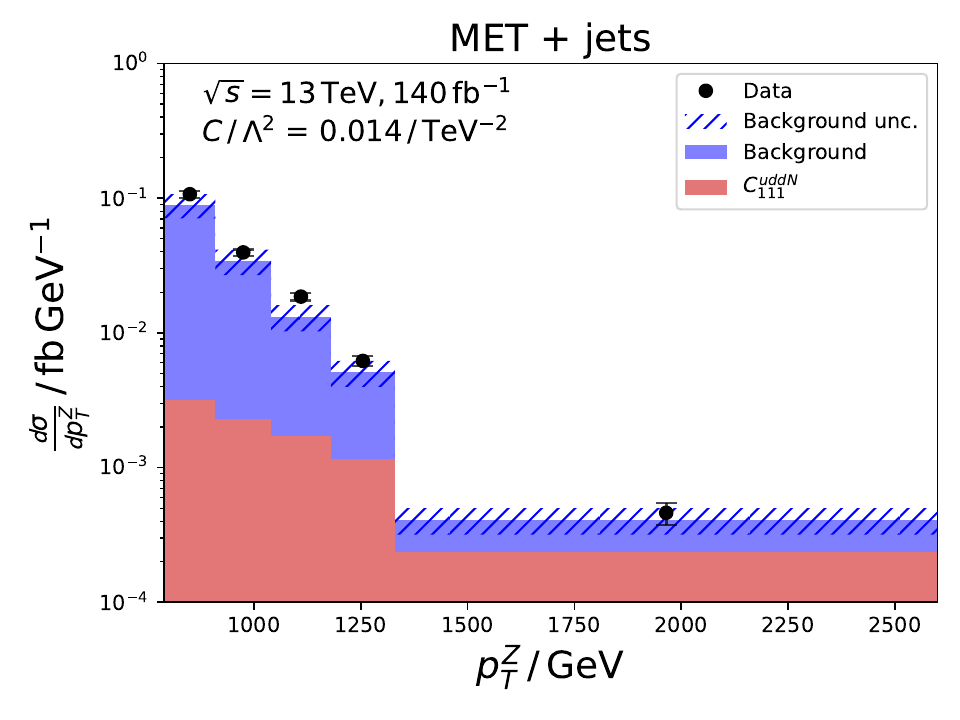}
    \includegraphics[width=0.48\linewidth]{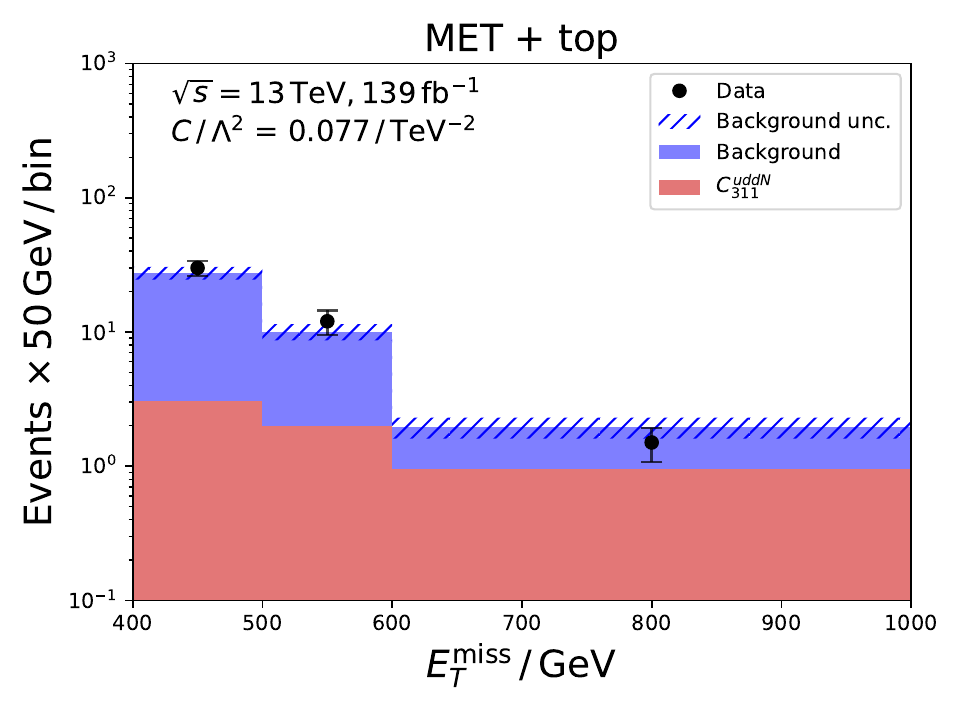}
    \includegraphics[width=0.48\linewidth]{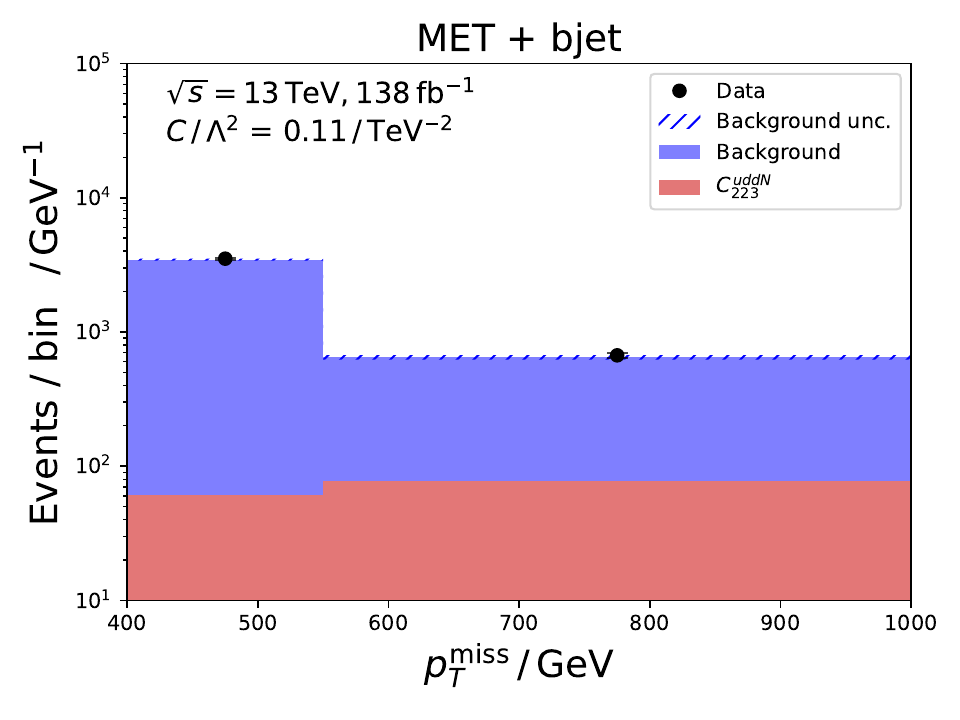}
    \caption{ Kinematic distributions of the MET observables listed in Tab.~\ref{tab:BNV_obs}. Shown are the highest bins of MET + jets (upper left), MET + b (lower plot) and MET + top (upper right).
    Black dots represent data and blue areas correspond to the background, both taken from the respective analyses \cite{ATLAS:2024vqf,ATLAS:2020yzc,CMS:2024rkj}.
    The red areas denote sample EFT signal contributions, derived based on the simulation pipeline described in Sec.~\ref{sec:recast}.
For each plot, we choose the specific notation, normalization and type of observable used in the respective analysis \footnote{We like to thank Monica D'Onofrio from the ATLAS collaboration, for confirming the normalization of the MET + top plot.}, to allow for easier comparison.}
    \label{fig:PTspectra}
\end{figure}

To validate the SM results, we generate the SM process $ p p \to \nu \bar \nu +\text{jets}$ for MET +jet and MET + bjet observables, $p p \to t \bar t$ and $ p p \to W + \text{jets}$ for the MET + top search and 
multijet events, with up to $4$ jets in the final state.
Afterwards we shower the events using \texttt{Pythia8} and merge the different jet multiplicities using MLM matching with $Q_{cut} = \SI{20}{\GeV}$.
 We reproduce the NLO QCD results from Ref.~\cite{ATLAS:2024vqf}, \cite{ATLAS:2020yzc} and \cite{CMS:2024rkj} by considering the LO process and up to one additional jet, where we find results for the different leading backgrounds to be within the uncertainties given by the ATLAS collaboration. 
Signal events are generated with the self implemented UFO model at LO with up to $2$ jets alongside the $N$ in the final state and are showered and merged following the procedure from the SM validation.
The constraints on WCs are extracted using the CLs method in the framework \texttt{pyhf} \cite{Heinrich:2021gyp}, where all WCs with $\text{CLs} \leq 0.1$ are excluded. 
The data, background as well as respective uncertainties are obtained from \cite{ATLAS:2024vqf,ATLAS:2020yzc,CMS:2022usq,CMS:2024rkj}.
We estimate theory uncertainties from PDF and scale variation for the MET + jets signal cross sections and find them to be of order $ \sim 30\%$.
Even though these relative uncertainties are sizeable, in absolute terms this is negligible compared to the background uncertainties and therefore we neglect all signal uncertainties.
To distinguish between MET and multijet events, we  check if the decay length fulfills $L_N \geq \SI{5}{\meter}$ for MET and $L_N \leq \SI{1}{\meter}$ 
for the multijet signature, where we use Eqn.~\eqref{eqn:diff_width} in combination with Eqn.~\eqref{eqn:decay_length}.
In Fig.~\ref{fig:PTspectra} we show the highest bins  of the kinematic distributions in the experimental analyses \cite{ATLAS:2024vqf,ATLAS:2020yzc,CMS:2024rkj}, summarized in Tab.~\ref{tab:BNV_obs}, 
and some representative signal distributions (red areas).
We focus on the high energy bins, as these are the most sensitive for EFT searches.

\subsubsection{DV sensitivities}
\label{sec:DV_sensitivities}
There are no DV searches that directly apply to our signature. 
Similar searches have been performed by ATLAS~\cite{ATLAS:2023oti}, however, explicitly looking for pair production of long-lived particles, which for our signature would only enter at $\mathcal{O}\left( \left(\frac{C}{\Lambda^2}\right)^4 \right)$ on cross section level.
Therefore we estimate the sensitivity for a DV search based on a zero background approach. 
In a realistic scenario we would expect a small SM background, which we leave to the experimental collaboration to take into account, as we only want a rough estimate of the sensitivity.
Similar EFT studies have been done for the FCC-ee in Ref. \cite{Bolton:2025tqw}.
See also \cite{Bittar:2024fau} for theory works on DV  in UV-mediator model 1 with flavorful Majorana singlets.
We produce event samples, as described in the previous section, where now we check if the decay length fulfills $\SI{1}{\meter}\geq L_N \geq \SI{0.1}{\mm}$ for $\mathcal{L}^{\text{int}} = \SI{139}{\femto\barn }^{-1}$.
Additionally we apply the cuts given in Tab. \ref{tab:DV_cuts}, based on Ref.~\cite{ATLAS:2023oti}, and exclude regions in the ($M_N,\Lambda\, / \,\sqrt{C}$) parameter space with $N_{\text{signal}} > 2.3$, corresponding to $90 \%$ C.L. limits, given zero background.
\begin{table}
    \centering
    \begin{tabular}{c c  }
      \multicolumn{2}{c}{Jet selection}  \\
      \toprule
      $N_{\text{jet}} $ & $\geq 4$ \\
       $P_T^i  $  & $ \geq \SI{150}{GeV}$ \\
       $|\eta_{\text{j}}^i|$ & $\leq 2.4$
    \end{tabular}
    \caption{Kinematic cuts applied on the signal region of the DV sensitivities. The cuts are loosely adopted from Ref.~\cite{ATLAS:2023oti}. 
   $N_{\text{jet}}$ denotes the number of jets and $\eta^i_j ,P_T^i  $ denote the pseudorapidity and the transverse momentum, respectively, for the $i$-th leading jet, with $i \leq 4 $.  }
    \label{tab:DV_cuts}
\end{table}

\section{Collider Constraints and Correlations\label{sec:res}}

In this section we present results  based on the low-energy and \highPt~ observables introduced in Sec.~\ref{sec:low_energy_theory} and Sec.~\ref{sec:highPt_theory}, respectively. 
Constraints are obtained from LHC data taken at $\sqrt{s}=13 \, \text{TeV}$ and $(138-140)\,  \text{fb}^{-1}$, given in
Tab.~\ref{tab:BNV_obs}.

\subsection{Constraints from MET observables \label{sec:METlimits}}
\subsubsection{One-dimensional fits}
\begin{figure}
    \centering
    \includegraphics[width = 0.48 \textwidth]{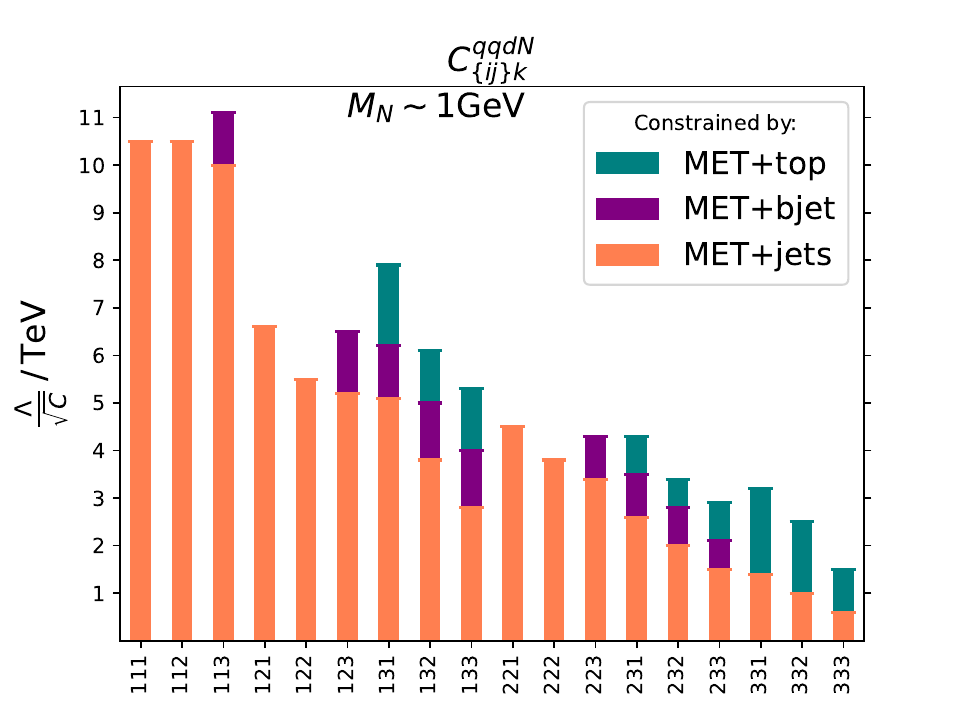}
    \includegraphics[width = 0.48 \textwidth]{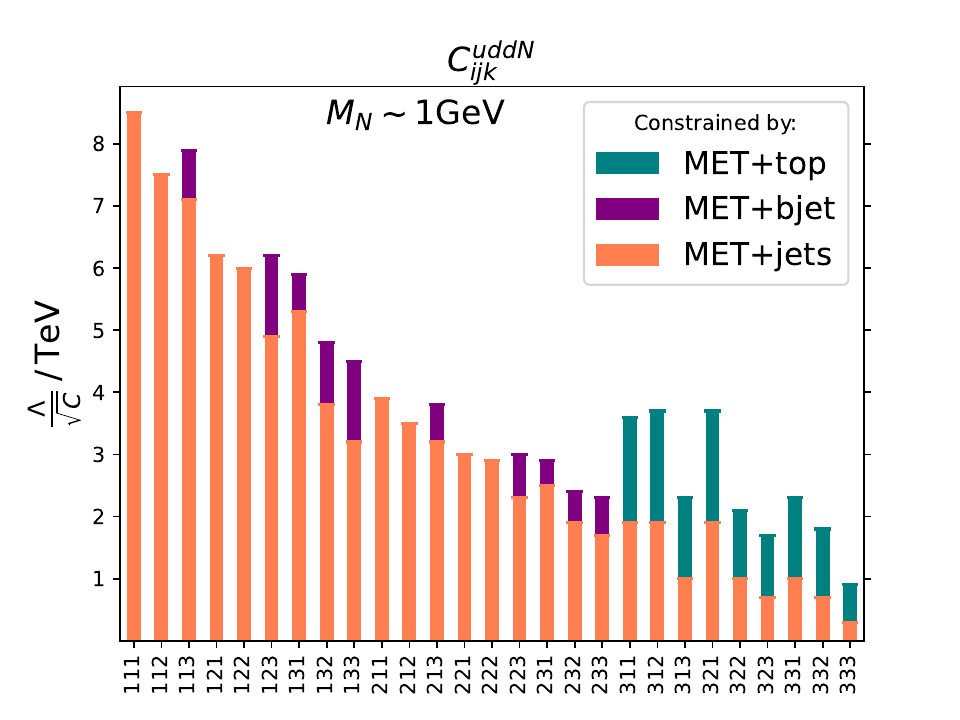}

    \caption{Constraints on $\Lambda \, / \, \sqrt{C}$ for BNV four-fermion operators $C^{qqdN}_{\{ij\}k}$(left) and $C^{uddN}_{ijk}$(right) derived from MET + jets, MET + top  and MET + bjet recasts, see Tab. \ref{tab:BNV_obs}.
    The bounds are given for $M_N = \SI{1}{\GeV}$, for which the $N$ particle  is always detector stable, see Sec.~\ref{sec:N_decay}. 
    However the bounds are approximately valid for larger ranges $M_N \gtrsim \SI{1}{\GeV} $, depending on the final states of $N$-decays, see Sec.~\ref{sec:N_decay} and Fig.~\ref{fig:CuddN_constraints}. Operators involving quarks from the heavier generations are subject to possible caveats due to the validity of the EFT expansion, see text. 
    \label{fig:NP_BNV}}
\end{figure}

In Fig. \ref{fig:NP_BNV} we show the one-dimensional constraints on $\Lambda \, / \sqrt{C}$ coming from MET+jets and MET + top searches for fixed $M_N = \SI{1}{\GeV}$.
Additionally the numerical values are listed in Tabs.~\ref{tab:bounds_numerical_CqqdN}  and \ref{tab:bounds_numerical_CuddN} in the App.~\ref{app:AddResults}. 
For this assignment of $M_N$ the particle is always at least detector stable for all WCs and above the proton mass.
We can see that bounds on lighter flavors are significantly stronger, as we expect the heavier quarks to be PDF suppressed.
The MET + top observable generally gives stronger constraints for transitions involving a top quark, due to a significantly reduced background relative to the fully inclusive analysis. Similarly for MET + bjet constraints are stronger on transitions involving bottom-quarks.
Operators involving only first and second generation quarks are constrained by the MET + jets observable, as this is fully inclusive and does not distinguish between different kinds of jets.
 The overall strongest bound with $\sim 11 \, \text{TeV}$ is achieved for $(i,j,k) = (1,1,3)$ with the MET + bjet analysis, as this involves both the production through two light quarks, which means it is maximally PDF enhanced, and the tagging of one bjet, which significantly reduces the background. 

Overall $C_{\{ij\}k}^{qqdN}$ is slightly better constrained than $C_{ijk}^{uddN}$, due to a combinatorial factor from the symmetry in the first two indices, which also is evident in Eqn.~\eqref{eqn:BNV_4F} and Eqn.~\eqref{eqn:BNV_partial_width}. However, for $ j = k $ the difference between the bounds is slightly reduced as now the interference term in Eqns.~\eqref{eqn:xsec_ud} and~\eqref{eqn:xsec_dd} also contributes to the total cross section. 

WCs involving heavier generations are generally less constrained, which leads to potential caveats for the EFT expansion.
The highest bin of the MET + jets analysis \cite{ATLAS:2024vqf} lies between $1330 - 2600 \, \si{\GeV}$, which corresponds to an average energy of $\sim \SI{2}{\TeV}$ which we take as a lower cutoff for the robustness of the bounds. For the MET + top analysis \cite{ATLAS:2020yzc} the highest bin is inclusive, starting at $\SI{600}{\GeV}$. Here  we use a cutoff of $\sim \SI{1}{\TeV}$. Similarly for the MET + bjet analysis, the highest bin lies between $550 - 1000 \, \si{\GeV}$, where we also take a cutoff of $\sim \SI{1}{\TeV}$.   The rather weak bounds obtained from the numerical  calculation that are below these cut-offs are marked in Tabs. \ref{tab:bounds_numerical_CuddN} and \ref{tab:bounds_numerical_CqqdN} with an $*$.
Nevertheless, there are scenarios where the EFT bounds may still hold even in these cases, as the effective energy scale $E$ 
for the EFT expansion can be as high as $E\sim\frac{4\pi\Lambda}{\sqrt{C}}$ for four-fermion operators~\cite{Gavela:2016bzc}.

Additional bounds from MET + bjet or MET + top on processes not involving explicit bottom- or top-quarks, due to misstaging, are found to be negligible compared to the fully inclusive ones and are therefore not considered in this work.

\subsubsection{Two-dimensional fits}
\begin{figure}
    \centering
    \includegraphics[width = 0.3\textwidth]{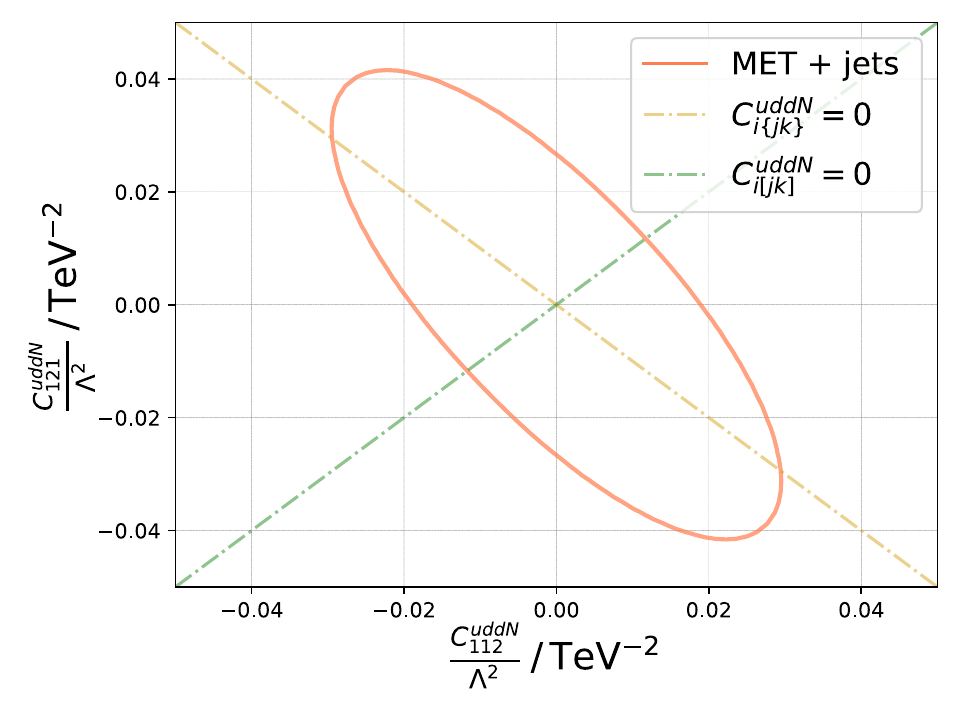}
    \includegraphics[width = 0.3\textwidth]{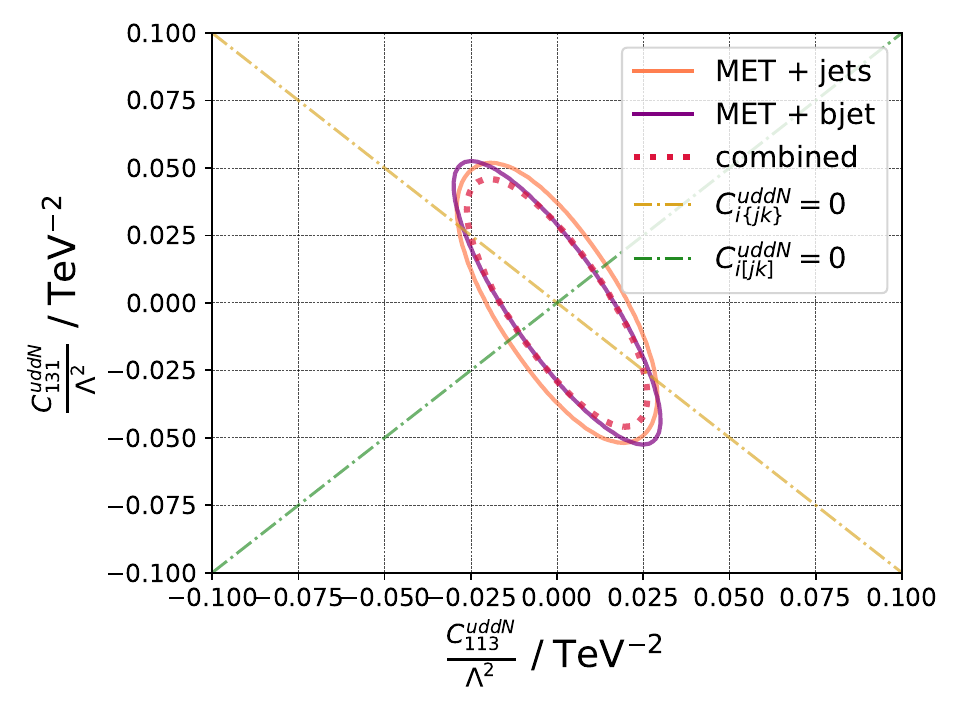}
    \includegraphics[width = 0.3\textwidth]{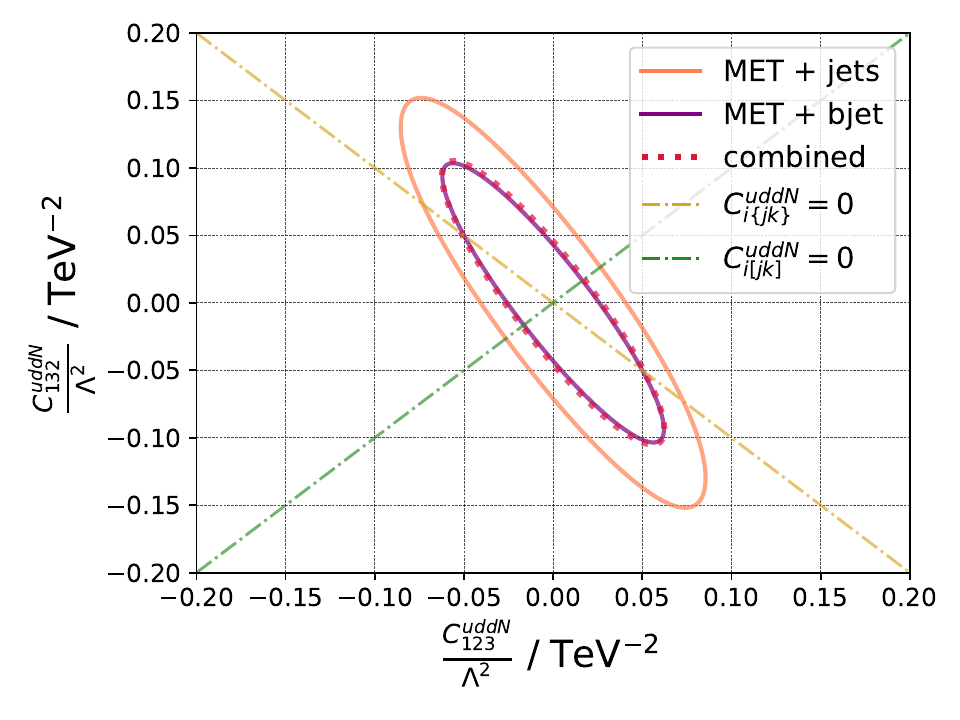}
    \includegraphics[width = 0.3\textwidth]{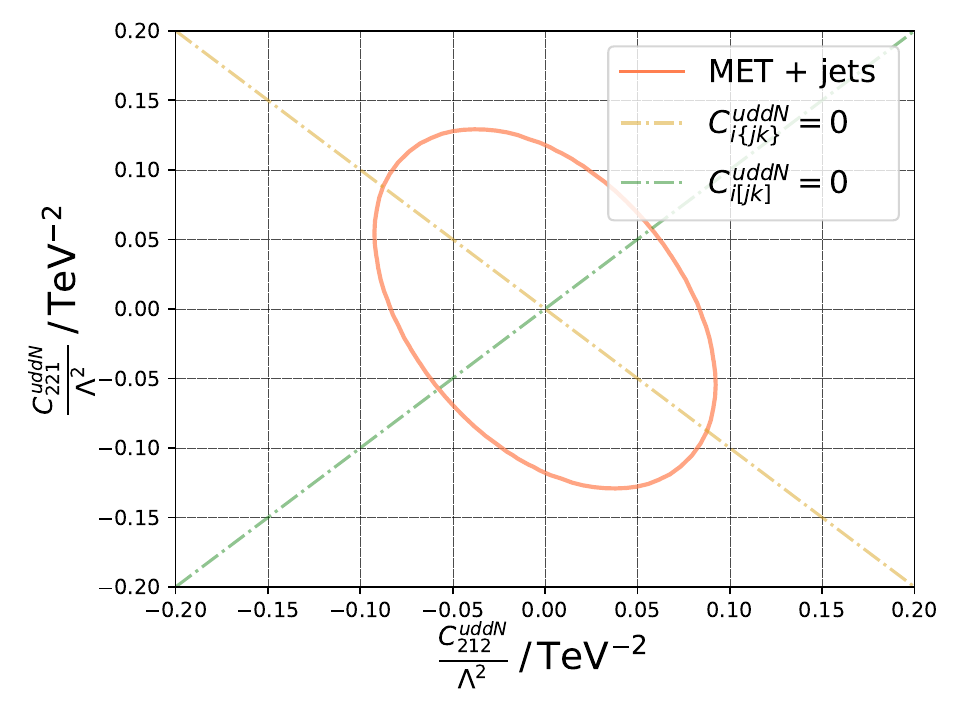}
    \includegraphics[width = 0.3\textwidth]{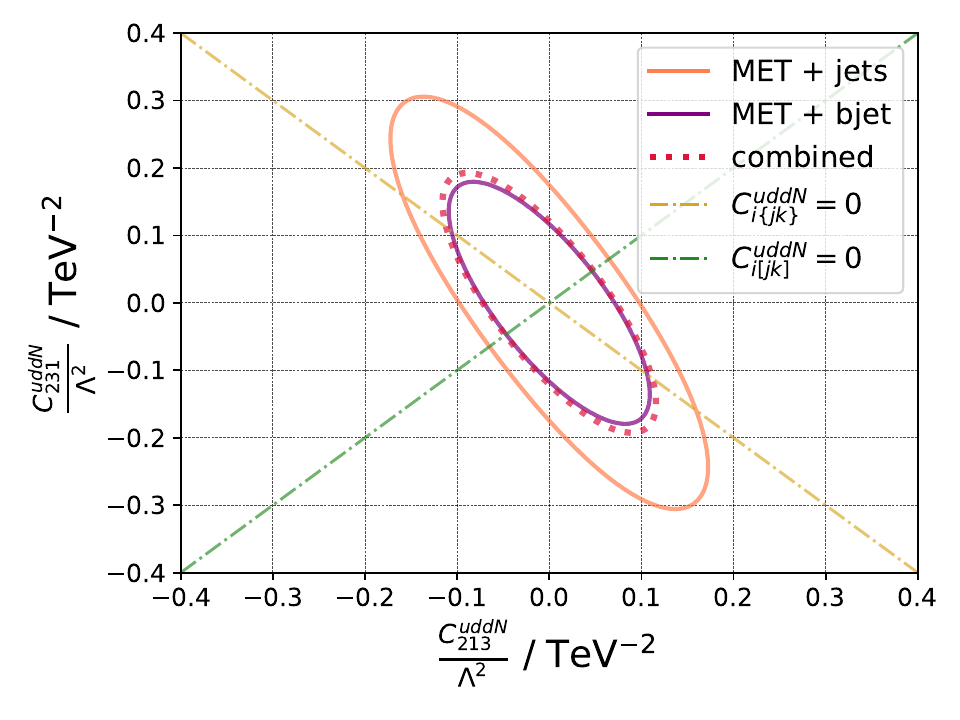}
    \includegraphics[width = 0.3\textwidth]{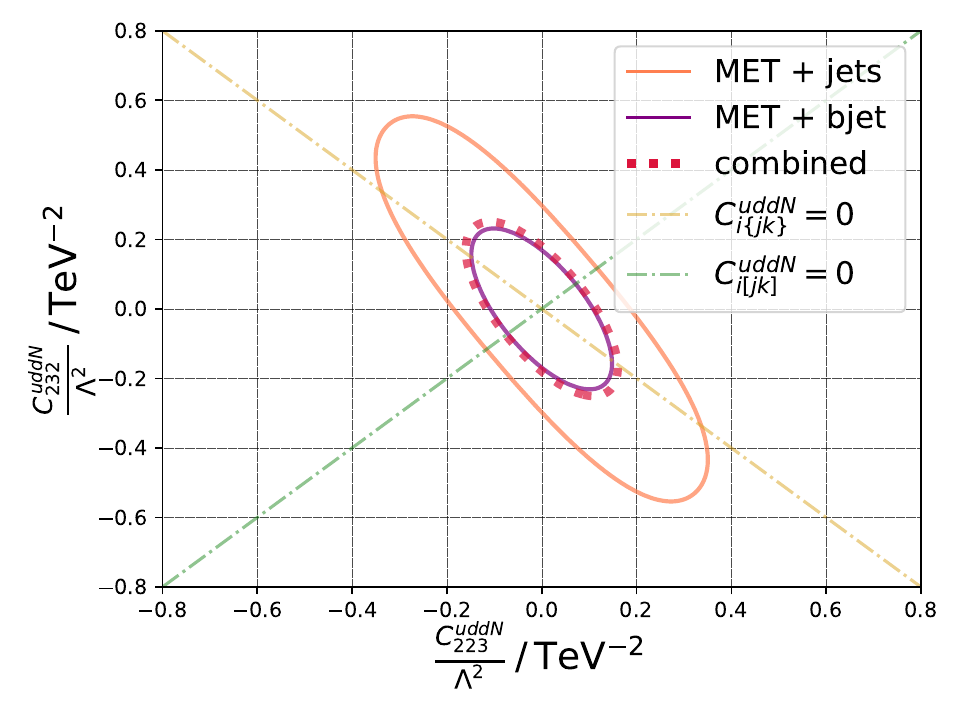}
    \includegraphics[width = 0.3\textwidth]{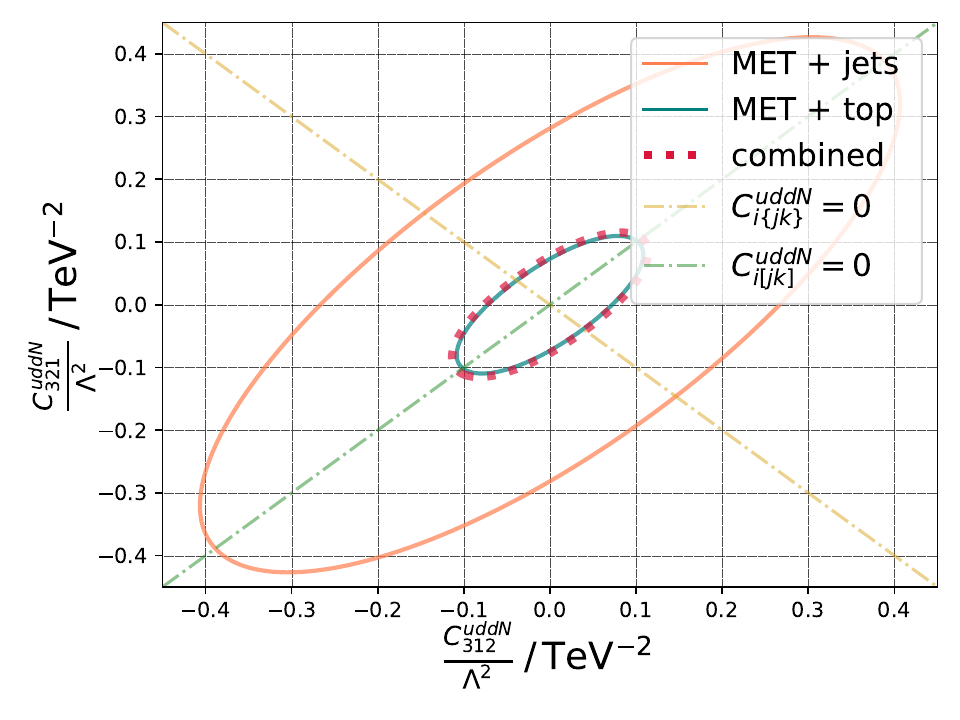}
    \includegraphics[width = 0.3\textwidth]{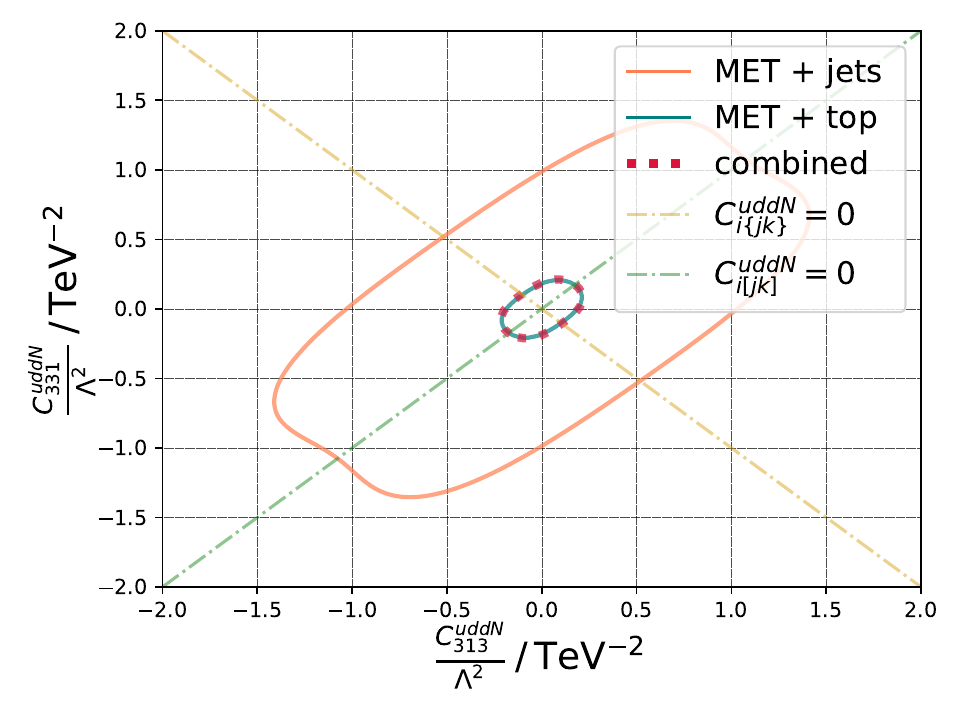}
    \includegraphics[width = 0.3\textwidth]{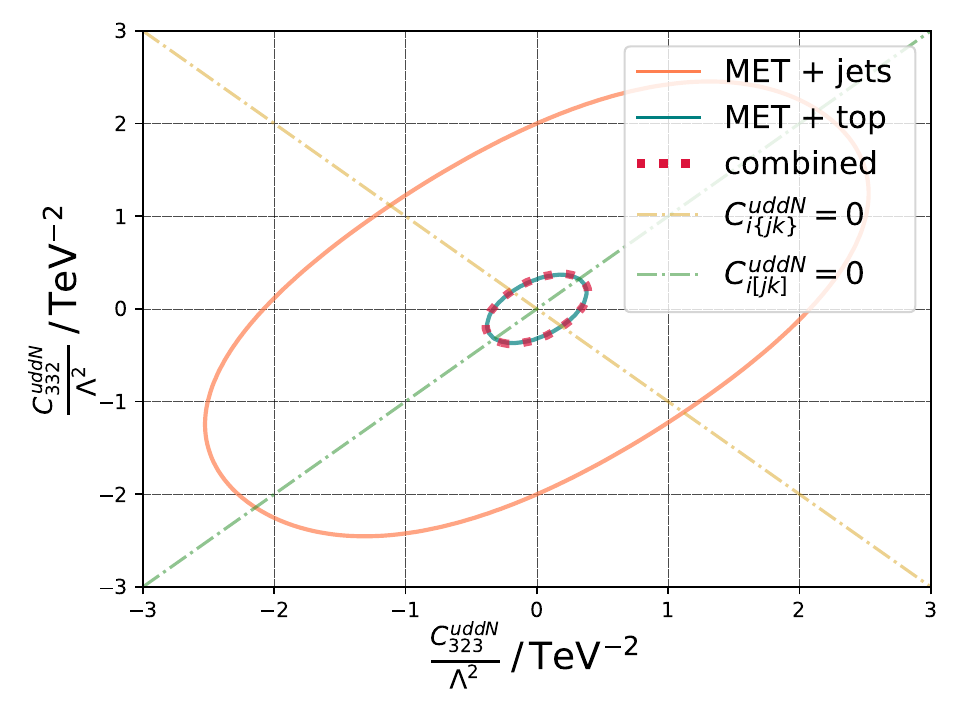}
    \caption{The $90 \% $ exclusion limits for the two-dimensional fits of different combinations of $C^{uddN}_{ijk}\, , j \neq k$, which allows for interference effects.
Shown are limits from MET + jets, MET + top ($i = 3$), MET + bjet ($ j,k = 3$) and the possible combination for $i = 1,2,3$ in the upper, middle and lower row, respectively. 
The green (yellow) dash-dotted lines show the points where $C^{uddN}_{i[jk]}$($C^{uddN}_{i\{jk\}}$) vanishes. }
    \label{fig:2D_fits_BNV}
\end{figure}

As one can see from Eqns.~\eqref{eqn:xsec_ud} and \eqref{eqn:xsec_dd} there is a non-trivial interference term for the operator $\mathcal{O}^{uddN}_{ijk}$ with $j \neq k$.
Therefore we consider a two-dimensional fit for fixed $i$ and the corresponding interference terms induced by different combinations of $j \neq k$.
In Fig.~\ref{fig:2D_fits_BNV} we can see the correlations in the $(C^{uddN}_{ijk}, C^{uddN}_{ikj} )$-plane, where we show the constraints by MET + jets, MET + top and by MET + bjet, as well as the combinations of  observables. Furthermore we show the numerical values obtained on flavor symmetric and antisymmetric WCs in Tab.~\ref{tab:asym_bounds}. This is motivated by the properties of the UV mediators given in Tab.~\ref{tab:UV_mediators}
which can lead to cancellation or summation of the contributions, yielding stronger or weaker bounds, depending on the sign of the interference term. Bounds for $j=k$ are given in Fig.~\ref{fig:NP_BNV} and Tab.~\ref{tab:bounds_numerical_CuddN}.
 Recall from Sec.~\ref{sec:mix} that in the presence of more than one coupling at a time strong meson mixing bounds apply in the UV models.

\begin{table}[]
    \centering
    \begin{tabular}{c| c| c | c |c }
    $ijk$ & $ \frac{|C^{uddN}_{i[jk]}|}{\Lambda^2} \,/ \, \si{\TeV}^{-2} $ & $\frac{\Lambda}{\sqrt{|C^{uddN}_{i[jk]}|}}\,/ \, \si{\TeV}$ & $ \frac{|C^{uddN}_{i\{jk\}}|}{\Lambda^2} \,/ \, \si{\TeV}^{-2} $ & $\frac{\Lambda}{\sqrt{|C^{uddN}_{i\{jk\}}|}}\,/ \, \si{\TeV}$  \\
    \toprule 
    $112$ & $0.028$  & $ 6$    &$0.010    $&$ 9.9$  \\
    $113$ & $0.025 $ & $ 6.3$    &$0.011  $&$ 9.5$  \\
    $123$ & $0.048 $ & $ 4.6$    &$0.021    $&$ 6.9$  \\
    $212$ & $0.091 $ & $ 3.3$    &$0.056    $&$ 4.2 $ \\
    $213$ & $0.11 $ & $ 3.0$    &$0.048  $&$ 4.6$  \\
    $223$ & $0.16 $ & $ 2.5$     &$0.075    $&$ 3.7 $ \\
    $312$ & $0.040 $ & $ 5.0$    &$0.11    $&$ 3.0 $ \\
    $313$ & $0.097$  & $ 3.2$    &$0.18    $&$ 2.4 $ \\
    $323$ & $0.22$  & $ 2.1$     &$0.32    $&$ 1.8 $ \\
    \end{tabular}
    \caption{One-parameter limits obtained on the anti-symmetric (second and third column) and symmetric (last two columns) part  of $\mathcal{O}^{uddN}_{ijk} $ from projecting the two-dimensional fit in Fig.~\ref{fig:2D_fits_BNV}.
    The bounds are given at a fixed mass $M_N = \SI{1}{\GeV}$, for which $N$ is always detector stable, see Sec.~\ref{sec:N_decay}. 
    However the bounds are approximately valid for larger masses, depending on possible final states of the decay of $N$, 
    see Sec.~\ref{sec:N_decay} and Fig.~\ref{fig:CuddN_constraints}.     }
    \label{tab:asym_bounds}
\end{table}

Generically one expects the interference terms to cause a tilting of the ellipses in the 2D-analyses. 
This leads to an increase of the allowed regions, as shown in Fig.~\ref{fig:2D_fits_BNV}.
We also learn that 
 the gain from the combination of MET-observables is very small in the $(C^{uddN}_{113},C^{uddN}_{131})$- and $(C^{uddN}_{213},C^{uddN}_{231})$-plots, and negligible for all other flavor combinations, 
 as one observable is always the dominant one.

For $i = 1,2$ we see in Fig.~\ref{fig:2D_fits_BNV}, that ellipses are skewed towards opposite sign WCs, due to the $ud$-channel contributing the most to the cross section.
As can be seen in Eqn.~\eqref{eqn:xsec_ud} the $ud$-channel has a positive interference term for same sign WCs, which leads to a larger cross section for same sign WCs and therefore to less stringent constraints for opposite sign WCs. 
This effect is slightly  stronger for MET + bjet, as we now need one final state bottom quark, which suppresses contributions for the $dd$-channel given in~\eqref{eqn:xsec_dd} even more, which explains the slightly slimmer ellipses for $j = 1$.

For $i = 3 $ we see in Fig.~\ref{fig:2D_fits_BNV} that the ellipses are skewed towards same sign WCs, as now the $dd$-channel, given in Eqn.~\eqref{eqn:xsec_dd}, is the dominant contribution, since initial state top quarks are strongly suppressed.
This leads to less stringent constraints in the same sign region of the correlation plots. 

\subsection{Combined constraints on $\Lambda$ and $M_N$ }
\begin{figure}
    \centering
    \includegraphics[width = 0.31 \textwidth]{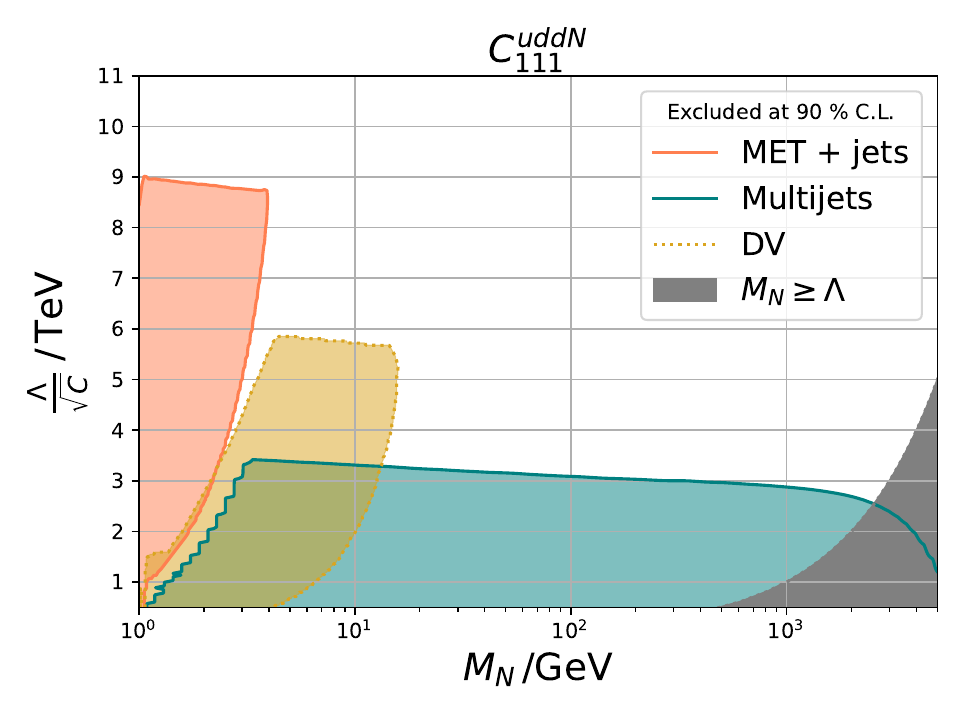}
    \includegraphics[width = 0.31 \textwidth]{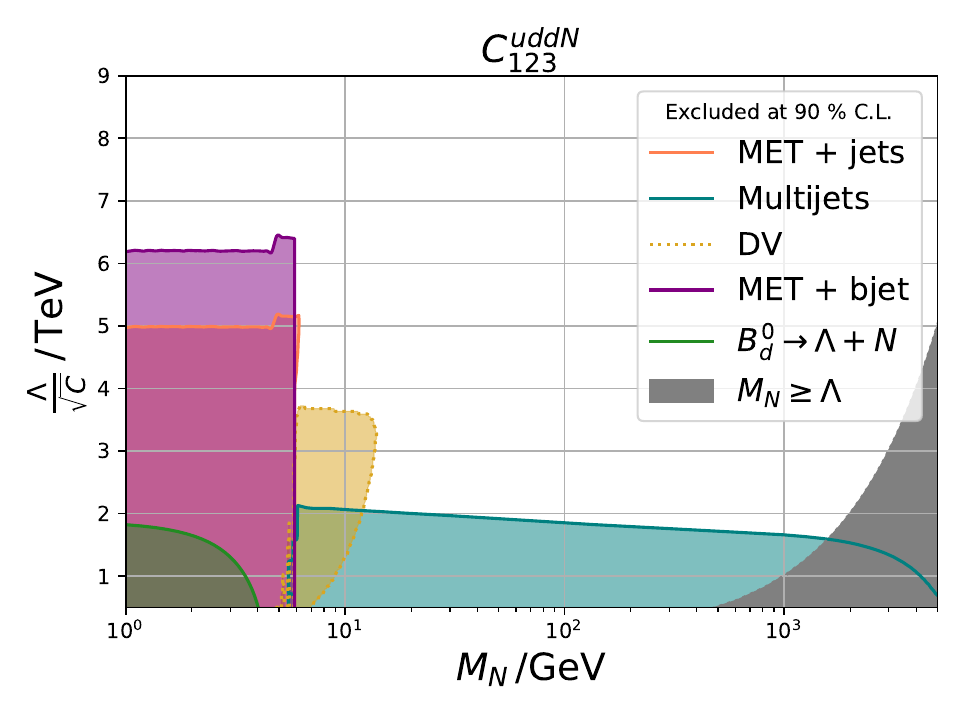}
    \includegraphics[width = 0.31 \textwidth]{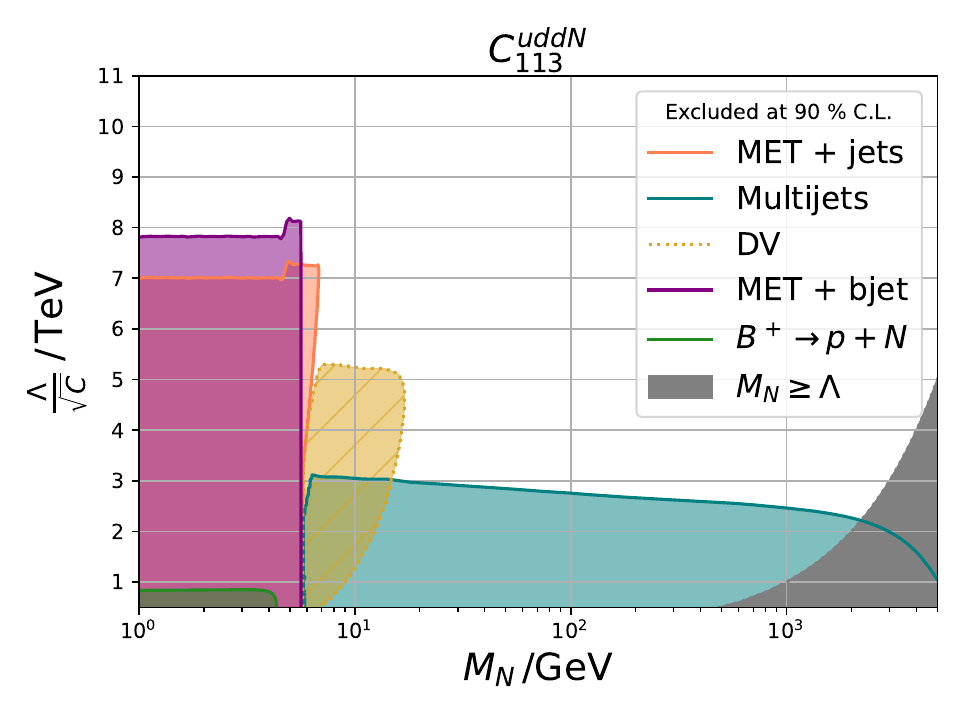}
    \includegraphics[width = 0.31 \textwidth]{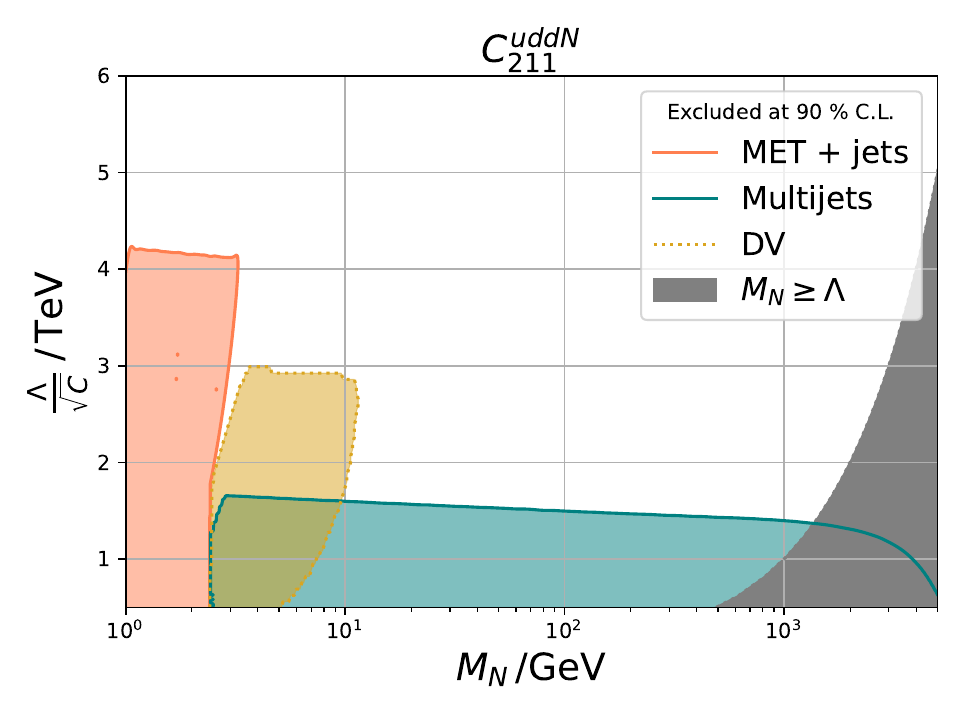}
    \includegraphics[width = 0.31 \textwidth]{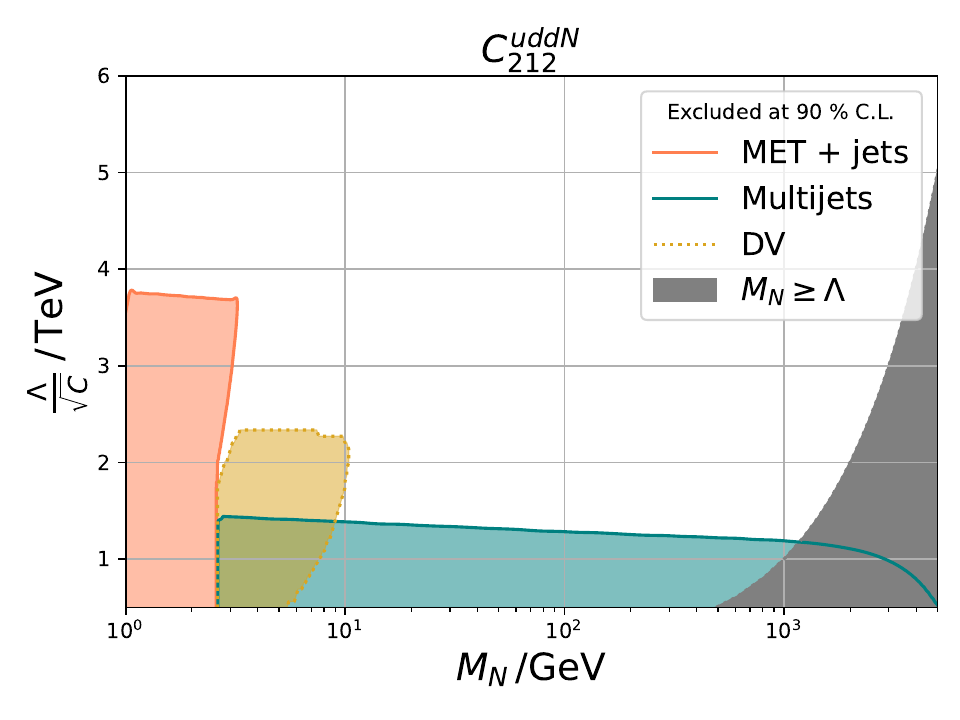}
    \includegraphics[width = 0.31 \textwidth]{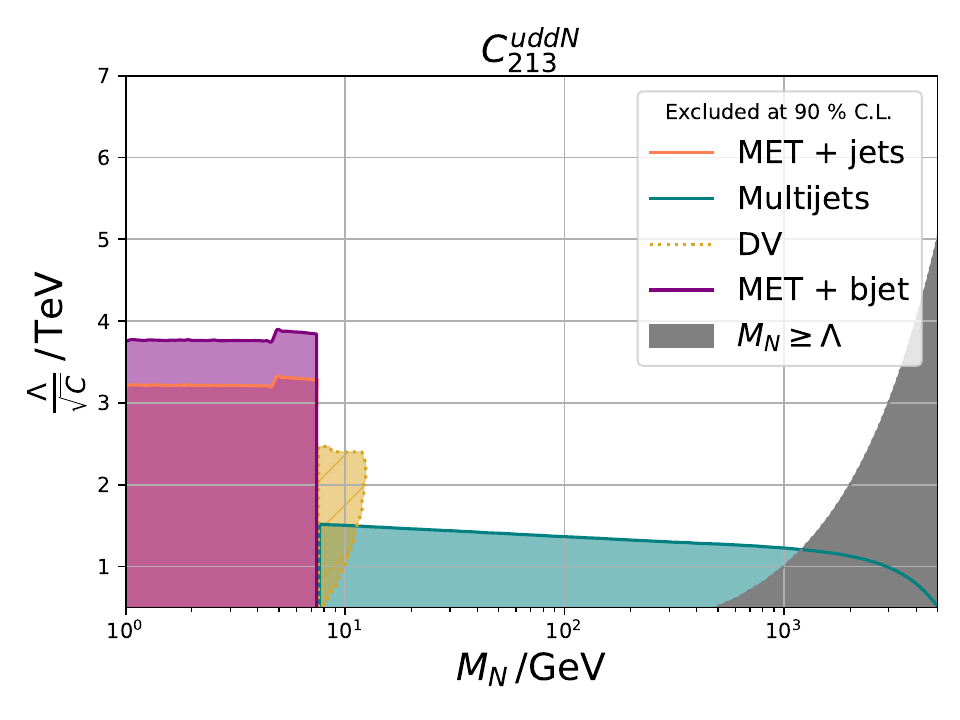}
    \includegraphics[width = 0.31 \textwidth]{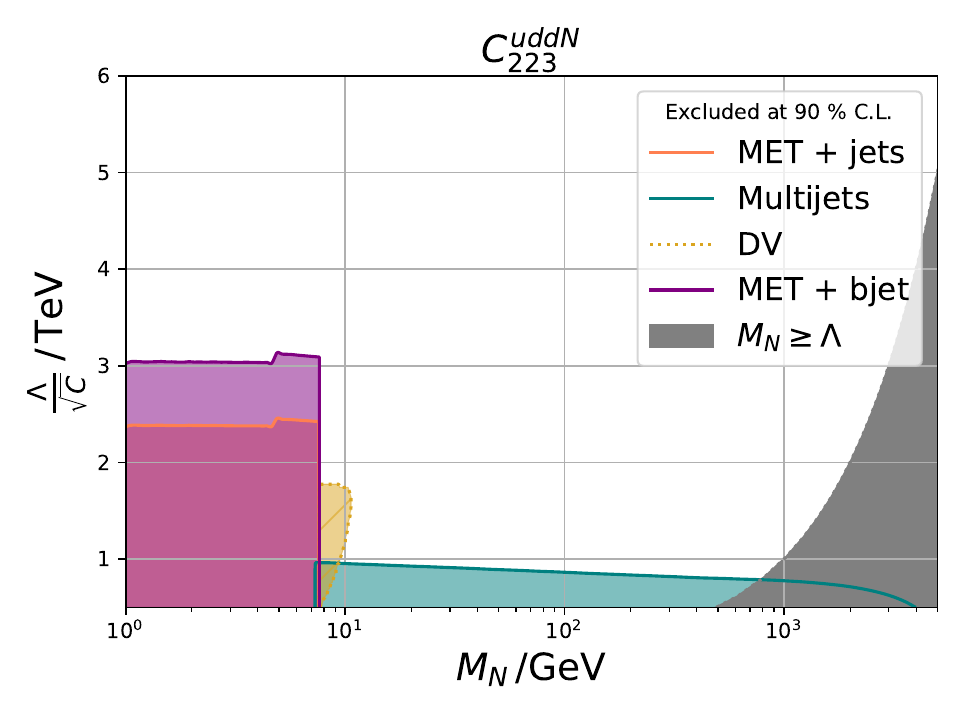}
    \includegraphics[width = 0.31 \textwidth]{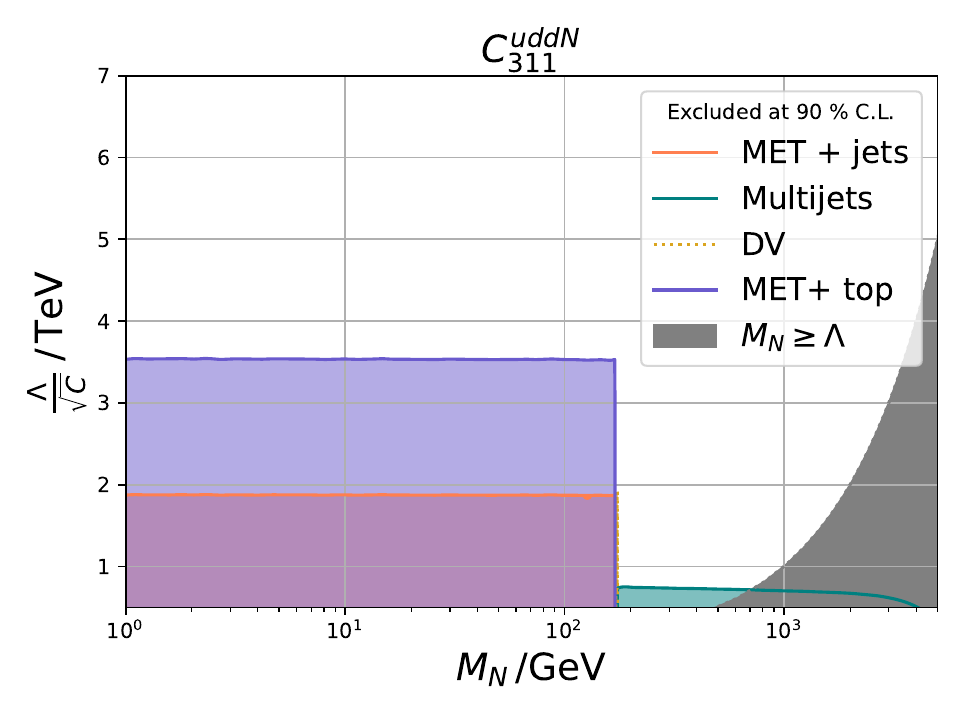}
    \includegraphics[width = 0.31 \textwidth]{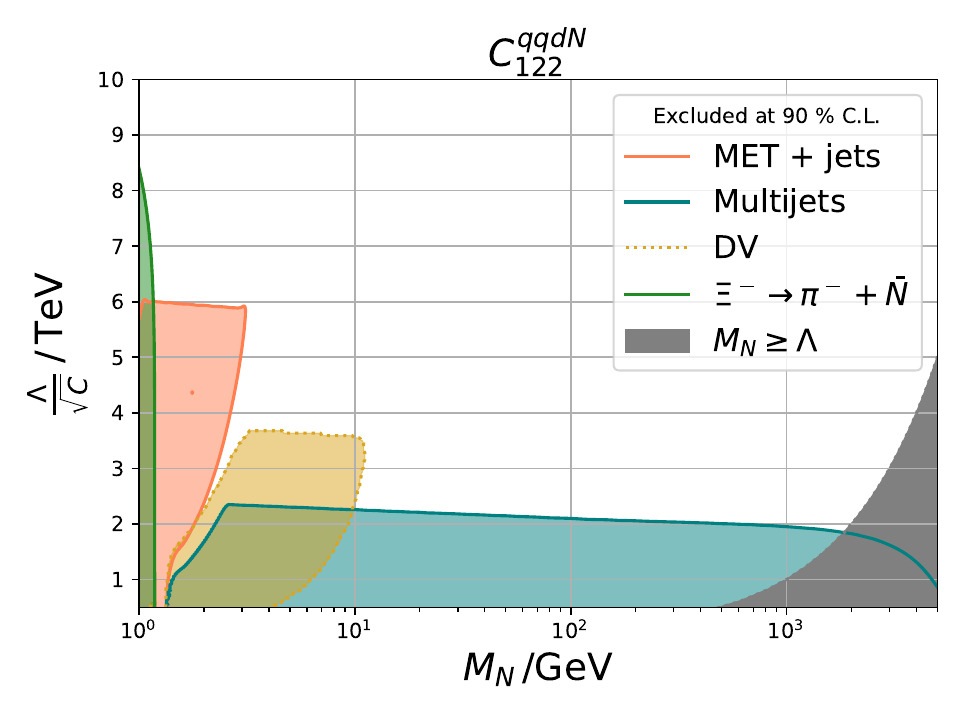}
     \caption{Constraints in the $(M_N,\Lambda / \sqrt{C})$-plane, based on MET-,Multijet-, DV-signatures and low-energy decays (dark green, where available) on $C^{uddN}_{ijk}$ and $C^{qqdN}_{\{ij\}k}$ for a sample of flavor combinations.
    MET and Multijets observables are summarized in Tab.~\ref{tab:BNV_obs}. The dotted lines  correspond to DV  sensitivities for current  colliders, see Sec.~\ref{sec:DV_sensitivities}.
 }
    \label{fig:CuddN_constraints}
\end{figure}

We combine previously derived results to constrain BNV operators in the $(M_N,\Lambda / \sqrt{C})$-plane.
This combines MET-, Multijet-, DV-signatures and low-energy decays to cover large parts of available parameter space.
Additionally we also show the region defined by $M_N \geq \Lambda / \sqrt{C}$, at which point our EFT loses its validity and the $N$ should be integrated out to obtain a new EFT.
In Fig.~\ref{fig:CuddN_constraints} we show the derived constraints in the $( M_N,\Lambda / \sqrt{C})$-plane for representative assignments of $i,j,k$.
Unlike the other flavors, the  $C^{uddN}_{311}$ plot (bottom row, mid panel) shows no DV exclusion region. This is caused by a large decay width, which scales with $M_N^5$ as in Eqn.~\eqref{eqn:BNV_partial_width}, which leads to a narrower DV-band, as can be seen in the corresponding scenario (bottom row) in Fig.~\ref{fig:decay_plots}.
This in combination with the cuts in Tab.~\ref{tab:DV_cuts} leads to no signal events in the DV region for $C^{uddN}_{311}$.
 Note also that if the $N$ couples to a dark sector, it may decay invisibly to those dark states. In this case $N$ decays to SM particles could be suppressed such that
the $N$ essentially always becomes MET.
A comparison of high-$p_T$ and rare decay data is given in Sec.~\ref{sec:comp}.

\subsection{Constraints on heavy mediator models }\label{sec:resultsUV}

We work out limits on the Yukawa interactions of LQ model 1 defined in Eqn.~\eqref{eqn:LQ_model_1}. 
We improve on earlier works~\cite{Allahverdi:2017edd} on monojet limits in model 1. Constraints for contact terms involving up-type quarks and $N$ can be obtained from Ref.~\cite{Hiller:2024vtr} by rescaling bounds for $p = p'$ and calculating the respective cross sections for light up-type quarks. Couplings involving top-quarks need an additional analysis and are beyond the scope of this work.  Additionally we take dijet bounds from Ref.~\cite{CMS:2018ucw}, which we also rescale by the respective cross sections predicted by the four-quark contact terms in Eqn.~(\ref{eqn:LQ_model_1}).
In Tab. \ref{tab:LQ_model_1_bounds} we list the derived bounds from both analyses.
While dijet searches constrain the Yukawa $\tilde y^{dd}$, and $N$-pair production probes
$y^{Nu}$, the constraints from $C^{uddN}_{i[jk]}$ in Tab.\ref{tab:asym_bounds}
provide complementary information
on the product of the two LQ Yukawas.

\begin{table}[]
    \centering
    \begin{tabular}{c|c| c}
         p p'  & $ \frac{|y_p^{Nu} y_{p'}^{Nu}|}{2M_{\Psi}^2}  /  \text{TeV}^{-2}$  & $ \frac{|\tilde y_p^{dd} \tilde y_{p'}^{dd}|}{M_{\Psi}^2} /  \text{TeV}^{-2}$  \\\toprule 
         $1 1$ & $0.0058$  & $0.22$\\
         $2 2$ & $0.11$  &$0.077$ \\
         $3 3$ & $-$ & $0.031$ \\
    \end{tabular}
    \caption{Limits on four-fermion contact terms, generated by LQ model 1 \eqref{eqn:LQ_model_1}, obtained by rescaling bounds from Ref.~\cite{Hiller:2024vtr} from $N$-pair production and \cite{CMS:2018ucw} for dijets. Limits for $y_3^{Nu} y_3^{Nu}$ contribute to top-quark which could be constrained using Ref.~\cite{ATLAS:2020yzc}, which is however beyond the scope of this work.}
    \label{tab:LQ_model_1_bounds}
\end{table}

\begin{figure}
    \centering
    \includegraphics[width=0.3\linewidth]{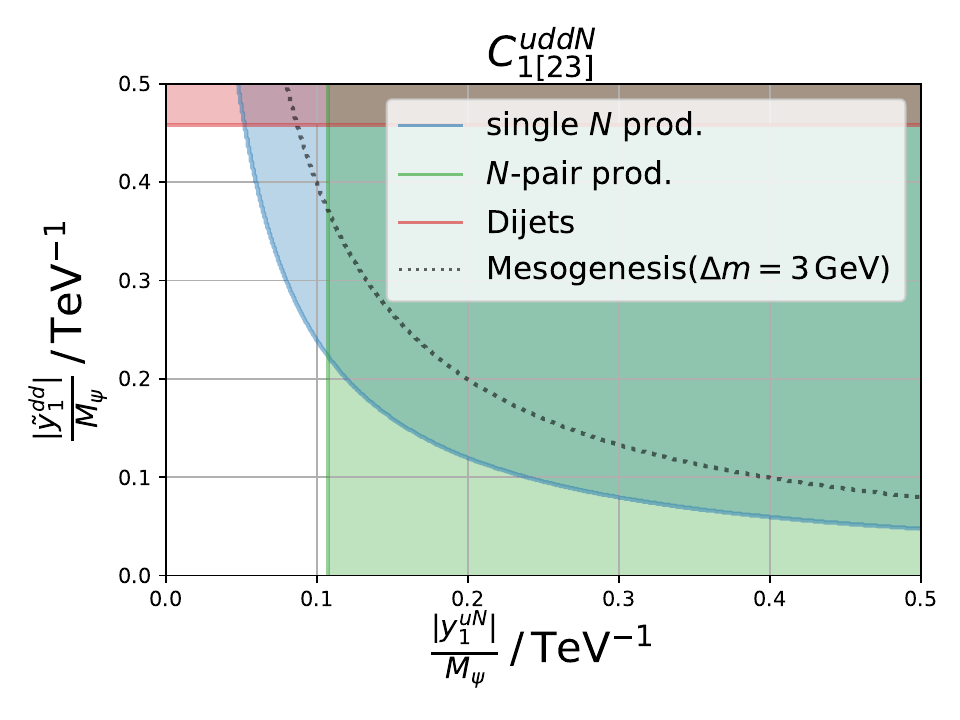}
    \includegraphics[width=0.3\linewidth]{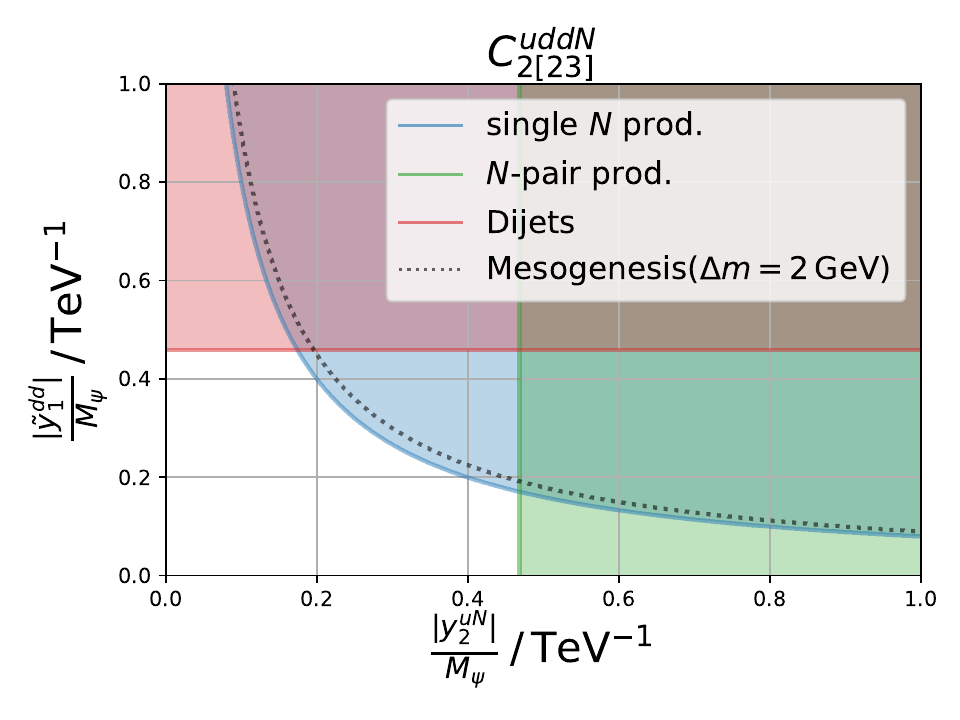}
    \includegraphics[width=0.3\linewidth]{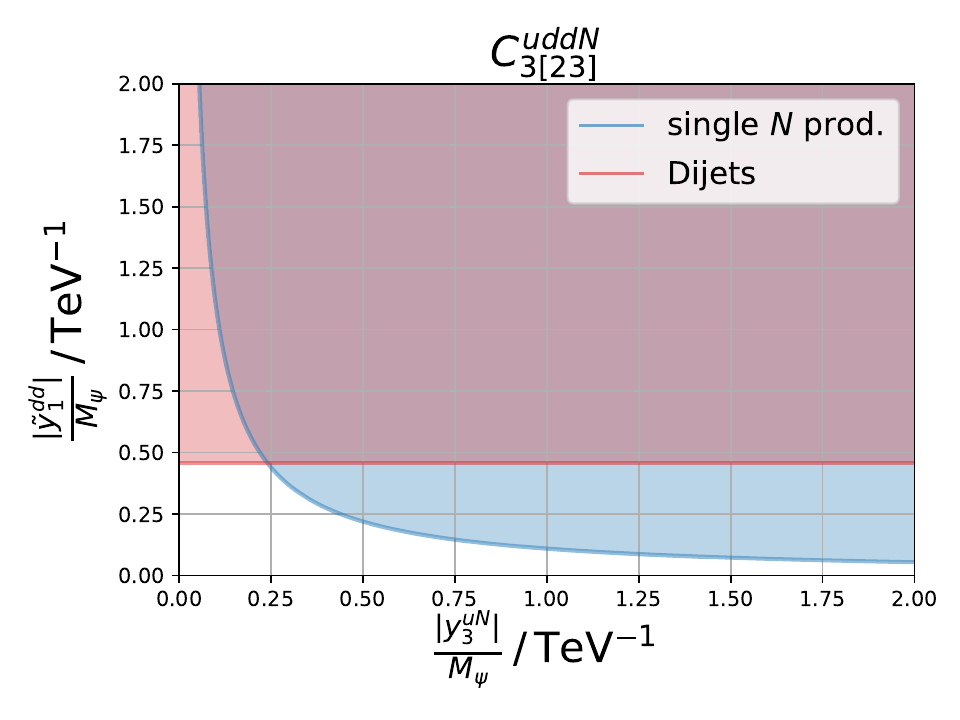}
    \includegraphics[width=0.3\linewidth]{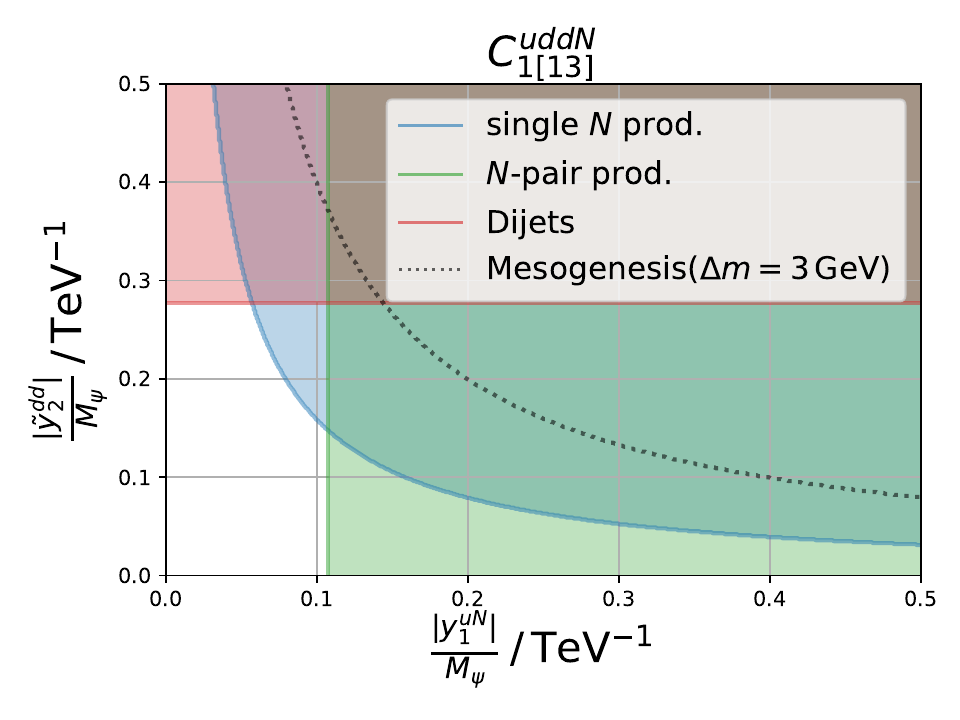}
    \includegraphics[width=0.3\linewidth]{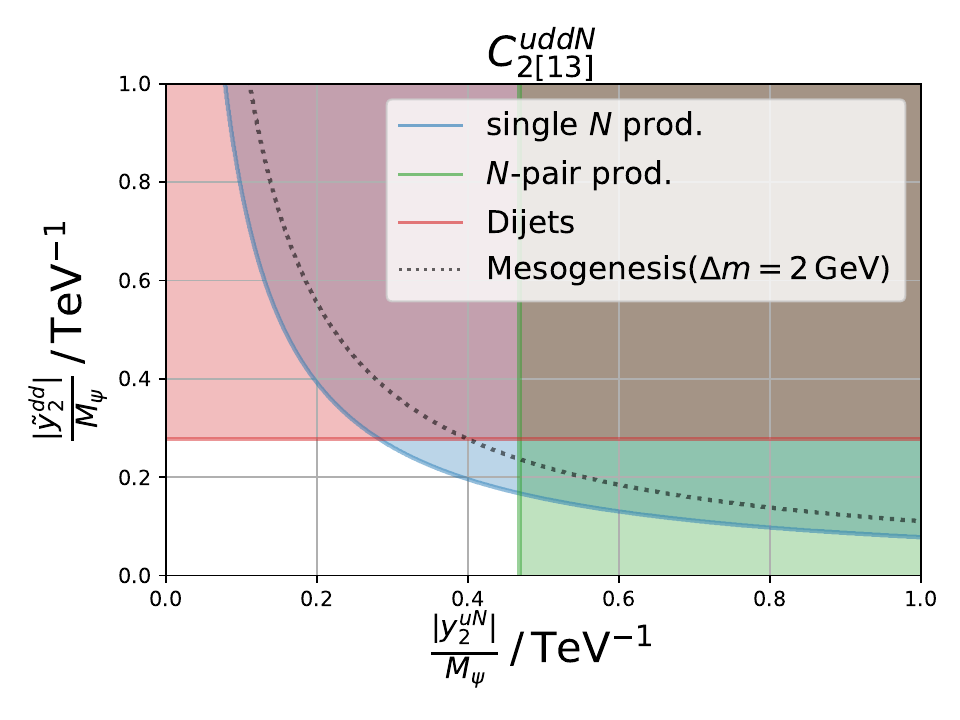}
    \includegraphics[width=0.3\linewidth]{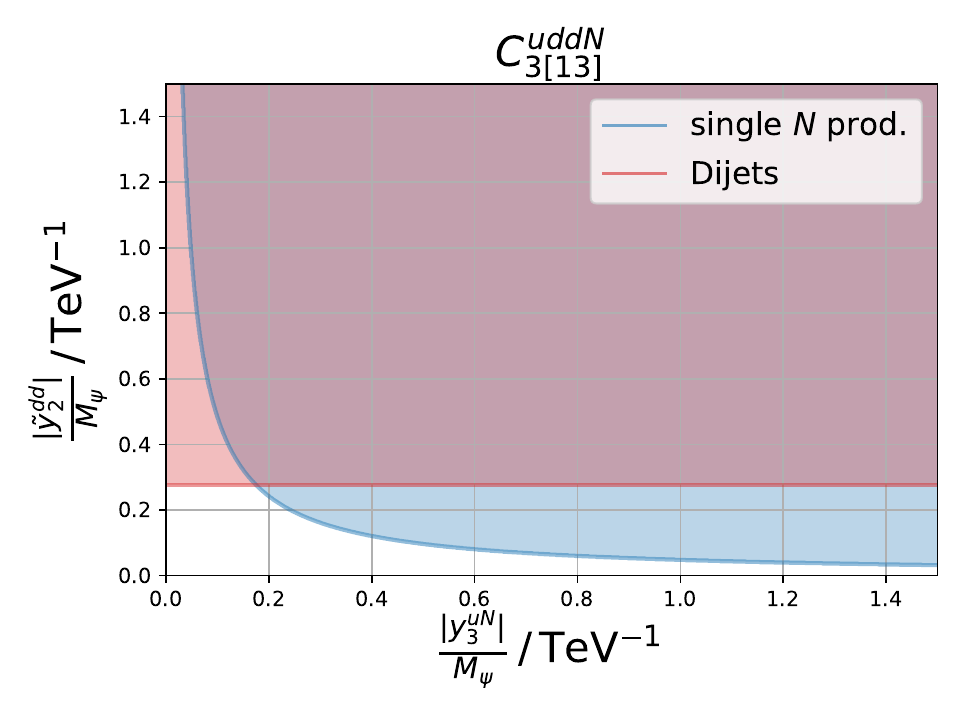}
    \includegraphics[width=0.3\linewidth]{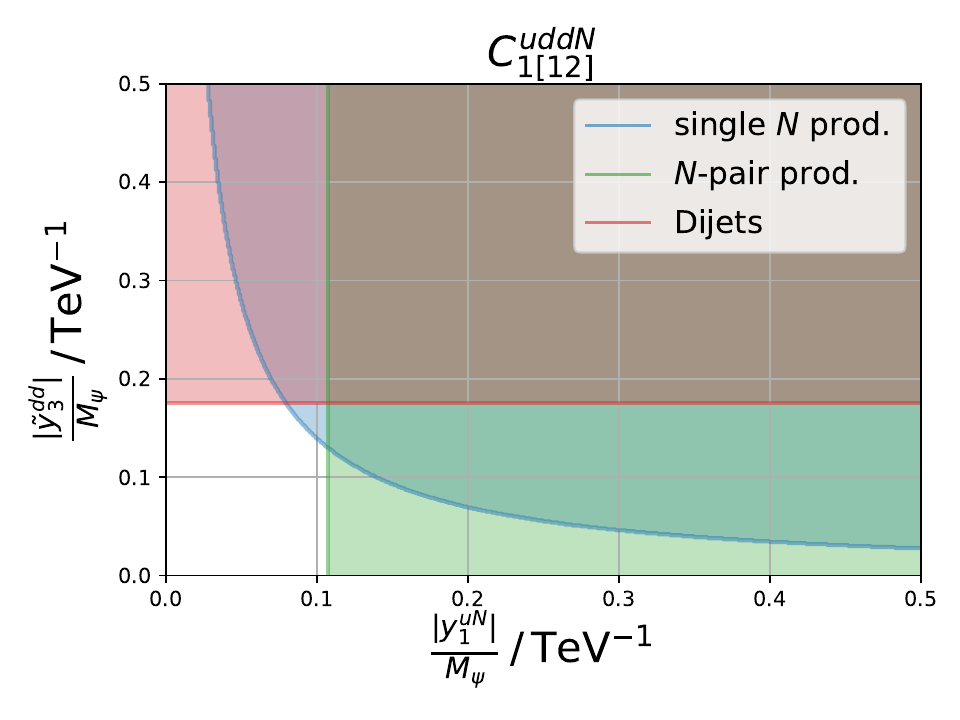}
    \includegraphics[width=0.3\linewidth]{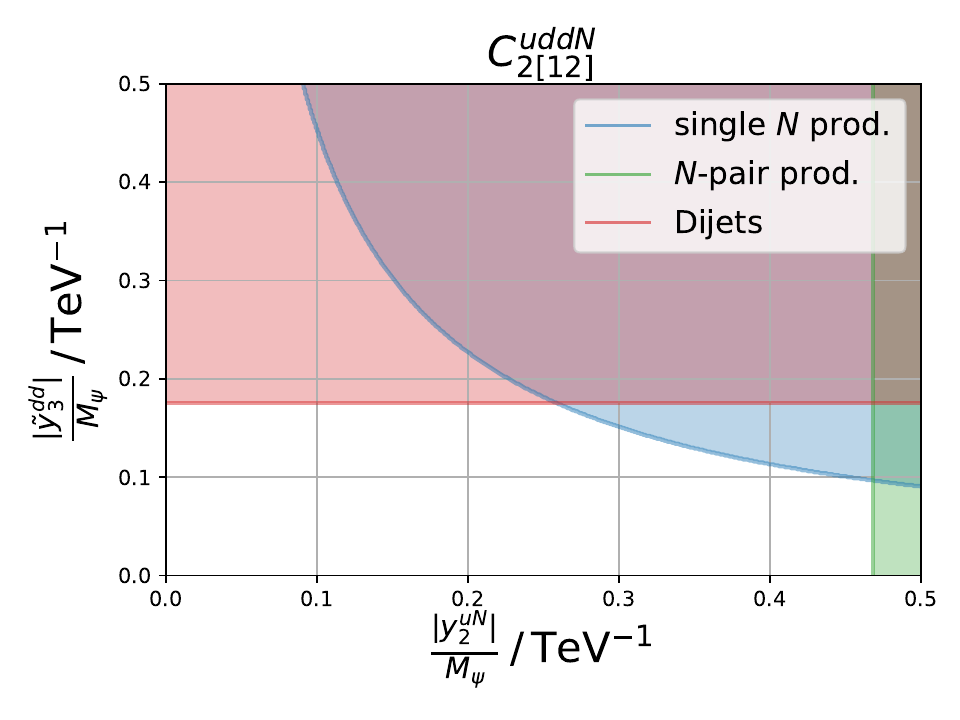}
    \includegraphics[width=0.3\linewidth]{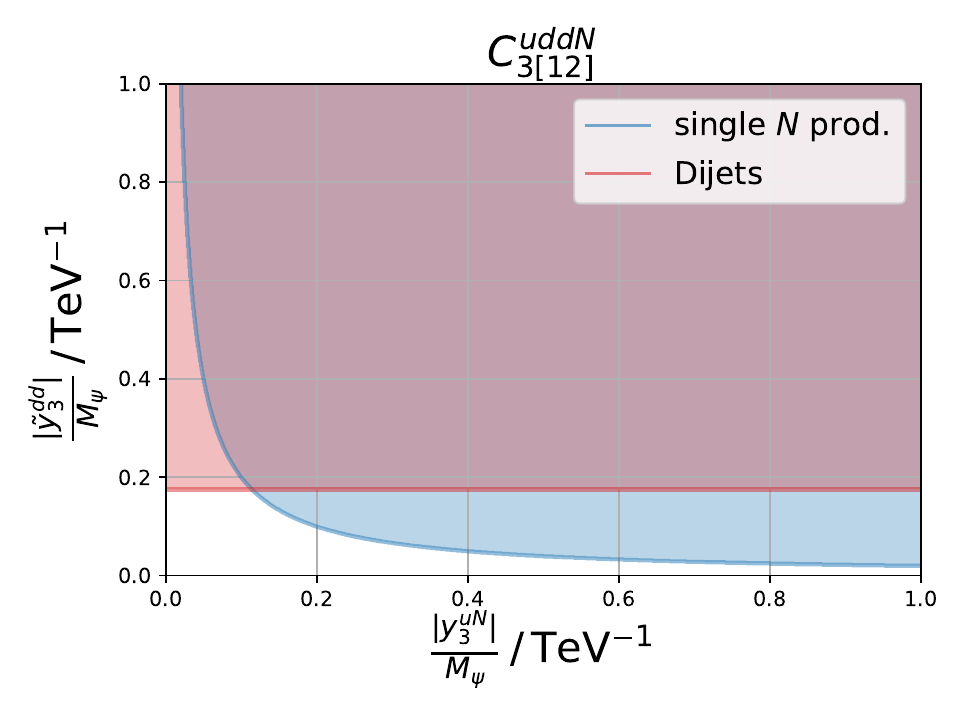}

    \caption{Excluded regions of couplings in LQ model 1, using constraints from Tab.~\ref{tab:asym_bounds} and 
    Refs.~\cite{Hiller:2024vtr,CMS:2018ucw}. Shown are the LQ couplings to $N$ and up-type quarks ($y^{Nu}_p$, x-axis) and the coupling to down-type quarks ($\tilde y^{dd}_p$, y-axis), for flavors $p=1,2,3$. The bounds are given for  $M_N = \SI{1}{\GeV}$, for which the $N$-particle is always detector stable, see Sec.~\ref{sec:N_decay}. 
    However the bounds are approximately valid for larger masses, depending on possible final states of the decay of $N$, see Sec.~\ref{sec:N_decay} and Fig.~\ref{fig:CuddN_constraints}.
    The dotted line corresponds to a lower limit, from $B$-meson branching ratios, needed to obtain successful Mesogenesis~\cite{Alonso-Alvarez:2021qfd}.}
    \label{fig:LQ_plot}
\end{figure}

In Fig. \ref{fig:LQ_plot} we show the excluded region in the $(y_p^{uN},\tilde y_{p'}^{dd})$-plane using dijets (red), $N$-pair production (green) and
single $N$-production (blue).
The bounds 
from single $N$-production and $N\bar N$-pair production are derived from the same observable, which would imply correlations in regions where the bounds overlap. In principle these bounds should be combined doing a full recast for both observables, which is beyond the scope of this work. However contributions towards the total cross section do not interfere, 
between single and pair production, which implies our bounds are  conservative.
From Fig.~\ref{fig:LQ_plot} 
one observes that the collider observables are apparently complementary. For all flavor-combinations
the single $N$-production improve the constraints, except
for three cases, $(y_3^{uN},\tilde y_{1}^{dd})$, $(y_3^{uN},\tilde y_{2}^{dd})$ and
$(y_2^{uN},\tilde y_{1}^{dd})$.

We also show the boundary (dotted black line) above which mesogenesis works with heavy scalar color triplets coupling to $b$-quarks and $N$ \cite{Alonso-Alvarez:2021qfd}. The latter
resembles model 1.  Interestingly, none of the 
four possible flavor combinations (upper left 2 by 2 block),
has a surviving area 
after the constraints from single-$N$ production. 
We recall that limits from single-$N$ production are for the four cases essentially independent of the $N$ mass in the relevant  range below $m_p < M_N \lesssim m_B$,
see Fig.~\ref{fig:CuddN_constraints}.
Moreover, for larger values of $M_N$, the difference 
$\Delta m \leq m_{B} -M_N-m_p-m_\pi \lesssim 3.3 \, \text{GeV}$ (no charm) or $\Delta m \leq m_{B} -M_N-m_{\Lambda_c}-m_\pi
\lesssim 1.9 \, \text{GeV}$ (with charm) gets smaller and the mesogenesis boundary moves further to the upper right,
into the exclusion region.
Main differences between this and  previous works \cite{Alonso-Alvarez:2021qfd} are about four times more data, including tagged bjets,
 see Tab.~\ref{tab:BNV_obs},  an indepth study of $N$-decays and notably higher $P_T$-bins.
 
\begin{figure}
    \centering
    \includegraphics[width=0.75\linewidth]{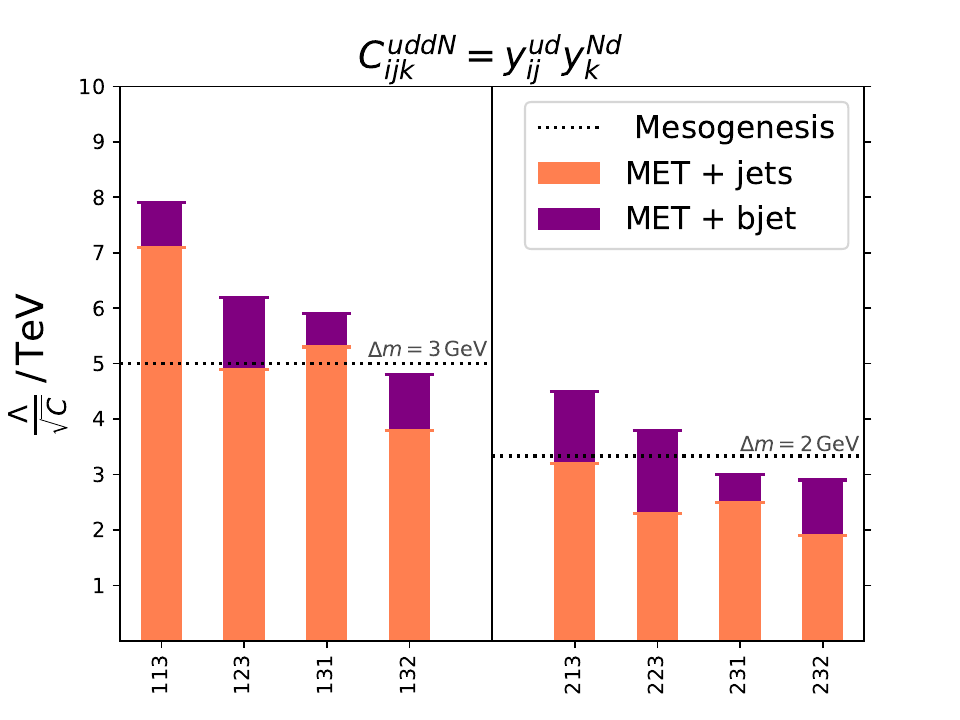}
    \caption{$90 \%$ lower limits on $\Lambda / \sqrt{C}$ for LQ model 2, with $C = C^{uddN}_{ijk} = y^{ud}_{ij} y^{Nd}_k$, from MET observables and the corresponding upper limit (black dotted lines) necessary for the Mesogenesis model detailed in Ref.~\cite{Alonso-Alvarez:2021qfd}. The $\Delta m $-values corresponds to the mass differences of the initial and final states for a $B$-meson decay, which is smaller for decays including charm final states ($2jk$, right hand side) relative to 
    light quark final states ($1jk$, left hand side). The MET+bjet  limits (purple) are  stronger than the MET+jet ones (orange).}
    \label{fig:Mesogenesis_model2}
\end{figure}

For LQ model 2 (\ref{eq:model2}), assuming $y^{QQ}=0$ as the doublet-Yukawa always induces meson mixing,  eight independent flavor combinations exist, since $C^{uddN}_{ijk}$ is not necessarily antisymmetric  or symmetric.
In Fig.~\ref{fig:Mesogenesis_model2} 
we show the lower limits on $\Lambda\, / \,\sqrt{C} $ from MET observables (Fig.~\ref{fig:NP_BNV}, Tab.~\ref{tab:bounds_numerical_CuddN}) 
as well as the upper limit (dotted lines) necessary for successful mesogenesis.
 We learn that in particular the MET + bjet analysis is very powerful in constraining  model 2, and narrows down viable mesogenesis
to the flavor combinations $cbs$, $cbd$ and $ubs$, corresponding to $j = 3$.

The proximity of constraints to the mesogenesis boundary in model 2 implies that either of the three observables, dijets $bbuu$ and $bbcc$ 
 or single-$N$, 
 or $N$-pair production $\bar dd \bar NN$ and $\bar ss\bar NN$, are therefore in the position to identify, or rule this out. 
 Given the indicated low scale of the mediator, 
 $\sim 3 \, \text{TeV}$ (with charm) and $\sim 5 \, \text{TeV}$ (no charm) for Yukawas $\lesssim 1$ dedicated resonance searches are also promising.

The sensitivity to the LQ mass
scales with the fourth root of the luminosity ratio, between current and future experiments.  
For HL-LHC ($ \mathcal{L}^{\text{int}} = \SI{3}{ab}^{-1}$) we expect an improvement of roughly a factor $\sim 2.2( \sim 3.0)$ for single and pair $N$-production (dijets~\cite{CMS:2018ucw}), which potentially can exclude the remaining parameter space in Fig.~\ref{fig:Mesogenesis_model2}. 
In the context of mesogenesis, the constraints obtained in this work also imply  that $b$ decays to $N$ take place with
the $ubs$, $cbd$ or $cbs$ current.
Indicating the flavor content in parentheses, these are, if we restrict to two-body decays, respectively, the $b$-meson decays 
$\mathcal{M}(b \bar q) \to \mathcal{B}( \bar u \bar s \bar q)$, $\mathcal{M}(b \bar q) \to \mathcal{B}( \bar c \bar d \bar q)$ and
$\mathcal{M}(b \bar q) \to \mathcal{B}( \bar c \bar s \bar q)$, and baryon decays 
$\mathcal{B}(bsq) \to \mathcal{M}( \bar u q)$ or $\mathcal{B}(buq) \to \mathcal{M}( \bar s q)$, 
$\mathcal{B}(bdq) \to \mathcal{M}( \bar c q)$ or $\mathcal{B}(bcq) \to \mathcal{M}( \bar d q)$, and
$\mathcal{B}(bsq) \to \mathcal{M}( \bar c q)$ or $\mathcal{B}(bcq) \to \mathcal{M}( \bar s q)$, where $q$ can be any flavor except top.
Sample modes include $B^0 \to \Lambda N$, already probed at $\sim 2.6 \,\text{TeV}/\sqrt{C}$, see Table \ref{tab:current_bounds}, and $\Lambda_b \to K^0 \bar{N}$ induced by $ubs$, and
$B^+ \to \Lambda_c N$, $\Lambda_b \to \bar D^0 \bar{N}$ via $cbd$.
For the least constrained $cbs$ current, none of the modes contains one of  the lowest mass $b$-baryons (charm baryons) in the initial (final) state.
The larger $B$ meson production rates  suggest decays $B^{0,+} \to \Xi^{0,+}_c N$, or $B_s \to \Omega^0_c N$ 
for further study. Further multi-body decays, such as $B^+\to \Lambda D^{+}N$, are also allowed in this scenario.

\subsection{EFT vs heavy mediator model}

 We investigate the validity of the EFT as opposed to an UV-model analysis. 
The EFT will cease to work if the mediator is too light. Collider  limits on leptoquark masses are in part
model-dependent, as they assume
branching ratios in specific, mostly  lepton-quark final states. On the other hand, pair production is rather universal as induced by QCD
but suffers from further phase space suppression.
Experimental lower limits  for quark-neutrino final states are, very roughly, around $0.9-1.3$ TeV \cite{ATLAS:2025gjo}.
We therefore do not consider states lighter than a TeV.
 On the other hand, for larger masses the resonance models are expected to decouple and limits should converge 
towards the EFT ones. We note that the UFO-files used in the model simulation also include leptoquark couplings to gluons.
These induce higher-dimensional operators not considered in the EFT analysis (\ref{eq:L-BNV_4F}). Numerically, they are of minor numerical importance here.

Generally, resonance effects are most pronounced in $\hat s$-channel production, which also require to take finite-width effects into account.
The width $\Gamma_{\Phi}$ of the mediator  with mass $M_\Phi$ is induced by both types of Yukawa couplings, here for model 2,
\begin{equation}
\label{eq:GoverM}
\Gamma_\Phi =\frac{M_{\Phi}}{16 \pi }\sum_{ijk} (2 (y^{ud}_{ij})^2 +(y_k^{dN})^2) \, ,  
\end{equation}
where the sum goes over all flavors but contains at least those Yukawas that
induce the MET + jet, $b$ or top signal. Using for simplicity $y^{ud}\simeq  y^{Nd}$ the leptoquark is unstable and wide $\Gamma_\Phi/M_\Phi \sim O(1)$ once
the corresponding Wilson coefficient $C = y^{ud} y^{Nd}$ is above $\sim 16 \pi/3$. 
\footnote{The coefficient $C$ is maximized for $y^{ud} =  y^{Nd}/\sqrt{2}$, which only marginally lifts the width constraint to 
$\sim 16 \pi/(2 \sqrt{2})$.}
In this regime, perturbation theory has broken down although the expansion is expected to receive
finite width corrections already for smaller widths. 
In addition, jacobian peaks appear for $P_T \lesssim M_\Phi/2$, see (\ref{eqn:BNV_hadronxsec}), which
enhances cross sections over EFT estimates. Therefore, for these processes, on general grounds
we expect the EFT analyses to  give weaker bounds, which makes those
conservative. 
On the other hand, $\hat t,\hat u$-channel mediator contributions exhibit better convergence between UV-model and EFT \cite{Allwicher:2024mzw}
but cross sections are smaller than the EFT ones.
In general, UV mediators contribute in  $\hat s$ and $\hat t,\hat u$-channels, however with different PLFs.
It is therefore a quantitative question, which depends on the model, the flavor combinations involved and the leptoquark mass,
whether the  EFT-limits hold for the heavy mediators.

For the mesogenesis scalar mediators in model 1 and 2 involving beauty in one of the couplings we note the following:
model 1 has  $\hat s$-channel contributions from $db$, $sb$-fusion, with $t$-jet (not relevant for mesogenesis) or an up-type jet  in the final state.
These contributions have strong PLF suppression, importantly, also  relative to the 
          $\hat t, \hat u$-channel ones, which stem from $ud$,$us,cd,cs$-fusion, with $b$-jet in final state, constrained by MET+$b$.
          The relative suppression of the $\hat s$-channel is stronger in $1[j3]$ than in $2[j3]$, $j=1,2$ flavor combinations.

Model 2 receives contributions from $\hat s$-channel via  fusion of an up-type and a down-type quark. The other channels contribute
 also via  down-type quark scattering. Comparing the channels and their PLF hierarchies, one finds that in combinations $i13$ and $i23$, $i=1,2$, which involve
$y^{Nb}$, the $\hat s$ channel is PLF enhanced. We therefore expect EFT limits to hold for these flavors also in model 2.
 On the other hand, the $\hat s$-channel is PLF suppressed in $i32$, involving $y^{Ns}$, and even more so in $i31$, with $y^{Nd}$.
As for model 1 a quantitative analysis is required to check  the validity of EFT constraints.

 \begin{figure}
    \centering
    \includegraphics[width=0.48\linewidth]{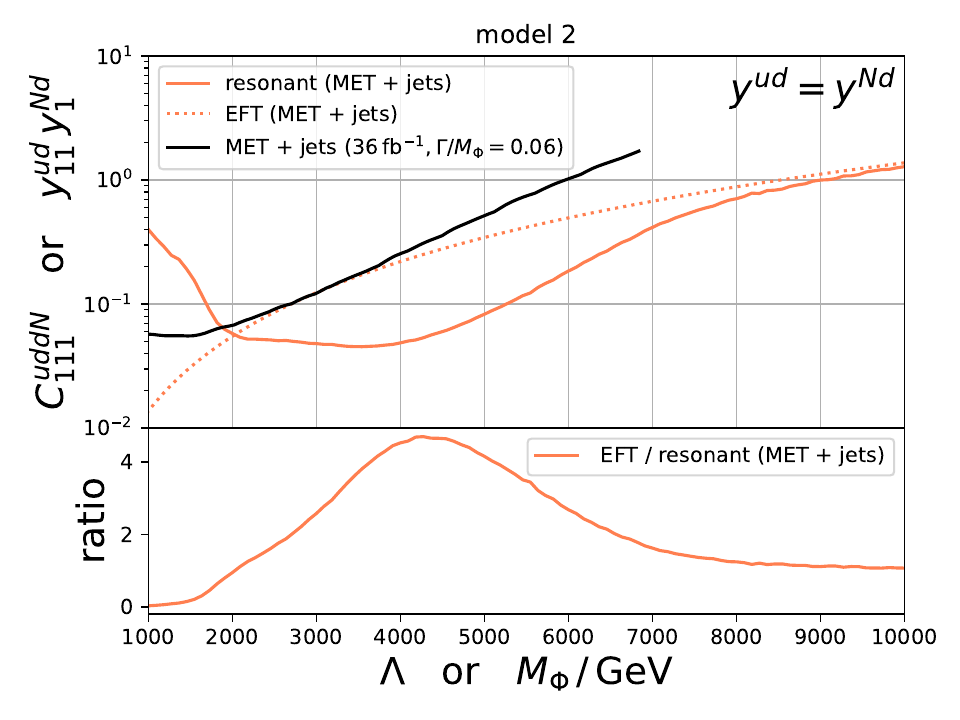}
    \includegraphics[width=0.48\linewidth]{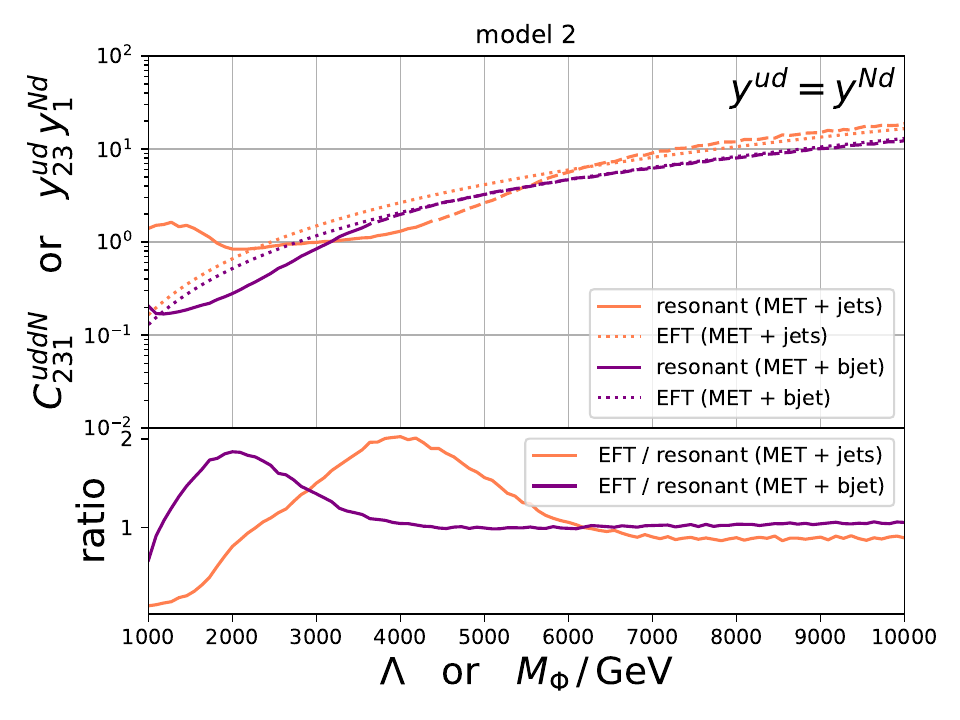}
    \includegraphics[width=0.48\linewidth]{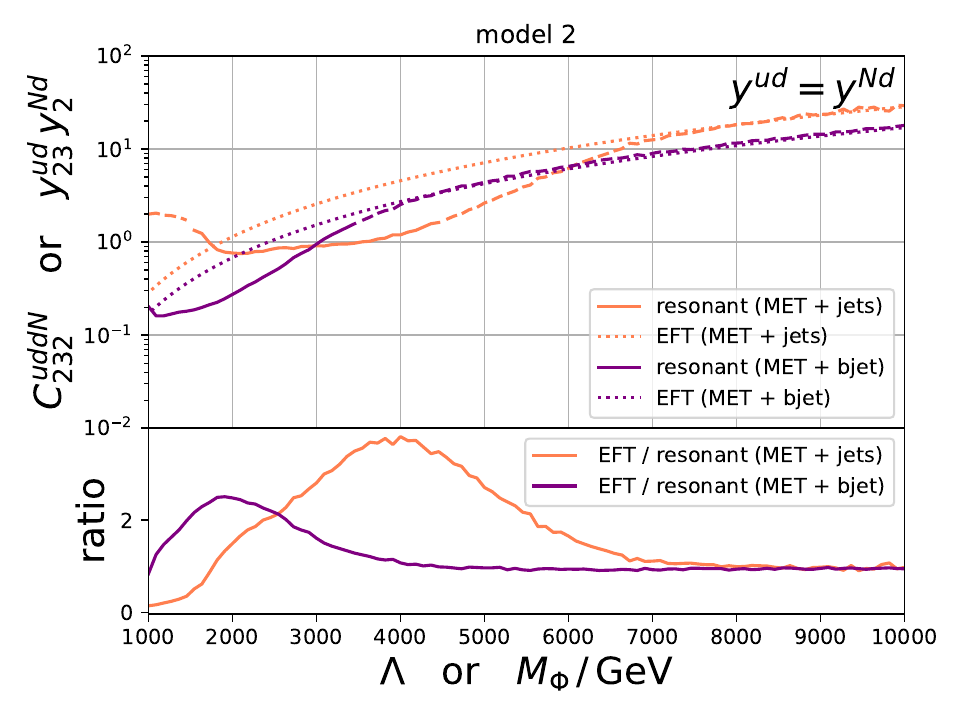}
 \includegraphics[width=0.48\linewidth]{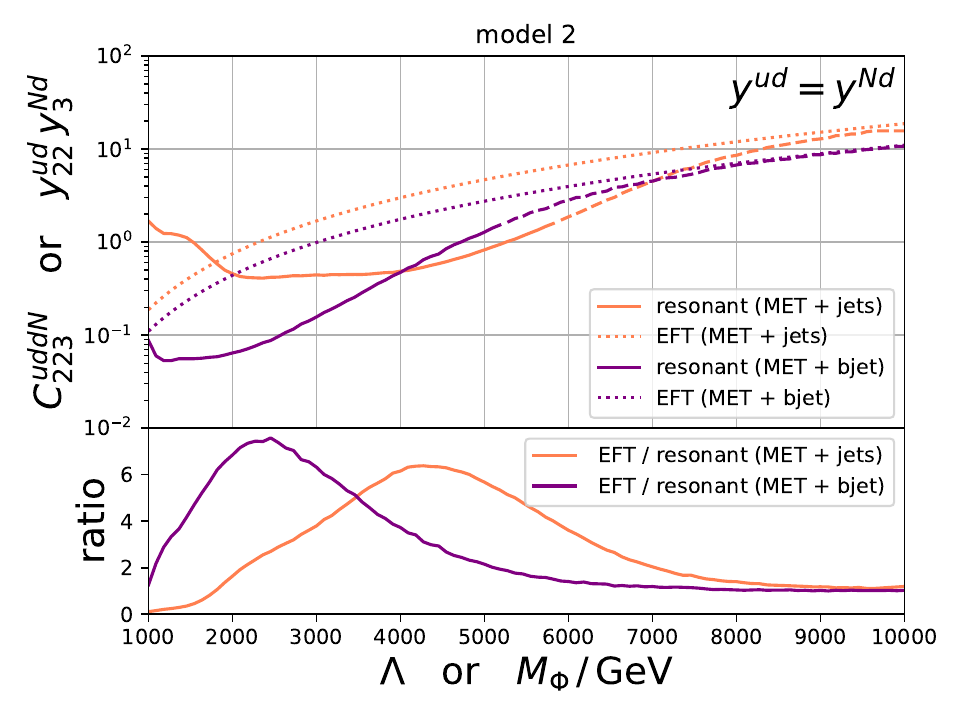}
    \includegraphics[width=0.48\linewidth]{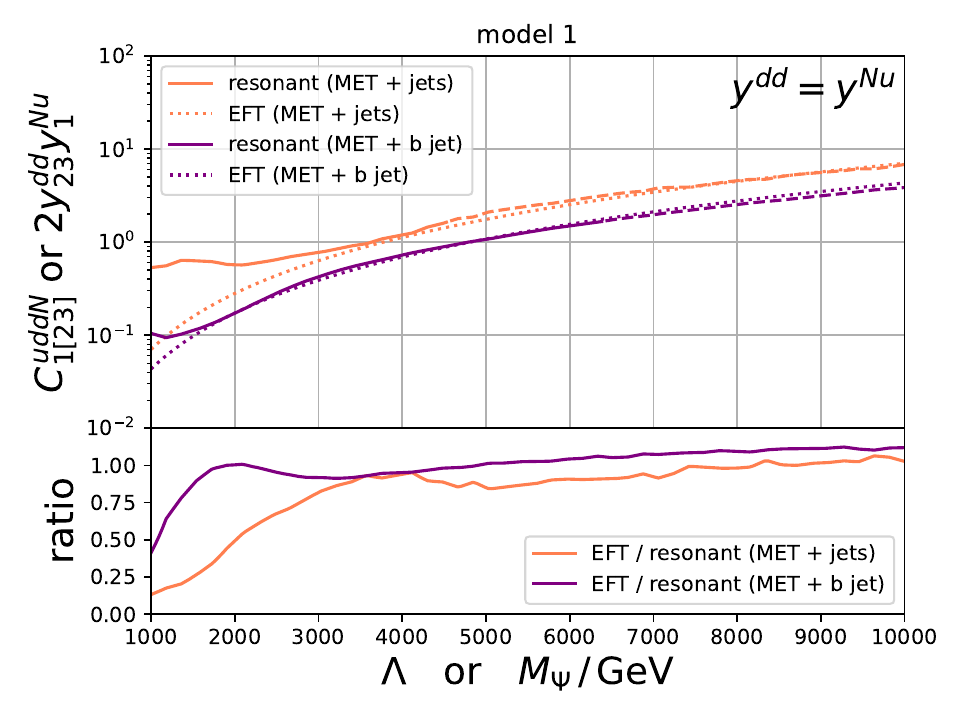}
    \includegraphics[width=0.48\linewidth]{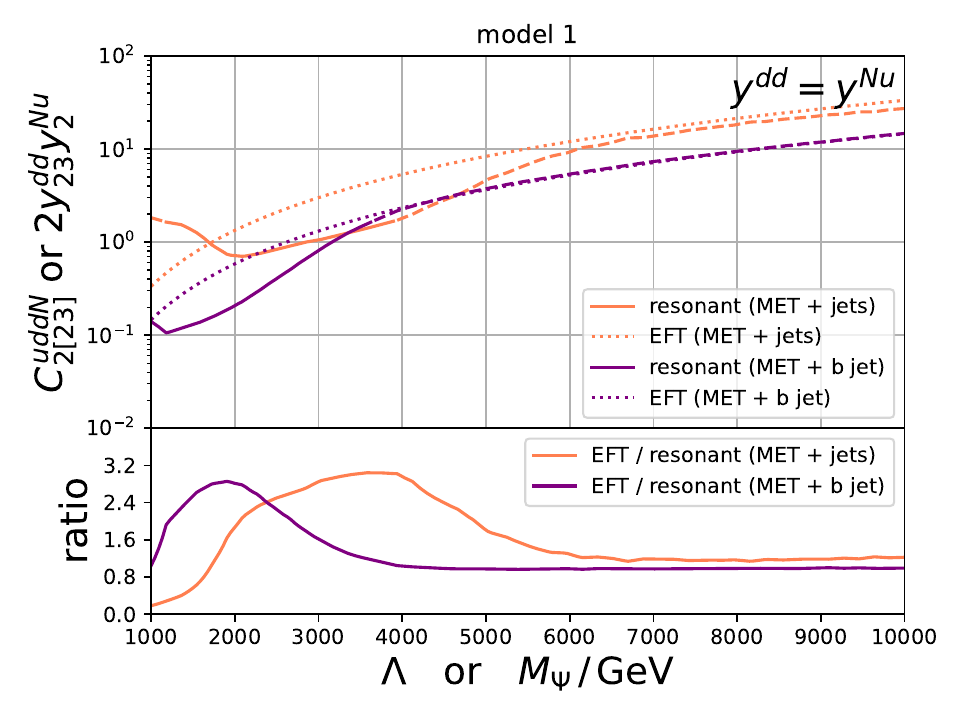}
    \caption{Comparison between the EFT results (dotted lines), obtained in  Sec.~\ref{sec:METlimits}
    and bounds on the resonant leptoquark models (solid/dashed lines) introduced in Tab.~\ref{tab:UV_mediators}.
    The upper four plots are for model 2, Eqn.~\eqref{eq:model2}, with  flavors 111 (upper left), 231 (upper right),
    232 (mid left) and 223 (mid right). The lower left (right) plot is for model 1 with flavors  $1[23]$ ($2[23]$).
    For light quarks (111) we also show previous results (black solid) from Ref.~\cite{Alonso-Alvarez:2021qfd}.
    We assume equal diquark and $Nq$ Yukawas, as the width depends on the sum of the square of the individual 
    couplings, see Eqn.\eqref{eq:GoverM}, and use the MET observables from Sec.~\ref{sec:recast}, where we only use the high energy bins. 
    Dashed lines correspond to $\Gamma_\Phi \, / \, M_{\Phi}\geq 0.1$, see text. }
    \label{fig:Resonant_comp}
\end{figure}

In Fig.~\ref{fig:Resonant_comp} we display limits on couplings versus mediator mass with
$\Gamma_\Phi/M_\Phi <0.1$ in solid lines, and mark those with larger width with dashed lines. The corresponding EFT-limits are shown by dotted lines.
Purple  (orange) lines refer to MET+b (MET+jets) constraints. The upper four plots are for model 2, while the lower two plots for model 1,
each corresponding to flavor combinations with the different PLF hierarchies between contributing channels, as detailed before.
Upper right (231), mid left (232) and mid right (223) have increasing $\hat s$ channel contributions, in that order.
Similarly, the $\hat s$ channel suppression is lifted from lower left ($1[23]$) to the lower right plot ($2[23]$).

Recall the highest bins used in this analysis, $P_T \sim 1.3-2.6$ TeV (MET+jets) for which the  jacobian peak covers masses up to $2.6-5.2$ TeV. 
For MET+$b$ the highest bin lies between $550 - 1000 \, \si{\GeV}$ which corresponds to  masses up to $1.1-2$ TeV.
The lower $P_T$-bins in the MET+b analysis therefore result in larger cross sections  at lower leptoquark  masses than the MET+jet one.
These features are visible in a strengthening of resonance limits compared to the EFT ones.
The combination of  MET+jets and MET+b searches, if applicable, covers a larger mass region than the individual ones.

Generally, as expected, 
the UV-mediator analysis gives stronger limits than the EFT one.
 An exception is model 1 for flavors with strongly suppressed $\hat s$-channel contributions, 
$udd=1 [23 ]$ (lower left plot), especially in the MET+jet channel (orange curves) for lighter $M_\psi$. Apart from that, the improvement of resonance model limits

is up to a factor of few, and most pronounced for flavors with larger $\hat s$-channels, such as model 2, 223 (mid right plot).
Improvements are for masses starting at $\sim 2$ TeV (MET+jets) and a TeV (MET+b).
For these mass ranges  limits derived in EFT are hence conservative, and apply. The MET+b search is in particular  a valuable probe of lighter mediators.
The strain on the mesogenesis in concrete mediator models is even larger than the EFT-analysis indicates. Dedicated LHC searches in leptoquark  models are hence highly motivated.

We also  compare limits in the EFT and model 2 for light quarks, 111 (upper left of Fig.~\ref{fig:Resonant_comp}),
where we show the previous results (solid black) from Ref.~\cite{Alonso-Alvarez:2021qfd}. 
The latter assumes fixed $\Gamma_\Phi \, / \, M_{\Phi} =0.06$, which fixes one Yukawa coupling in terms of the other, as can be seen in Eqn.~\eqref{eq:GoverM}, 
instead of assuming a fixed ratio between them.
Comparing this to the present model 2 range (solid orange) one notes that
it is not only the increased luminosity, but also the high-$P_T$ bin that matters.
 In the low mass region ($ M_\Phi \lesssim \SI{2}{\TeV}$), we see that the analysis in Ref.~\cite{Alonso-Alvarez:2021qfd} gives stronger bounds.
 This is an artifact of restricting our analysis to the highest-energy bins, see Fig.~\ref{fig:PTspectra}.
 For larger masses up to $ M_\Phi \sim \SI{4}{\TeV}$ the limits on $y^{ud}y^{Nd}$ are approximately flat in the mediator mass. 
 Here our analysis improves upon Ref.~\cite{Alonso-Alvarez:2021qfd}, due to the lower $P_T$ bins in the latter.
 This also goes against the naive statistical scaling of the analysis in Ref.~\cite{Alonso-Alvarez:2021qfd}. 
 In the high mass region ($M_\Phi \sim \SI{9}{TeV}$) we see that the bound derived in this work converges to the EFT bound from Sec.~\ref{sec:res} and scales as $\propto M_\Phi^4$.

In Fig.~\ref{fig:Resonant_comp} we notice that the convergence to the EFT value is slightly off for model 2 231 (upper right) and the model 1 plots (lower row). This small discrepancy may be traced back to the residual numerical instabilities affecting the recast for these UV models.\footnote{In particular, we suspect it comes from the phase-space integration for the resonant-mediator models in MadGraph, which we find to be sensitive to the sampling strategy used.}

\section{Exploring charm decays}
\label{sec:charm_decays}

Charm physics offers complementary information from the up-quark sector
on flavor in and beyond the SM. We explore  two-body decays of charm baryons
induced by the operators  (\ref{eqn:BNV_4F}).
In Sec.~\ref{sec:QCDF} we work out the $\Lambda_c \to \pi(K) + \bar N$  decay amplitudes and estimate decay rates  
within QCD factorization (QCDF).  Specifically, we compute the hard-scattering contribution to the $\Lambda_c \to \pi,K$ form factors, see App.~\ref{app:QCDF} for details.
This method has large uncertainties and moreover, gives a contribution that is $\alpha_s$-suppressed relative to the soft form factors.
The latter are non-perturbative objects that require non-perturbative methods, such as lattice QCD, or LCSR, neither of which is presently available
for $\Lambda_c \to \pi,K$.
We therefore view our form factor results as rather lower estimates of the full ones. Further study is clearly desirable. We also compute
upper limits on branching ratios using the \highPt constraints of Sec.~\ref{sec:res}.
We work out the reach at current and future experiments
in Sec.~\ref{sec:charm-reach} and discuss the complementarity of flavor and collider constraints in Sec.~\ref{sec:comp}.

\subsection{$\Lambda_c \to \pi(K) + \bar N $  in QCDF \label{sec:QCDF}}

We parametrize the matrix elements  $\Lambda_c(p+q) \to P(p) + \bar{N}(q) $, with $P=\pi,K$, as 
\begin{equation}
\label{eqn:charm_matrix_element}
\begin{aligned}
    i \mathcal{M}_{\pi}  =&  2 \frac{C^{qqdN}_{ \{1 2\} 1}}{\Lambda^2} \left( \bar v_N^C(q)  \bra{P(p)} \mathcal{\tilde O}^{3,L} \ket{\Lambda_c(p+q)} \right) + 
    \frac{C^{uddN}_{2 1 1}}{\Lambda^2}  \left( \bar v_N^C (q)  \bra{P(p)} \mathcal{\tilde O}^{3,R}\ket{\Lambda_c(p+q)} \right) \, ,  \\
    i \mathcal{M}_K  =&  2 \frac{C^{qqdN}_{\{1 2\} 2}}{\Lambda^2}  \left( \bar v_N^C(q)  \bra{P(p)} \mathcal{\tilde O}^{1,L} \ket{\Lambda_c(p+q)} \right) + 
    \frac{C^{uddN}_{2 1 2}}{\Lambda^2}  \left( \bar v_N^C (q)  \bra{P(p)} \mathcal{\tilde O}^{1,R}\ket{\Lambda_c(p+q)} \right)  \\
    +& 2 \frac{C^{qqdN}_{\{2 2\} 1}}{\Lambda^2}  \left( \bar v_N^C(q)  \bra{P(p)} \mathcal{\tilde O}^{2,L}\ket{\Lambda_c(p+q)} \right) + 
    \frac{C^{uddN}_{221}}{\Lambda^2}  \left( \bar v_N^C (q)  \bra{P(p)} \mathcal{\tilde O}^{2,R} \ket{\Lambda_c(p+q)} \right) \,, 
\end{aligned}
\end{equation}
with operators
\begin{equation}
\label{eqn:BNV_low_energy}
    \begin{aligned}
       \mathcal{\tilde O}^{1,X} &=  \epsilon_{\alpha \beta \gamma}    P_R s^{\gamma}\left(\bar c^{C,\alpha} P_X d^{\beta} \right)  \,,\\
       \mathcal{\tilde O}^{2,X} &=  \epsilon_{\alpha \beta \gamma}    P_R d^{\gamma}\left(\bar c^{C,\alpha} P_X s^{\beta} \right) \,,\\
       \mathcal{\tilde O}^{3,X} &=  \epsilon_{\alpha \beta \gamma}    P_R d^{\gamma}\left(\bar c^{C,\alpha} P_X d^{\beta} \right) \,,
    \end{aligned}
\end{equation}
with $X = R,L$.
The relevant reduced matrix elements can be parametrized as 
\begin{equation}
\label{eqn:LambdaCFormFactor}
\begin{aligned}
    i \mathcal{\tilde M}^{I,X} \equiv  \bra{ P(p)} \mathcal{\tilde O}^{I,X}\ket{\Lambda_c(p+q)} &=  P_R \left( F^{I,X}(q^2) + \frac{\slashed{q}}{M_{\Lambda_c}} \tilde F^{I,X}(q^2) \right) u_{\Lambda_c}(p+q) \,, \\
    \end{aligned}
\end{equation}
where we introduced twelve independent form factors $\{F^{I,X}, \tilde F^{I,X}\}$, one for each $X=R,L$ and $I=1,2,3$.
There is no contribution proportional to $P_L$, as $ P_L \cdot \mathcal{\tilde O}^{R,L}_{ijk} = 0$ and terms proportional to $\slashed{p}$ can be reabsorbed into $F^{I,X}, \tilde F^{I,X}$ by the equation of motion of $u_{\Lambda_c}$. 

As there is currently no input available for the form factors defined in Eqn.~\eqref{eqn:LambdaCFormFactor}, we estimate them within 
QCDF and the heavy quark expansion.
This allows us to estimate form factors at large hadronic recoil, $q^2 \approx  0$, see App.~\ref{app:QCDF} for details. 
We obtain, consistent with helicity arguments, the form factor relations, 
\begin{equation}
    \label{eqn:form_factor_relations}
    \begin{aligned}
    \tilde F^{I,L} & = - F^{I,L} \, , \quad     \tilde F^{I,R} =0 \, , \quad I =1,2,3  \, , \\
            F^{1,L} &=   F^{2,L}  \, , 
      \end{aligned}
\end{equation}
which can be used to reduce the twelve form factors to five independent ones.
Numerically,\footnote{Only the relative signs between different form factors matter and we do not make any final claims on the absolute sign.} 
\begin{equation}
\label{eqn:form_factors_numeric}
    \begin{aligned}
        F^{1,L}(0) & = -7.2\, \pm 0.7 ^{+0.3}_{-0.7} \, \cdot  \SI{e-3}{\GeV}^2 \, , \\ 
        F^{3,L}(0) & =  -1.2\, \pm 0.1 ^{+0.1}_{-0.5} \, \cdot \SI{e-2}{\GeV}^2   \, , \\
    \end{aligned}
\end{equation}
where the first (symmetric) uncertainties stem from input parameters in Tab.~\ref{tab:input_values_LCDA} and the second (asymmetric) uncertainties from variation of $\tau$, details are given in App.~\ref{app:QCDF}.  
See~\cite{Xing:2025pfw,Zheng:2024tkj,Li:2024htn}  for similar calculations of the  left-handed form factors in $b$-baryon decays.
Form factors $F^{I,R}$ arise at higher twist and are subject to endpoint singularities. Regulating them induces order one uncertainties,
see Fig.~\ref{fig:form_factors}.
Quantitatively, the $F^{I,R}$ can exceed the ones in Eqn.~\eqref{eqn:form_factors_numeric}. 
Since the uncertainties for $F^{I,R}$ are larger than for $F^{I,L}$, predictions for $C^{uddN} \cdot \mathcal{\tilde O}^{I,R}$ have larger uncertainties than
for $C^{qqdN} \cdot \mathcal{\tilde O}^{I,L}$. Further theory works is desirable.

Using the form factor relations Eqn.~\eqref{eqn:form_factor_relations} the matrix elements can be simplified to
\begin{equation}
\label{eqn:LambdaCFormFactor-simple}
\begin{aligned}
    \bra{ P(p)} \mathcal{\tilde O}^{I,L}\ket{\Lambda_c(p+q)} &=   P_R F^{I,L}(q^2)\left(1 -   \frac{\slashed{q}}{M_{\Lambda_c}}  \right) u_{\Lambda_c}(p+q) \,, \\
    \bra{ P(p)} \mathcal{\tilde O}^{I,R}\ket{\Lambda_c(p+q)} &=   P_R  F^{I,R}(q^2)   u_{\Lambda_c}(p+q)  \,. \\
    \end{aligned}
\end{equation}
Using the  Dirac equation $\bar v_N^C (q) \slashed{q} = M_N \bar v_N^C (q) $, the total matrix element in Eqn.~\eqref{eqn:charm_matrix_element} reads
\begin{equation}
\begin{aligned}
    i \mathcal{M}_K =& \bar v_N^C (q)  \Bigg( 2 \frac{C^{qqdN}_{\{ 12\} 2} + C^{qqdN}_{\{2 2 \} 1}}{\Lambda^2}  F^{1,L}(q^2) \left(P_R - P_L\frac{M_N}{M_{\Lambda_c}} \right) \\
&+P_R \frac{C^{uddN}_{2 1 2}}{\Lambda^2}  F^{1,R}(q^2) +P_R \frac{C^{uddN}_{2 2 1}}{\Lambda^2}  F^{2,R}(q^2)   \Bigg) u_{\Lambda_c}(p+q) \, , \\
 i \mathcal{M}_\pi =&  \bar v_N^C (q)  \Bigg( 2\frac{C^{qqdN}_{\{ 1 2 \} 1}}{\Lambda^2}  F^{3,L}(q^2) \left(P_R - P_L\frac{M_N}{M_{\Lambda_c}} \right) +  P_R\frac{C^{uddN}_{2 1 1}}{\Lambda^2}  F^{3,R}(q^2)  \Bigg) u_{\Lambda_c}(p+q) \,.
\end{aligned}
\end{equation}

The corresponding decay widths, with $F^{I,X} =F^{I,X}( M_N^2)$, read 

\begin{equation}
    \begin{aligned} \label{eq:Lc-gamma}
        \Gamma( \Lambda_c \to \pi + \bar{N} ) =& \frac{\mathcal{N}_\pi}{\Lambda^4}  \Bigg\{ 4 {C^{qqdN}_{\{12\}1}}^2 {F^{3,L}}^2 \left( \left( M_{\Lambda_c}^2 +M_N^2 -M_\pi^2 \right) \left( 1 + \frac{M_N^2}{M_{\Lambda_c}^2}\right) - 4  M_N^2 \right)  \\
        &+ {C^{uddN}_{211}}^2 {F^{3,R}}^2 \left( M_{\Lambda_c}^2 + M_N^2 - M_\pi^2 \right) \\
        &+ 4 C^{uddN}_{211} C^{qqdN}_{\{12\}1} F^{3,R} F^{3,L} \left( M_{\Lambda_c}^2 - M_N^2 - M_\pi^2 \right)   \Bigg\} \,,\\
        \Gamma( \Lambda_c \to K + \bar{N} ) =& \frac{\mathcal{N}_K}{\Lambda^4}  \Bigg\{ 4 \left( C^{qqdN}_{\{ 12\} 2} + C^{qqdN}_{\{2 2 \} 1} \right)^2 {F^{1,L}}^2 
        \left( \left( M_{\Lambda_c}^2 +M_N^2 -M_K^2 \right) \left( 1 + \frac{M_N^2}{M_{\Lambda_c}^2}\right) - 4  M_N^2 \right)   \\
         &+ \left(C^{uddN}_{212} F^{1,R}+ C^{uddN}_{221} F^{2,R}\right)^2 \left( M_{\Lambda_c}^2 + M_N^2 - M_K^2 \right) \\
         &+ 4 \left( C^{qqdN}_{\{ 12\} 2} + C^{qqdN}_{\{2 2 \} 1} \right)F^{1,L} \left(C^{uddN}_{212} F^{1,R}+ C^{uddN}_{221} F^{2,R}\right)\left( M_{\Lambda_c}^2 - M_N^2 - M_K^2 \right)    \bigg\}\,,
    \end{aligned}
\end{equation}
with prefactor 
\begin{equation}
    \mathcal{N}_P = \frac{1}{32 \pi M_{\Lambda_c}^3}  \lambda^{1/2}\left(M_{\Lambda_c}^2,M_N^2,M_P^2 \right) \,, 
\end{equation}
where $\lambda^{1/2}$ is the square root of the Källén function.
To extrapolate the form factors in Eqn.~\eqref{eqn:form_factors_numeric} to $q^2=M_N^2$, we use
\begin{equation}
    F^{I,X}(q^2) = \frac{1}{1 - \frac{q^2}{M_{\text{pole}}^2}} F^{I,X}(0)\,,
\end{equation}
where $M_{\text{pole}}$ corresponds to the lowest mass state of the respective operator. For $\Lambda_c \to \pi(K) + \bar N $ we use $M_{\Sigma^0_c}$($M_{\Xi^0_c}$).
Both the derivation of the form factor relations and the numerical values of the form factors have been obtained using heavy quark methods. For the decays at hand, there are 
therefore several caveats, that make further theory studies desirable: First, charm is not very much separated from the QCD scale,
and second, we need the form factors at $q^2$ within 1 and 4 GeV$^2$, which makes the $\Lambda_c$-decay products not very energetic.
In absence of other means we make theses approximations to illustrate the achievable rates.

\begin{figure}
    \centering
      \includegraphics[width=0.49\linewidth]{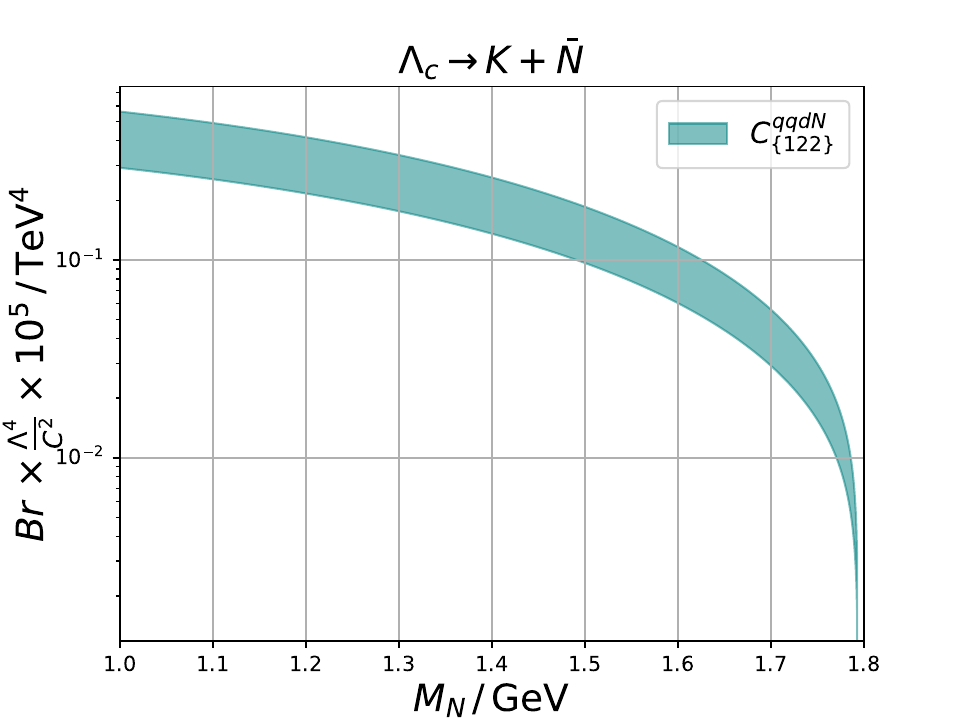}
      \includegraphics[width=0.49\linewidth]{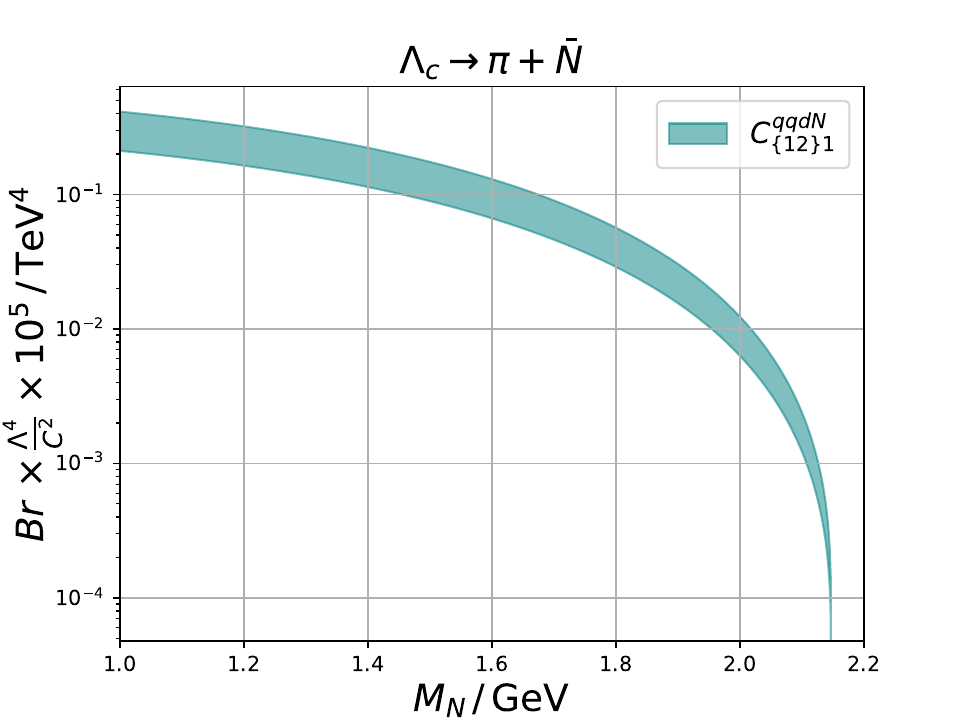}
    \caption{Branching ratios for $\Lambda_c \to \pi^+ (K^+) + \bar N $ decays in units of $C^2/\Lambda^4 \text{ TeV}^4$.
    The envelope is obtained using the maximal and minimal form factor values in 
    Eqn.~\eqref{eqn:form_factors_numeric}.}
    \label{fig:branching_ratios_LambdaC}
\end{figure}

In Fig.~\ref{fig:branching_ratios_LambdaC} we show the branching ratios of $\Lambda_c \to K (\pi) + \bar{N}$  in units of $C^2/\Lambda^4 \text{ TeV}^4$ 
for the doublet operator, for which theory uncertainties are more modest and we define\footnote{This differs from the standard definition of a totally symmetric tensor.} the totally symmetric WC 
\begin{equation}
    C^{qqdN}_{\{1 2 2\}} \equiv \frac{1}{2} \left( C^{qqdN}_{\{ 12\} 2} + C^{qqdN}_{\{2 2 \} 1} \right)\,.
\end{equation}
which is the only independent combination probed by $\Lambda_c \to K + \bar{N}$.
With  the limits on single WCs, obtained in Sec.~\ref{sec:res} through MET searches, we find branching ratios as large as
\begin{equation}
\label{eqn:charm_BR_UL}
    \begin{aligned}
        &\text{Br}\left(\Lambda_c \to K + \bar N   \right)\bigg\vert_{C^{qqdN}_{\{122\}}} \lesssim  \num{5.1e-9}  \,, \\
        &\text{Br}\left(\Lambda_c \to \pi + \bar N \right)\bigg\vert_{C^{qqdN}_{\{12\}1}} \lesssim  \num{2.8e-9}  \,.
    \end{aligned}
\end{equation}
Taking into account the limits on $C^{uddN}$ and the form factors from Fig.~\ref{fig:form_factors}, the $\Lambda_c \to (K, \pi) + \bar N $
branching ratios induced by $\mathcal{O}^R$ can be about one order of magnitude larger than those in (\ref{eqn:charm_BR_UL}). In view of the substantial uncertainties we refrain from further
phenomenology from $\mathcal{O}^R$.

\subsection{Reach of $\Lambda_c$ decays \label{sec:charm-reach}}

To estimate the reach of the $\Lambda_c \to \pi(K) + \bar N$ branching ratios at future charm experiments, we assume benchmarks listed in Tab.~\ref{tab:charm_benchmark},
zero background and use the significance 
\begin{equation}
    Z =  N_{\Lambda_c} \times \epsilon \times \text{Br}(\Lambda_c \to \pi(K) + \bar N ) \,,
\end{equation}
where $Z = 2.3$  corresponds to $90 \%$ C.L. limits, given zero background and $\epsilon$ denotes the signal efficiency. 
As dedicated studies per experiment are not available, to make progress, we use $\epsilon \sim  0.01$ 
\footnote{  Ref.~\cite{Wang:2026vxh} obtains an order of magnitude larger efficiency for the STCF.}
which is supported by charm to invisibles in the FCC-ee environment \cite{DiCanto:2025fpk}.

\begin{figure}
    \centering
    \includegraphics[width=0.44\linewidth]{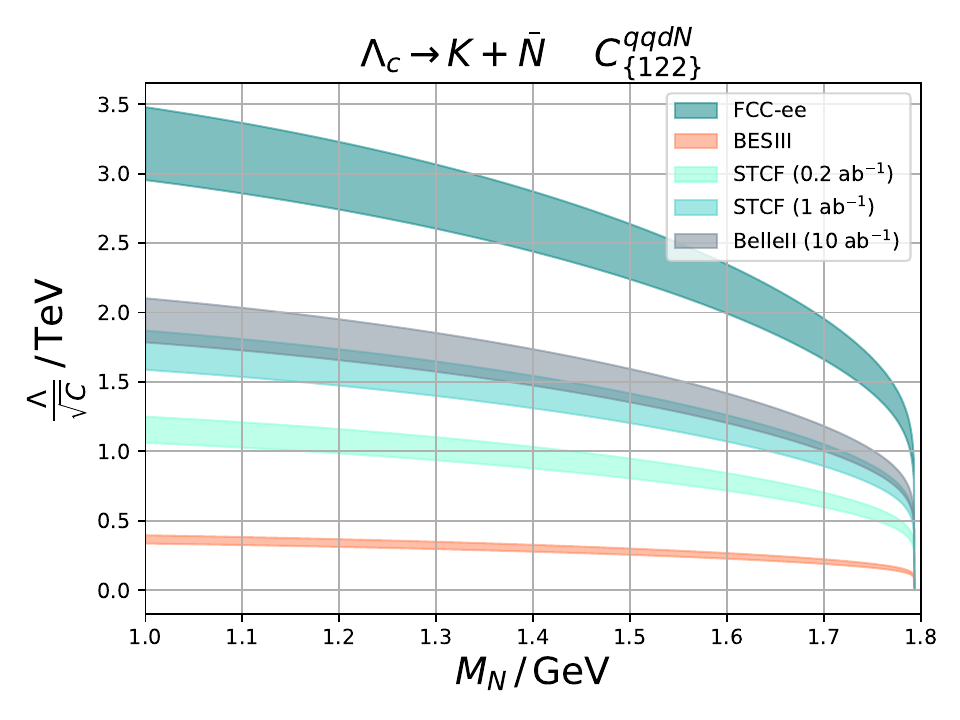}
    \includegraphics[width=0.44\linewidth]{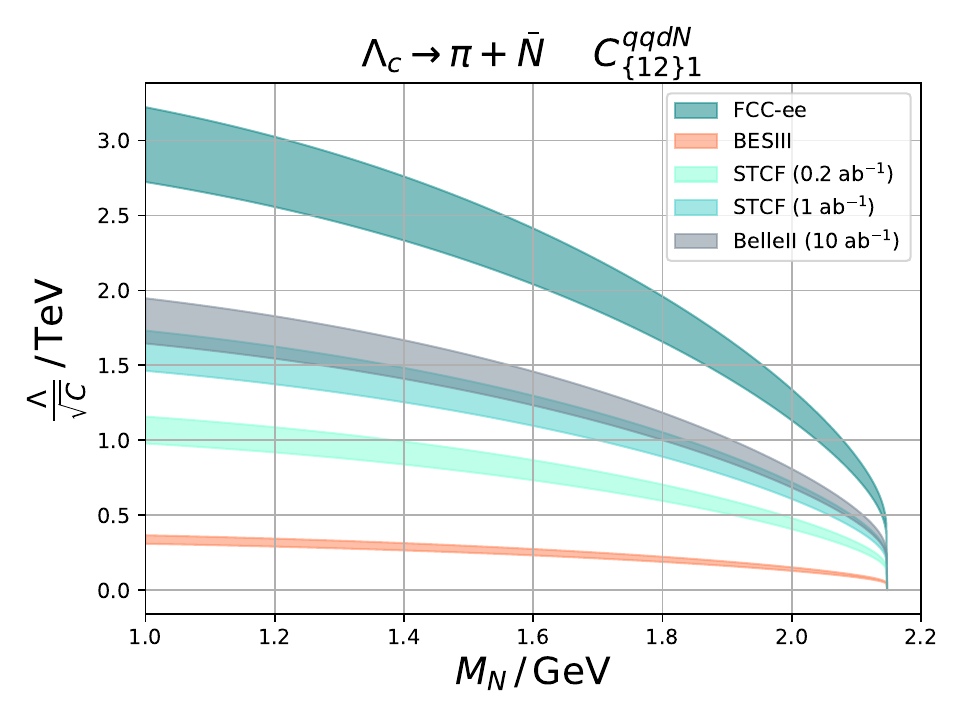}
    \caption{One parameter sensitivities on $\frac{\Lambda}{\sqrt{C}}$ at $90 \%$ C.L.  from four-fermion operators \eqref{eqn:BNV_4F} 
    for $\Lambda_c \to K + \bar N $  (left) and $\Lambda_c \to \pi + \bar N $ (right) at experimental setups summarized in Tab.~\ref{tab:charm_benchmark}.
    The bands correspond to  maximal and minimal form factor values in 
    Eqn.~\eqref{eqn:form_factors_numeric}.}
    \label{fig:LambdaC_sens}
\end{figure}

\begin{table}[]
    \centering
    \begin{tabular}{c c c c }
        Experiment & $\mathcal{L}^{\text{int}} \,/ \, \text{ab}^{-1}$ & $N_{\Lambda_c}$ & Ref. \\ \toprule
        Belle II & $10 $ & $\sim \num{8e8}$ & \cite{Achasov:2023gey}${}^{\dagger}$ \\
        BESIII  &  $\sim 0.02$ &$ \sim \num{1e6}$ &  \cite{Achasov:2023gey} \\
        STCF    & $0.2(1) $ & $\sim \num{1e8}(\sim \num{5e8}) $& \cite{Achasov:2023gey} \\
        FCC-ee  &  & $ \num{6e9} $& \cite{DiCanto:2025fpk}\\
    \end{tabular}
    \caption{Benchmarks used for the sensitivity analysis of $\Lambda_c \to K, \pi + \bar N$ in Fig.~\ref{fig:LambdaC_sens}. $^\dagger$We naively rescaled the
     $50 \text{ab}^{-1}$  event rate by $1/5$.}
    \label{tab:charm_benchmark}
\end{table}
Sensitivities of $\Lambda_c $-decays are shown in Fig.~\ref{fig:LambdaC_sens} in terms of $\frac{\Lambda}{\sqrt{C}}$. 
The FCC-ee sensitivities for couplings involving charm quarks probe scales up to $\SI{3.2}{\TeV}(\SI{3.5}{\TeV})$ for decays to $\pi(K)$. 
The uncertainties in the  $\Lambda_c$-branching ratios are quite sizable, however, being SM null tests, any signal would correspond to NP.

\subsection{Complementarity high and low $p_T$ \label{sec:comp}}

\begin{figure}
    \centering
    \includegraphics[width=0.48\linewidth]{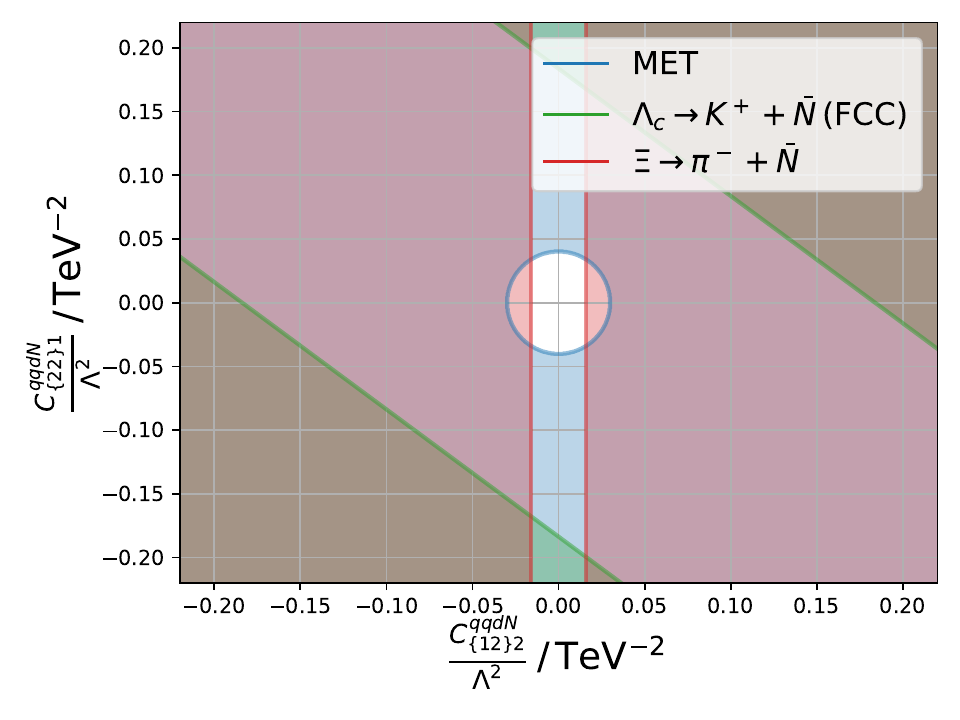}
    \includegraphics[width=0.48\linewidth]{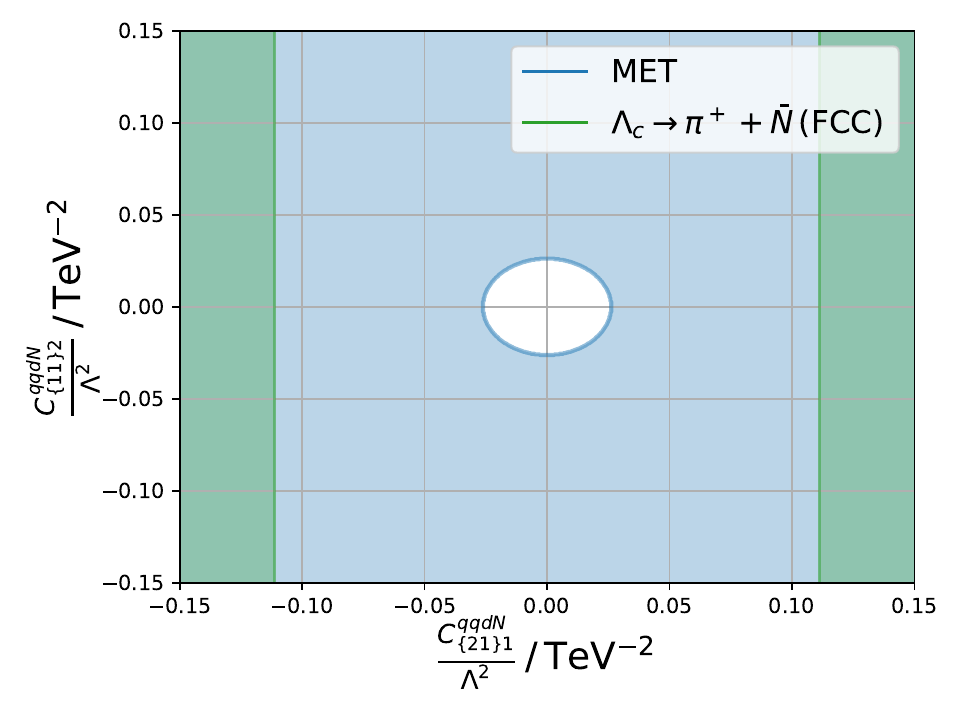}
    \caption{Bounds in the  ($C^{qqdN}_{\{12\}2},C^{qqdN}_{\{22\}1} $)-plane (left) and ($C^{qqdN}_{\{12\}1},C^{qqdN}_{\{11\}2}  $)-plane (right) derived 
    through MET + jet in Fig.~\ref{fig:NP_BNV} and $\Xi$ decays bounds in Fig.~\ref{fig:brs}  based on \cite{Alonso-Alvarez:2021oaj,BESIII:2025sfl} and $\Lambda_c$ FCC-ee sensitivities in Fig.~\ref{fig:LambdaC_sens}. Rare decay limits only apply if the singlet $N$ is sufficiently light and kinematically accessible. }
    \label{fig:lowEnergy_highPt}
\end{figure}
In Fig.~\ref{fig:lowEnergy_highPt} we show a direct comparison between the MET + jets bounds, from Fig.~\ref{fig:NP_BNV}, limits from $\Xi$ decays  in Fig.~\ref{fig:brs}  and $\Lambda_c$ FCC-ee sensitivities in Fig.~\ref{fig:LambdaC_sens} at fixed $M_N = \SI{1}{\GeV}$. 
While the high-$p_T$ data constrain the sum of the squares of the two WCs, in $\Lambda_c$-decays they interfere, see Eqn.~(\ref{eq:Lc-gamma}), whereas 
hyperon decays only probes one of them. Despite this complementarity Fig.~\ref{fig:lowEnergy_highPt} shows that the MET + jet bounds generally are more constraining, where $C^{qqdN}_{\{12\}2}$ is more constrained by $\Xi$-decays. Generally $\Lambda_c$ sensitivities estimated for the FCC-ee are weaker, however, cover larger range of $M_N$ than the hyperons. We also stress that an observation of such an apparently BNV-decay would
cleanly prove BNV.

\section{Reach in top decays \label{sec:top}}

\begin{figure}
    \centering
    \includegraphics[width=0.48\linewidth]{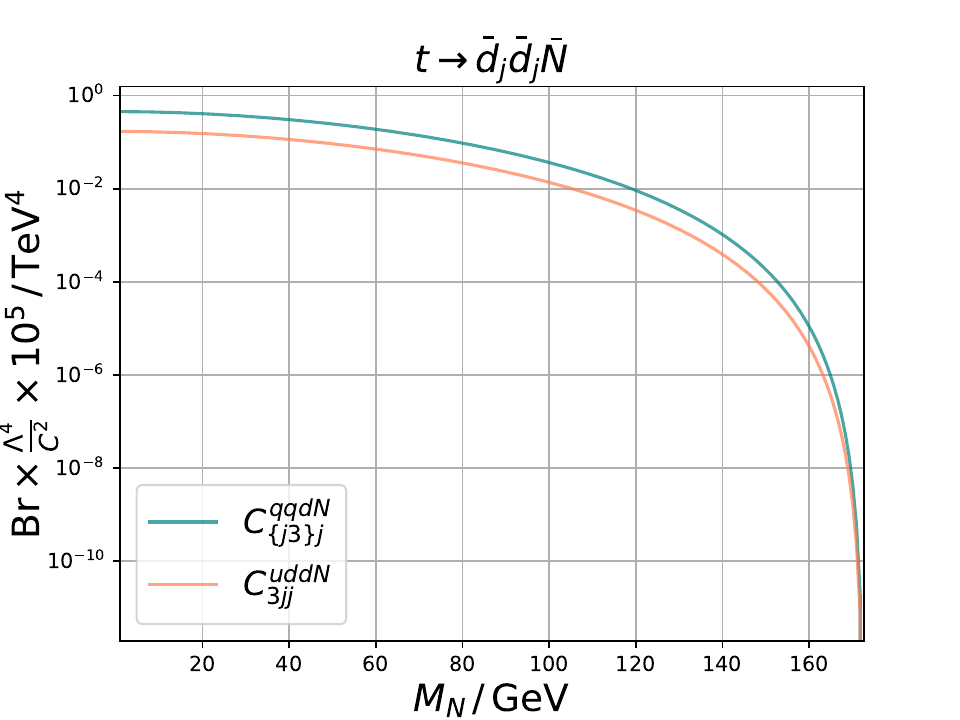}
    \includegraphics[width=0.48\linewidth]{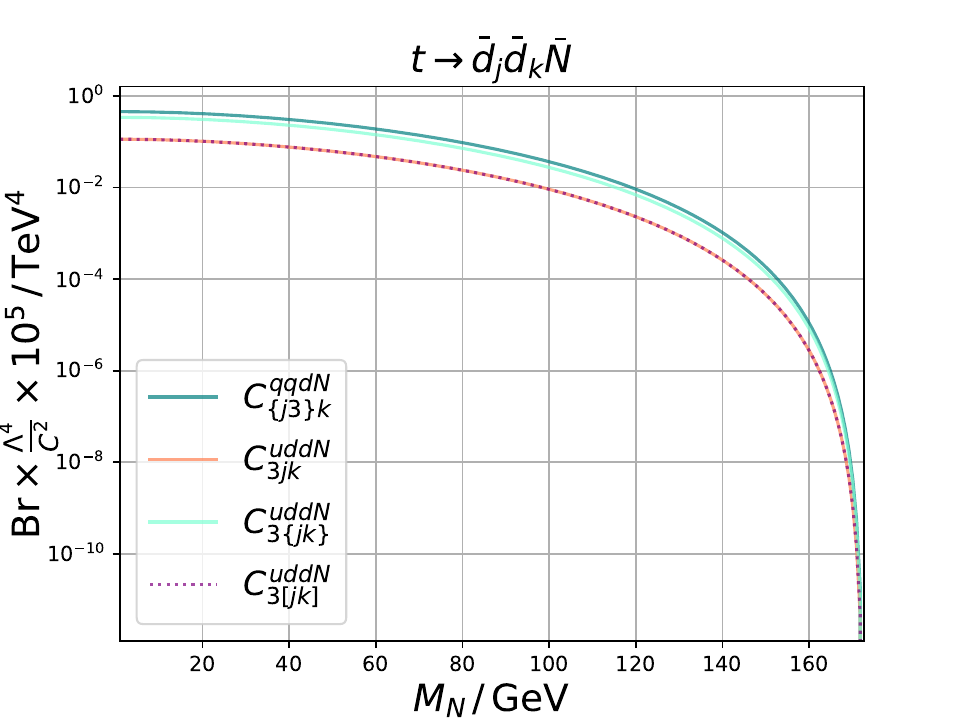}
    \caption{Predictions for top-quark branching ratios in units of $\frac{C^2}{\Lambda^4}$ to two down-type quarks and the singlet $N$.
    Shown are the cases of two identical down-type quarks $t \to \bar d_j \bar d_j \bar N $ (left) and different down-type quarks $t \to \bar d_j \bar d_k \bar N \, ,  j \neq k$ (right) induced by different combinations of WCs.}
    \label{fig:top_BR}
\end{figure}
\begin{table}[]
    \centering
    \begin{tabular}{c c  c c c }
           $\text{Br}\left(t \to \bar d_j \bar d_j \bar{N} \right)$ & $C^{qqdN}_{\{j3\}j}$ & $C^{uddN}_{3jj}$ & – & – \\ \toprule 
            $11$ & $\num{1.8e-9}$ &$ \num{8.8e-07}$ & $–$ &$–$  \\ 
            $22$ & $\num{3.7e-8}$ &$ \num{8.9e-08}$ & $–$ &$–$  \\ 
            $33$ & $\num{8.8e-7}$ &$ \num{4.7e-6}$ & $–$ &$–$  \\
             $\text{Br}\left(t \to \bar d_j \bar d_k \bar{N} \right)$& $C^{qqdN}_{\{j3\}k} $& $C^{uddN}_{3jk}$ & $C^{uddN}_{3\{jk\}}$ & $C^{uddN}_{3[jk]}$  \\ \toprule 
             $12$ &\num{4.1e-09} &\num{8.7e-8} &   \num{8.9e-11}    &\num{2.8e-6} \\
             $21$ &\num{1.1e-08} &\num{8.4e-8} &   \num{8.9e-11}   &\num{2.8e-6} \\
             $13$ &\num{7.3e-09} &\num{1.2e-6} &   \num{5e-9}     &\num{1.8e-6} \\
             $31$ &\num{4.5e-08} &\num{9.6e-7} &   \num{5e-9}    &\num{1.8e-6} \\
             $23$ &\num{6.5e-08} &\num{4.5e-6} &   \num{5.8e-9}  &\num{3.6e-9} \\
             $32$ &\num{1.2e-07} &\num{3.8e-6} &   \num{5.8e-9}  &\num{3.6e-9} \\
    \end{tabular}
    \caption{Upper limits ($90 \% $C.L.) for branching ratios of $t \to \bar d_j \bar d_k \bar N $ assuming one WC at a time given in the upper rows.
     The bounds are given for $M_N = \SI{1}{\GeV}$, for which the $N$-particle is always detector stable, see Sec.~\ref{sec:N_decay}. 
    However the bounds are approximately valid for larger masses, depending on possible final states of the particles, see Sec.~\ref{sec:N_decay} and Fig.~\ref{fig:CuddN_constraints}.}
    \label{tab:top_UL}
\end{table}
As couplings involving top-quarks are generally constrained less by the \highPt bounds in Sec.~\ref{sec:highPt_theory}, the measurement of top-quark decays $t \to  \bar N  \bar d \bar d'$ could lead to a potential improvement.
Similar searches have been carried out for BNV top-quark decays to charged leptons and two antiquarks~\cite{CMS:2013zol,CMS:2024dzv}, excluding small branching ratios such as \cite{CMS:2024dzv},
\begin{equation} \label{eq:top-limits}
\begin{aligned}
    &\text{Br}\left(t \to \mu^+ \bar u \bar d \right) \lesssim     \num{2e-9} \, ,  \\
    &\text{Br}\left(t \to e^+ \bar c \bar b \right) \lesssim     \num{2e-6} \, , 
\end{aligned}
\end{equation}
corresponding to the most stringent and least stringent $ 95 \%  $ C.L. experimental upper limit, respectively.

The three-body decay width can be obtained from Eqn.~\eqref{eqn:BNV_partial_width} by crossing symmetry, accounting for a factor $3$ due to color averaging,  as
\begin{equation}
    \label{eqn:top_partial_width }
    \begin{aligned}
    \Gamma(t \to \bar N \bar d_j \bar d_k) =& \frac{ m_t^5}{3072 \pi^3 \Lambda^4 \left( 1 + \delta_{jk}\right)} \left( 4  \left({C^{qqdN}_{\{j3\}k}}^2 +  {C^{qqdN}_{\{3k\}j}}^2\right)  +  {C^{uddN}_{3jk}}^2 + {C^{uddN}_{k3j}}^2  + C^{uddN}_{3jk} C^{uddN}_{3kj}\right)  \\
                                & \times \left( 1 - x^8 + 8 \left( x^6 - x^2 \right) - 24 x^4 \log x \right) \, , \\
    \end{aligned}
\end{equation}
where $x = M_N \, / \, m_t$  and the down quarks are assumed to be massless.
In Fig.~\ref{fig:top_BR} top quark branching ratios for $t \to \bar d \bar d' \bar N $ are shown for identical and different down-quark types, as well as different combinations of WCs.
In Tab.~\ref{tab:top_UL} we give upper limit on branching ratios for top-decays, using the  constraints obtained in Sec.~\ref{sec:res}.
Depending on the mass of $N$, branching ratios as large
as $\sim \text{few} \times 10^{-6}$ can arise, corresponding to $t \to \bar N \bar b \bar b$. In view of the strong limits from the LHC (\ref{eq:top-limits}), study at
the LHC for BNV with MET in top decays is encouraged, particularly because the operators probed by the charged-lepton limits in Eqn.~(\ref{eq:top-limits}) are already severely constrained by proton-stability bounds (see Ref.~\cite{Gisbert:2024sjw}), whereas the operators \eqref{eqn:BNV_4F} involving the $N$-particle are not.

 At the HL-LHC the expected number of $t \bar t$ events is large, $\sim 10^{10}$. Searches in, for instance,  $bbb \ell^++$ MET and
$\bar b\bar b \bar b \ell^-+$ MET, where the opposite, tagging top is decaying to
$W^+b$ and $W^- \bar b$, would therefore imply up to few $\times 10^4$ events, modulo detection efficiencies.

\section{Conclusions \label{sec:con}}

We study extensions of the SM in which MET is linked to baryon number, 
in the form of a  light SM-singlet fermion $N$, with  mass above the proton one.
We consider a minimal framework where $N$ is vector-like, to allow for a mass term, and
carries baryon number, to evade baryon-oscillation constraints and to decouple it from the lepton sector.
We stress, however, that the LHC-limits we obtain on operators hold in much more general settings.

Using an EFT that only keeps the SM plus $N$ as light degrees of freedom we find that existing LHC data constrain scales up to 10 TeV, and if sensitive to third generation quarks the MET+$b$ searches reach  11 TeV. The limits depend on the quark flavors involved, and are detailed in Tab.~\ref{tab:bounds_numerical_CuddN} and Tab.~\ref{tab:bounds_numerical_CqqdN} for the two operators that arise at dimension six, Eqns.~\eqref{eqn:BNV_4F}.
The operators are induced at tree-level from three mediators, Tab.~\ref{tab:UV_mediators}.
In these models, also $N$-pair production and four-quark operators are induced.
Combination of observables provides synergy, resulting in improved  constraints, see 
Fig.~\ref{fig:LQ_plot}.
Interestingly, constraints on the mesogenesis model with heavy color triplet scalars \cite{Alonso-Alvarez:2021qfd} are severe.
While there is none left in model 1, see 
Fig.~\ref{fig:LQ_plot}, a small region of parameter space remains viable in model 2 in flavor combinations $cbs, cbd$ and $ubs$, 
see Fig.~\ref{fig:Mesogenesis_model2}. 
Corresponding $B$-decays include $B^0 \to \Lambda N$, $B^+ \to \Lambda_c N$, and 
$B^{0,+} \to \Xi^{0,+}_c N$, or $B_s \to \Omega^0_c N$.
About a factor two improvement in reach, as expected from $3 \, \text{ab}^{-1}$ at HL-LHC, can close the window
with dijets, single-$N$, or $N$-pair production. Given the mass scale $\sim  y \cdot 3\, \text{TeV}$ (with charm) and $\sim y \cdot 5 \, \text{TeV}$ (charmless) for couplings $y$, 
dedicated  resonance searches are also promising.

Findings are largely independent of the $N$ mass, as long as it appears as MET. However,
once the lifetime reduces due to larger couplings or larger mass, displaced vertices appear.
In absence of a corresponding search we utilize a related LHC analysis \cite{ATLAS:2023oti}.
Fig.~\ref{fig:CuddN_constraints} 
illustrates the potential of such a measurement, once done fully-fledged and model-specific: it allows to probe different parameter space and has very small background.
 For Majorana-$N$, the decay width is a factor 2 larger while the production cross section 
is the same as in the Dirac case. While this shortens the decay length and therefore in principle the applicability of search criteria, see e.g. 
Figs.~\ref{fig:decay_plots} and \ref{fig:CuddN_constraints},
we expect the limits to be similar.

Interactions also induce exclusive  rare decays of type baryon (meson) to meson (baryon) plus MET if kinematically allowed.  While except for strange hadrons  less sensitive than the high $p_T$ searches they could prove BNV.
We work out signatures in charm, $\Lambda_c \to (\pi,K) + \mathrm{invisible}$.
Branching ratios have sizable uncertainties which invite further theory study. Constraints from high $p_T$ require  high luminosity flavor facilities 
such as a  Tera-$Z$-factory FCC-ee, or  CEPC.
Synergies arise with hyperon decays and MET-collider searches, illustrated in Fig.~\ref{fig:lowEnergy_highPt}.

We also investigate the reach in top decays, see Sec.~\ref{sec:top}.
Branching ratios as large
as $\sim \text{few} \times 10^{-6}$ can arise, especially for $t \to N \bar b \bar b$, or into a single $b$,
see Tab.~\ref{tab:top_UL}.
This motivates further searches at
the LHC with MET in top decays, particularly because the operators probed by existing searches in top decays with charged leptons are already severely constrained by proton-stability bounds, whereas the operators considered here are not.

\acknowledgments

We are grateful to Hector Gisbert for collaboration during the initial stages of this project.
We also like to thank
Alexandre Carvunis, Bhupal Dev, Miguel Escudero and Dominik Suelmann for useful discussions,
 and in particular Chaja Baruch, Gilly Elor, Jared Goldberg, Omer Shtaif and Yotam Soreq for drawing our attention to lepton cuts in the MET+top analysis. ARS has been supported by the Spanish Government (Agencia Estatal de Investigación
MCIN/AEI/10.13039/501100011033) Grant No. PID2023-146220NB-I00.

\appendix

\section{Additional MET-Results}
\label{app:AddResults}
In Tab.~\ref{tab:bounds_numerical_CuddN} and Tab.~\ref{tab:bounds_numerical_CqqdN}  we give numerical values for the bounds obtained 
on the WCs in Sec.~\ref{sec:res}  from the MET +jets, MET + bjets, and MET + top  observables.
\begin{table}[]
    \centering
    \resizebox{0.6\columnwidth}{!}{
    \begin{tabular}{c| c| c | c |c | c | c }
    &\multicolumn{2}{c|}{MET + jets } & \multicolumn{2}{c|}{MET + top}& \multicolumn{2}{c}{MET + bjet}   \\
    $ijk$ & $ \frac{|C^{uddN}_{ijk}|}{\Lambda^2} \,/ \, \si{\TeV}^{-2} $ & $\frac{\Lambda}{\sqrt{|C^{uddN}_{ijk}|}}\,/ \, \si{\TeV}$ & $ \frac{|C^{uddN}_{ijk}|}{\Lambda^2} \,/ \, \si{\TeV}^{-2} $ & $\frac{\Lambda}{\sqrt{|C^{uddN}_{ijk}|}}\,/ \, \si{\TeV}$   & $ \frac{|C^{uddN}_{ijk}|}{\Lambda^2} \,/ \, \si{\TeV}^{-2} $ & $\frac{\Lambda}{\sqrt{|C^{uddN}_{ijk}|}}\,/ \, \si{\TeV}$\\
    \toprule 
    $111$ &0.014  &8.5   &- &-                &- &-       \\
    $112$ &0.018  &7.5   &- &-                &- &-       \\
    $113$ &0.020   &7.1   &- &-                &0.016 &7.9     \\
    $121$ &0.026  &6.2   &- &-                &- &-       \\
    $122$ &0.028  &6.0   &- &-                &- &-       \\
    $123$ &0.041  &4.9   &- &-                &0.026 &6.2       \\
    $131$ &0.036  &5.3   &- &-                &0.029 &5.9       \\
    $132$ &0.071  &3.8   &- &-                &0.043 &4.8       \\
    $133$ &0.10   &3.2   &- &-                &0.050 & 4.5      \\
    $211$ &0.065  &3.9   &- &-                &- &-       \\
    $212$ &0.081  &3.5   &- &-                &- &-       \\
    $213$ &0.096  &3.2   &- &-                &0.070 & 3.8       \\
    $221$ &0.11   &3.0   &- &-                &- &-       \\
    $222$ &0.12   &2.9   &- &-                &- &-       \\
    $223$ &0.19   &2.3   &- &-                &0.11 & 3.0       \\
    $231$ &0.17   &2.5   &- &-                &0.12 & 2.9      \\
    $232$ &0.28   &$1.9^*$   &- &-                &0.17 & 2.4      \\
    $233$ &0.34   &$1.7^*$   &- &-                &0.19 & 2.3      \\
    $311$ &0.28   &$1.9^*$   &0.077  &3.6          &-  &-   \\
    $312$ &0.28   &$1.9^*$   &0.074  &3.7          &-  & -   \\
    $313$ &1.0    &$1.0^*$   &0.19  &2.3          &-  & -   \\
    $321$ &0.27   &$1.9^*$   &0.075  &3.6          &-  & -   \\
    $322$ &1.0    &$1.0^*$   &0.23  &2.1          &-  & -   \\
    $323$ &2.0    &$0.7^*$   &0.34  &1.7          &-  & -   \\ 
    $331$ &0.92   &$1.0^*$   &0.18  &2.3          &-  & -   \\
    $332$ &1.8    &$0.7^*$   &0.33  &1.8          &-  & -   \\
    $333$ &8.5    &$0.3^*$   &$1.1^*$  &$0.9^*$  &-  & -     \\
    \end{tabular}
    }
    \caption{One-parameter limits obtained on the BNV four-fermion operator $\mathcal{O}^{uddN}_{ijk} $  based on the MET + jets, MET+ top and MET + bjet observables.
    The value $M_N = \SI{1}{\GeV}$ is fixed, but some bounds are valid for larger ranges, see discussion in Sec.~\ref{sec:res}.
    A corresponding $\Lambda$-reach plot is presented in Fig.~\ref{fig:NP_BNV}. Bounds marked with $^*$ are outside of the validity of the EFT, see discussion of Fig.~\ref{fig:NP_BNV}. }
    \label{tab:bounds_numerical_CuddN}
\end{table}
\begin{table}[]
    \centering
    \resizebox{0.6\columnwidth}{!}{
    \begin{tabular}{c| c| c | c |c | c | c }
    &\multicolumn{2}{c|}{MET + jets } & \multicolumn{2}{c|}{MET + top}& \multicolumn{2}{c}{MET + bjet}  \\
    $ijk$ & $ \frac{|C^{qqdN}_{\{ij\}k}|}{\Lambda^2} \,/ \, \si{\TeV}^{-2} $ & $\frac{\Lambda}{\sqrt{|C^{qqdN}_{\{ij \}k}|}}\,/ \, \si{\TeV}$ & $ \frac{|C^{qqdN}_{\{ij\}k}|}{\Lambda^2} \,/ \, \si{\TeV}^{-2} $ & $\frac{\Lambda}{\sqrt{|C^{qqdN}_{\{ij\}k}|}}\,/ \, \si{\TeV}$ & $ \frac{|C^{qqdN}_{\{ij\}k}|}{\Lambda^2} \,/ \, \si{\TeV}^{-2} $ & $\frac{\Lambda}{\sqrt{|C^{qqdN}_{\{ij\}k}|}}\,/ \, \si{\TeV}$   \\
    \toprule 
    111& 0.0090 & 10.5 & -    &-    & -     &-  \\
    112& 0.0090 & 10.5 & -    &-    & -     &-  \\
    113& 0.010  & 10.0 & -    &-    & 0.0081      &11.1    \\
    121& 0.023 & 6.6 & -     &-    & -     &-  \\
    122& 0.033 & 5.5 & -     &-    & -     &-  \\
    123& 0.037 & 5.2 & -     &-    &  0.024    &6.5  \\
    131& 0.038 & 5.1 & 0.016 &7.9  &  0.026    &6.2   \\
    132& 0.070  & 3.8 & 0.027 &6.1  &  0.040    &  5.0  \\
    133& 0.13  & 2.8 & 0.036 &5.3  &  0.062    & 4.0   \\
    221& 0.050  & 4.5 & -     & -   & -     & -   \\
    222& 0.070  & 3.8 & -     & -   & -     & -   \\
    223& 0.088 & 3.4 & -     & -   &  0.053     & 4.3   \\
    231& 0.15  & 2.6 & 0.054 &4.3  & 0.084 & 3.5  \\
    232& $0.25^*$  &$ 2.0^*$ & 0.089  &3.4  & 0.13 & 2.8  \\
    233& $0.43^*$  &$ 1.5^*$ & 0.12  &2.9  & 0.22 & 2.1  \\
    331& $0.52^*$  &$ 1.4^*$ & 0.096  &3.2  & - & -   \\
    332& $0.91^*$  &$ 1.0^*$ & 0.16  &2.5  & - & -   \\
    333& $3.1^*$  &$ 0.6^*$  & 0.45  &1.5  & - & -   \\
    \end{tabular}
    }
    \caption{One-parameter limits obtained on the BNV four-fermion operator $\mathcal{O}^{qqdN}_{ijk} $  based on the MET + jets, MET+ top and MET + bjet observables.
    The value $M_N = \SI{1}{\GeV}$ is fixed, but some bounds are valid for larger ranges, see discussion in Sec.~\ref{sec:res}.
    A corresponding $\Lambda$-reach plot is presented in Fig.~\ref{fig:NP_BNV}. Bounds with $^*$ are possibly outside of the 
    validity of the EFT, see discussion of Fig.~\ref{fig:NP_BNV}.}
       \label{tab:bounds_numerical_CqqdN}
\end{table}

\section{$\Lambda_c$-decay amplitudes in QCDF }
\label{app:QCDF}

Here, we describe the calculation of the form factors defined in Eqn.~\eqref{eqn:charm_matrix_element}.
We start by introducing the framework, follow up with the calculation of the matrix element and show numerical results.

\subsection{QCDF framework}
We estimate the decay amplitude of $\Lambda_c(p + q) \to P(p) + \bar N(q) $, where $P = \pi, K$ in QCDF and 
the heavy quark expansion. We compute the hard-spectator interaction from the exchange of a gluon with the spectator quark and 
quarks connected to the operators defined in Eqn.~\eqref{eqn:BNV_low_energy}. Diagrams are shown in Fig.~\ref{fig:LambdaC_Pi}.
The $\Lambda_c \to P$ soft-form factors from the Feynman-mechanism at $\alpha_s^0$ require a non-perturbative computation such as lattice QCD or sum rules, neither of 
which is presently available.

Within HQET  $M_{\Lambda_c} = m_c+\bar \Lambda+\mathcal{O}(1/m_c)$, where $\bar \Lambda$ corresponds to a scale of order $\Lambda_{\text{QCD}}$ that
stands for binding energy. The $\Lambda_c$ is assumed to be at rest with velocity  $v^{\mu}$, $v^2=1$.
The heavy charm quark  has momentum $p_c^\mu=m_c v^\mu + k^\mu$, with a residual momentum $k\sim \Lambda_{\text{QCD}}$.
Let the pseudoscalar meson (singlet fermion) scatter along the $n^{\mu}$  ($\bar n^\mu$) direction,
with $2 v^{\mu} = n^{\mu} + \bar n^{\mu}$, where  
\begin{equation}
    \begin{aligned}
        n^{\mu} = ( 1,0,0,1) \, , \quad
        \bar n^{\mu} = (1,0,0,-1) 
    \end{aligned}
\end{equation} 
are the light-cone coordinate vectors with $n^2=\bar n^2 =0$ and $n \cdot \bar n=2$.
This allows one to expand any momentum vector $p_x$ as 
\begin{equation}
    \begin{aligned}
    p_x^{\mu} &= \frac{n^{\mu}}{2}  ( \bar n \cdot p_x) + \frac{\bar n^{\mu}}{2}  ( n \cdot p_x) + p^{\mu}_{T}\\
     &\equiv ( n \cdot p_x , \bar n \cdot p_x , p^{\mu}_{T}) \, ,
    \end{aligned}
\end{equation}
with transverse component $ p^{\mu}_{T}$. We neglect any transverse ones $p_T^{\mu} \sim \Lambda_{\text{QCD}}$.
Furthermore we neglect all  masses except $m_c$. This leads to the representation of  external momenta 
\begin{equation}
    \begin{aligned}
        (p + q)^{\mu} &= \left( m_c, m_c,0 \right) \, ,\\
        p^{\mu} &= \left( 0, m_c,0 \right)\,, \\
        q^{\mu} &= \left( m_c,0 ,0 \right)\,. \\
    \end{aligned}
\end{equation}
In the heavy quark expansion it is customary to normalize states as $ \ket{\Lambda_c(p + q)}  = \sqrt{M_{\Lambda_c}} \ket{\Lambda_c(v)}$.
The light-cone distribution amplitudes (LCDAs) \cite{Ali:2012pn,Ball:2008fw} of the $\Lambda_c$ and  $P$ are defined by
\begin{equation}
    \label{eqn:LCDA}
    \begin{aligned}
     \bra{0} [u(t_1 n )]^{\alpha}_A  [d(t_2 n )]^\beta_B  [c(0)]^\gamma_C   \ket{ \Lambda_c( v )} =& \frac{\epsilon^{\alpha \beta \gamma }}{6}\sum_i \tilde \alpha_i [\Gamma_i]_{BA} [u_{\Lambda_c}]_C \\
    &\times \int \mathrm{d} u \mathrm{d} \omega \omega \exp\left(-i \omega(t_1 u +t_2 \bar u) \right)  \psi^{(i)}(u,\omega )\,,\\
        \bra{P(p)} [\bar u^{\alpha}(t\bar n) ]_A [d_l(0)^{\beta}]_B \ket{0} =& \frac{\delta^{\alpha \beta}}{3} \frac{i f_P}{8} \left( \bar n \cdot p\right) [\slashed{n} \gamma_5]_{BA} \int_0^1 \mathrm{d} x \exp\left(ixt \bar n \cdot p\right) \Phi_P(x)
        \end{aligned}
\end{equation}
where $l = d,s$ for $\pi,K$  and we  work in light-cone gauge in which the wilson lines, appearing between the non-local quark currents, are equal to unity.
The distribution function $\Psi^{(i)}(u,\omega)$ ($\Phi^P(x)$) describes the baryon (meson) along the light-cone direction, where $u$($x$) are the momentum fractions of the up-quark in the baryon (meson). For $\Lambda_c$ the variable $\omega$ describes the total momentum of the light quarks carried along the light-cone, as well as $\bar u = 1 - u$ the momentum fraction of the down-quark.   
The twist expansion \cite{Ali:2012pn,Aliev:2025cko} in Eqn.~\eqref{eqn:LCDA} reads
\begin{equation}
    \label{eqn:twist_expansion}
    \begin{aligned}
    &\tilde \alpha_1 = \frac{1}{8} f^{(2)}_{\Lambda_c}, &\quad \Gamma_1 &= \slashed{\bar n} \gamma_5 C^{-1} \, , \\
    &\tilde \alpha_2 = \frac{1}{4} f^{(1)}_{\Lambda_c}, &\quad \Gamma_2 &=  \gamma_5  C^{-1} \, , \\
    &\tilde \alpha_3 = \frac{1}{8} f^{(1)}_{\Lambda_c}, &\quad \Gamma_3 &= i \sigma_{\mu \nu} \bar n^{\mu} n^{\nu}  \gamma_5   C^{-1} \, , \\
    &\tilde \alpha_4 = \frac{1}{8} f^{(2)}_{\Lambda_c}, &\quad \Gamma_4 &= \slashed{ n} \gamma_5  C^{-1}\,\, ,
    \end{aligned}
\end{equation}
where $f_{\Lambda_c} \equiv  f^{(1)}_{\Lambda_c} \approx f^{(2)}_{\Lambda_c} $ are the decay constants of the $\Lambda_c$ and $\tilde \Gamma_i = \Gamma_i C \gamma_5$ is useful to make $C,\gamma_5$ explicit in the Dirac chains. 
The LCDA $\psi^{(1)}$ is leading twist whereas $\psi^{(2)},\psi^{(3)}$ are twist-3 and $\psi^{(4)}$ is twist-4.
The QCD interaction vertex is given by 
\begin{equation}
    \mathcal{L}_{\text{int}}  = g_s \bar q_{\alpha} \gamma^{\mu} q_\beta  A^a_{\mu}  T^a_{\alpha \beta}\,,
\end{equation}
with the replacement $\gamma^{\mu} \rightarrow v^{\mu} $ for the heavy charm quark.
The gluon propagator in light-cone gauge ($A\cdot n = 0$), the light quark propagator and heavy charm propagator, respectively, are given by 
\begin{equation}
\label{eqn:propagators}
\begin{aligned}
    G^{\mu \nu}(x,y) &= \int \frac{\mathrm{d}^4 p}{(2 \pi)^4} \frac{-i }{p^2} \exp(-ip(x-y))  \tilde \eta^{\mu \nu}(p) \,, \quad 
     \tilde \eta^{\mu \nu} (p)=  \eta^{\mu \nu} - \frac{ n^{\mu} p^{\nu} + n^{\nu} p^{\mu} }{p \cdot n} \, ,\\
    S(x,y) &=\int \frac{\mathrm{d}^4 p}{(2 \pi)^4} \frac{i \slashed{p}}{p^2} \exp(-ip(x-y))\,, \\
    \tilde S(x,y) & = \int \frac{\mathrm{d}^4 k}{(2 \pi)^4} \exp(-i(k+ m_c v)(x-y)) \frac{i P_+}{n \cdot k}\, , 
\end{aligned}    
\end{equation}
where we stripped the color factors, and used $P_+ = \frac{1+ \slashed{v}}{2}$ and residual momentum $k$.

In the heavy quark limit ($q^{\mu} \approx M_{\Lambda_c} \frac{\bar n^{\mu}}{2}$), the matrix element 
in Eqn.~\eqref{eqn:LambdaCFormFactor}
can be written as 
\begin{equation}
\label{eqn:LambdaCFormFactor2}
\begin{aligned}
   i \mathcal{\tilde M}^{I,X} 
    &= P_R \left( F^{I,X}(0) + \tilde F^{I,X}(0) \frac{\slashed{\bar n}}{2} \right)\sqrt{M_{\Lambda_c}}u_{\Lambda_c}(v) \\
    &= P_R \left( F^{I,X}(0) + \tilde F^{I,X}(0)  - \frac{\slashed{n}}{2} \tilde F^{I,X}(0)  \right)\sqrt{M_{\Lambda_c}}u_{\Lambda_c}(v) \,, \\
    & \equiv \mathcal{P}^{I,X}  \sqrt{M_{\Lambda_c}}u_{\Lambda_c}(v) \, , 
    \end{aligned}
\end{equation}
where we used $\frac{\slashed{\bar n}}{2} u_{\Lambda_c}(v) = (1 - \frac{\slashed{ n}}{2}) u_{\Lambda_c}(v)$ and defined $\mathcal{P}^{I,X}$ as the matrix element without the $u_{\Lambda_c}(v)$-spinor \footnote{The factor $\sqrt{M_{\Lambda_c}}$ cancels in the final expression, as the LCDAs in Eqn.~\eqref{eqn:LCDA} already assume the state $\ket{\Lambda_c(v)}$ to be normalized correctly.} .
This allows one to project out the form factors with traces
\begin{equation}
\label{eqn:form_factors_traces}
    \begin{aligned}
        F^{I,X} + \tilde F^{I,X}&=\frac{1}{2} \text{Tr} [ P_R \mathcal{P}^{I,X}  ] \,,\\
         \tilde F^{I,X}  &= -\frac{1}{2} \text{Tr} [  \slashed{\bar n} P_R \mathcal{P}^{I,X}  ]\,.\\
    \end{aligned}
\end{equation}
\subsection{Matrix element}
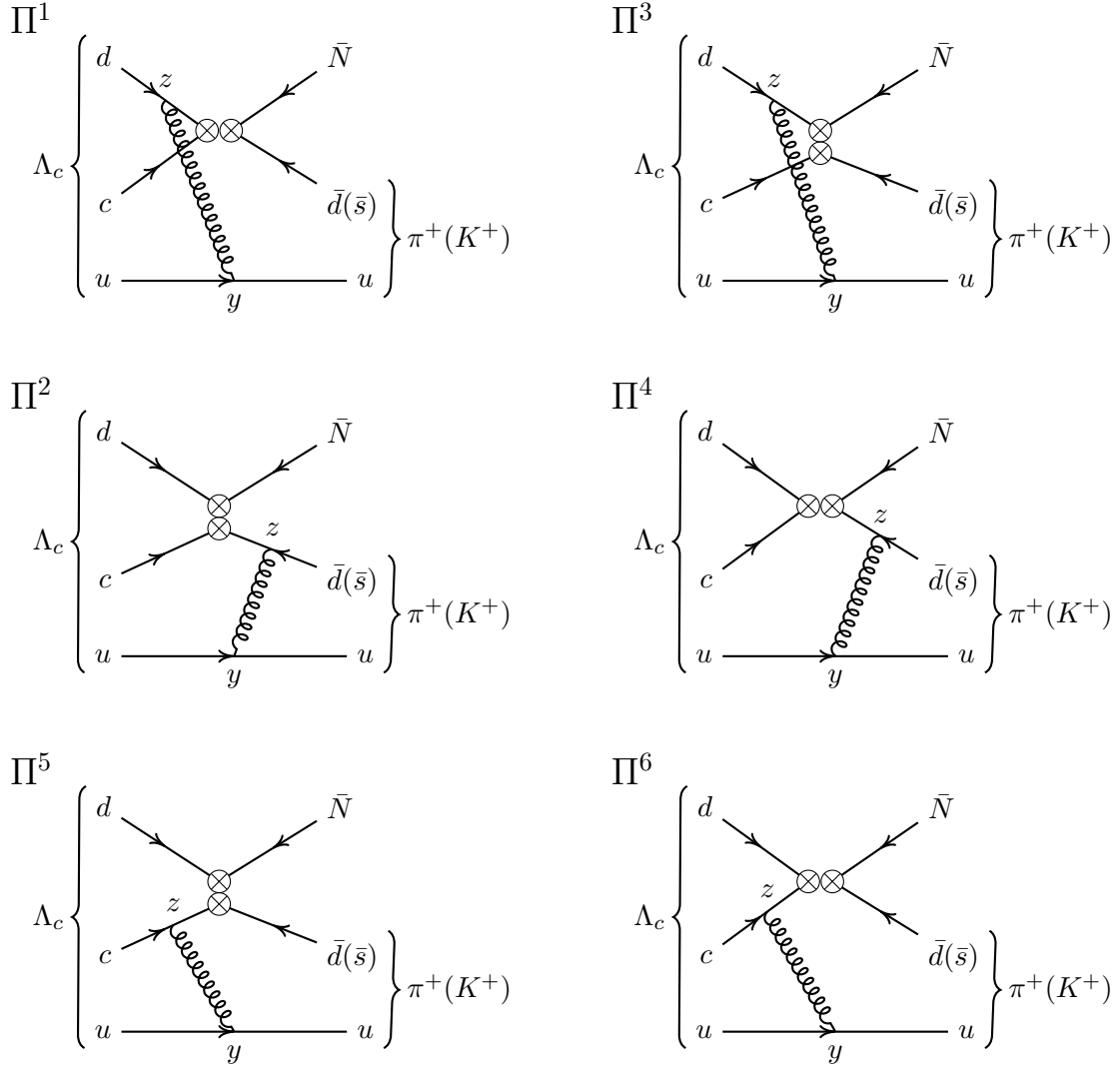
\begin{figure}
    \centering
    \input{feynman/LambdaC.tex}
    \caption{Diagrams for $\Lambda_c \to \pi (K) + \bar N$ decays.  Each diagram corresponds to one  $\Pi^i$, see Eqn.~\eqref{eqn:Pi_definition}.
   } 
    \label{fig:LambdaC_Pi}
\end{figure}

The $\Lambda_c \to \pi \bar N$ matrix element, at order $\mathcal{O}\left(\alpha_s\right)$, is given by 
\begin{equation}
\label{eqn:CtoPi_matrix_element}
    \begin{aligned}
        i \mathcal{\tilde M}^3 =& \int d^4 z d^4 y \bra{\pi} \mathcal{T}\{ \mathcal{O}^{3,X} , \mathcal{L}_{\text{int}}(z) ,\mathcal{L}_{\text{int}}(y) \} \ket{\Lambda_c } \\        
        =& - g_s^2 \int d^4 z d^4 y G^{\mu \nu}(y,z) \bra{\pi} [P_R d(0)^{\gamma}]_A \epsilon_{\alpha \beta \gamma} c(0)^{T,\alpha} C P_X d(0)^{\beta} \\
        &\times \bar q(z)^{\delta_1} \gamma^\mu T^a_{\delta_1 \delta_2} q(z)^{\delta_2} \bar u(y)^{\epsilon_1} \gamma^{\nu }  T^a_{\epsilon_1 \epsilon_2}u(y)^{\epsilon_2} \ket{\Lambda_c}\\
         =&-  g_s^2\Bigg([P_R]_{AB} [CP_X]_{CD} [\gamma^{\mu}]_{EF} [\gamma^{\nu}]_{GH} \left( \Pi^1+ \Pi^2 + \Pi^3 + \Pi^4 \right) \\
          &+[P_R]_{AB} [CP_X]_{CD} [v^{\mu}]_{EF} [\gamma^{\nu}]_{GH} ( \Pi^5 + \Pi^6) \Bigg) \,,
    \end{aligned}
\end{equation}
where we introduced $\Pi^{1,2,3,4,5,6}$ as dirac space tensors, which represent all possible contractions and are part of the diagrams shown in Fig.~\ref{fig:LambdaC_Pi}.
Explicitly they are given by 
\begin{equation}
\label{eqn:Pi_definition}
    \begin{aligned}
        \Pi^1 &=-\int d^4 z d^4 y G^{\mu \nu}(y,z) S_{DE}(0,z)  \epsilon_{\alpha \beta \gamma } T^a_{\beta\delta_2} T^a_{\epsilon_1\epsilon_2}\bra{\pi} \bar u^{\epsilon_1}_G(y) d^{\gamma}_B(0) \ket{0} \bra{0} u^{\epsilon_2}_H(y) d^{\delta_2}_F(z) c^{\alpha}_C(0) \ket{\Lambda_c} \\
        \Pi^2 &= \int d^4 z d^4 y G^{\mu \nu}(y,z) S_{DE}(0,z) \epsilon_{\alpha \beta \gamma } T^a_{\beta\delta_2} T^a_{\epsilon_1\epsilon_2}\bra{\pi} \bar u^{\epsilon_1}_G(y) d^{\delta_2}_F(z) \ket{0} \bra{0} u^{\epsilon_2}_H(y) d^{\gamma}_B(0) c^{\alpha}_C(0) \ket{\Lambda_c} \\
        \Pi^3 &= \int d^4 z d^4 y G^{\mu \nu}(y,z) S_{BE}(0,z) \epsilon_{\alpha \beta \gamma } T^a_{\gamma \delta_2 }T^a_{\epsilon_1\epsilon_2}\bra{\pi} \bar u^{\epsilon_1}_G(y) d^{\beta}_D(0) \ket{0} \bra{0} u^{\epsilon_2}_H(y) d^{\delta_2}_F(z) c^{\alpha}_C(0) \ket{\Lambda_c} \\
        \Pi^4 &=-\int d^4 z d^4 y G^{\mu \nu}(y,z) S_{BE}(0,z) \epsilon_{\alpha \beta \gamma } T^a_{\gamma \delta_2 }T^a_{\epsilon_1\epsilon_2}\bra{\pi} \bar u^{\epsilon_1}_G(y) d^{\delta_2}_F(z) \ket{0} \bra{0} u^{\epsilon_2}_H(y) d^{\gamma}_D(0) c^{\alpha}_c(0) \ket{\Lambda_c} \\
        \Pi^5 &=\int d^4 z d^4 y G^{\mu \nu}(y,z) \tilde S_{CE}(0,z) \epsilon_{\alpha \beta \gamma } T^a_{\alpha \delta_2 }T^a_{\epsilon_1\epsilon_2}\bra{\pi} \bar u^{\epsilon_1}_G(y) d^{\beta}_D(0) \ket{0} \bra{0} u^{\epsilon_2}_H(y) d^{\gamma}_B(0) c^{\delta_2}_F(z) \ket{\Lambda_c} \\
       \Pi^6 &= - \int d^4 z d^4 y G^{\mu \nu}(y,z) \tilde S_{CE}(0,z) \epsilon_{\alpha \beta \gamma } T^a_{\alpha \delta_2 }T^a_{\epsilon_1\epsilon_2}\bra{\pi} \bar u^{\epsilon_1}_G(y) d^{\gamma}_B(0)\ket{0} \bra{0} u^{\epsilon_2}_H(y) d^{\beta}_D(0) c^{\delta_2}_F(z) \ket{\Lambda_c} \\
    \end{aligned}
\end{equation} 
with relative signs from fermion permutations. Using \eqref{eqn:LCDA} we can deduce the color factors  
\begin{equation}
    \begin{aligned}
       \Pi^1  &\propto \epsilon_{\alpha \beta \gamma } T^a_{\beta\delta_2} T^a_{\gamma\epsilon_2} \epsilon_{\epsilon_2 \delta_2 \alpha} = \text{Tr} [T T]  \,,\\ 
        \Pi^2  &\propto \epsilon_{\alpha \beta \gamma } T^a_{\beta\delta_2} T^a_{\delta_2\epsilon_2} \epsilon_{\epsilon_2 \gamma \alpha} = 2\text{Tr} [T T]\,, \\ 
       \Pi^3 & \propto \epsilon_{\alpha \beta \gamma } T^a_{\gamma\delta_2} T^a_{\beta \epsilon_2} \epsilon_{\epsilon_2 \delta_2 \alpha} = - \text{Tr} [T T]\,, \\ 
       \Pi^4  &\propto \epsilon_{\alpha \beta \gamma } T^a_{\gamma\delta_2} T^a_{\delta_2\epsilon_2} \epsilon_{\epsilon_2 \gamma \alpha} = -2\text{Tr} [T T]\,, \\
        \Pi^5 & \propto \epsilon_{\alpha \beta \gamma } T^a_{\alpha\delta_2} T^a_{\beta \epsilon_2} \epsilon_{\epsilon_2 \gamma \delta_2} = - \text{Tr} [T T]\,, \\ 
       \Pi^6  &\propto \epsilon_{\alpha \beta \gamma } T^a_{\alpha\delta_2} T^a_{\gamma\epsilon_2} \epsilon_{\epsilon_2 \beta \delta_2} = \text{Tr} [T T]\,, \\
    \end{aligned}
\end{equation}
which give  relative signs between $\Pi^1$ and $\Pi^3$,  $\Pi^2$ and $\Pi^4$, and $\Pi^5$ and $\Pi^6$, respectively, and $\text{Tr} [T T] =4$. 
Using $\ket{\Lambda_c(p+q)} =\sqrt{M_{\Lambda_c}}\ket{\Lambda_c(v)}$, inserting the LCDAs in Eqn.~\eqref{eqn:LCDA} and performing position and 
momentum integrals, we can write each contribution as  
\begin{equation}
    \begin{aligned}
        \Pi^1 &= -\sum_i \! \int d x d u d \omega \omega \, \tilde \eta^{\mu \nu}(p_g)  \frac{\alpha_i f_p\Phi_{\pi}(x) }{48 \omega^2 u x^2 m_c }     [\slashed{p}^{(1)}_q]_{DE} [\slashed{n} \gamma_5]_{BG} [\Gamma_i]_{F H}\psi^{(i)}(u,\omega)  \sqrt{M_{\Lambda_c}}[ u_{\Lambda_c}(v)]_{C}\,, \\
        \Pi^2 &=\sum_i \! \int d x d u d \omega \omega \, \tilde \eta^{\mu \nu}(p_g) \frac{\alpha_i f_p\Phi_{\pi}(x) }{24 \omega^2 u^2 x  m_c }   [\slashed{p}^{(2)}_q]_{DE} [\slashed{n}\gamma_5]_{FG}    [\Gamma_i]_{B H}\psi^{(i)}(u,\omega)  \sqrt{M_{\Lambda_c}}[ u_{\Lambda_c}(v)]_{C} \,, \\
        \Pi^5 &= - \sum_i \! \int d x d u d \omega \omega \, \tilde \eta^{\mu \nu}(p_g)\frac{\alpha_i f_p \Phi_{\pi}(x)}{48 \omega^2 \bar u \bar x x u  m_c }    [P_+]_{CE} [\slashed{n}\gamma_5]_{DG}   [\Gamma_i]_{B H}\psi^{(i)}(u,\omega) \sqrt{M_{\Lambda_c}}[ u_{\Lambda_c}(v)]_{F} \,,
    \end{aligned}
\end{equation}
and $\Pi^3, \Pi^4,\Pi^6$ are related to $\Pi^1, \Pi^2,\Pi^5$, respectively, by interchanging the Dirac indices $D \leftrightarrow B $ .

The internal kinematics is given by
\begin{equation} \label{eq:internal}
    \begin{aligned}
        p_g &= ( \omega u , -x m_c ,0 ) \,, \\ 
        p_q^{(1)} &= 
            (-\omega ,x m_c   , 0 ) & \text{for}\quad  \Pi^{1,3} \,, \\
             p_q^{(2)} &=
            (-u \omega,m_c  , 0 ) &  \text{for} \quad \Pi^{2,4} \,, \\
            k =& ( - \bar u \omega ,x m_c , 0 )  &  \text{for} \quad \Pi^{5,6} \,,  
    \end{aligned}
\end{equation}
for the gluon and quark propagators, respectively.
Connecting the Dirac chains from Eqn.~\eqref{eqn:CtoPi_matrix_element} with the $\Pi^{1,2,3,4,5,6}$ we can write the total matrix element as 
\begin{equation}
    i \mathcal{\tilde M}^3 = -\sum_i \!\int d x d u d \omega \frac{ \alpha_s \pi\alpha_i f_p \Phi_{\pi}(x) \psi^{(i)}(u,\omega)}{3 \omega u x } \left( \frac{-4}{x} \left( \Sigma^1 + \Sigma^3  \right) + \frac{8}{u} \left( \Sigma^2 + \Sigma^4  \right) \right)  \sqrt{M_{\Lambda_c}}u_{\Lambda_c}(v)\,,
\end{equation}
with Dirac structures 
\begin{equation}
\label{eqn:sigma_def}
    \begin{aligned}
        \Sigma^1 &= -P_R \slashed{n} \gamma^{\mu}  \tilde \Gamma_i \gamma^{\nu} \slashed{p}^{(1)}_q P_X \tilde \eta_{\mu \nu}(p_g)   \,, \\
          \Sigma^2 
          &=- P_R \tilde \Gamma_i  \gamma^{\mu} \slashed{n} \gamma^{\nu} \slashed{p}^{(2)}_q P_X  \tilde \eta_{\mu \nu} (p_g) \,, \\
        \Sigma^3 %
        &= -P_R \slashed{p}^{(1)}_q \gamma^{\mu} \tilde \Gamma_i  \gamma^{\nu} \slashed{n}  P_X  \tilde \eta_{\mu \nu}(p_g) \,, \\
        \Sigma^4 %
         & = -P_R \slashed{p}^{(2)}_q \gamma^{\mu} \slashed{n} \gamma^{\nu}  \tilde \Gamma_i P_X \tilde \eta_{\mu \nu}(p_g)  \,, \\
         \Sigma^5 %
         &= -P_R \tilde \Gamma_i \gamma^{\mu} \slashed{n} P_X P_+ v^{\nu}\tilde \eta_{\mu \nu}(p_g)  \,,\\
         \Sigma^6 %
         &= -P_R  \slashed{n} \gamma^{\nu} \tilde \Gamma_i P_X P_+ v^{\mu} \tilde \eta_{\mu \nu}(p_g) \,.
    \end{aligned}
\end{equation}
Using kinematical identities one can show that $\gamma^{\mu} \slashed{n} v^\nu \tilde\eta_{\mu \nu}=0 $ and similarly 
$\slashed{n} \gamma^{\nu} v^{\mu} \tilde \eta_{\mu \nu}=0$. Therefore, $\Sigma^5 u_{\Lambda_c}(v) =\Sigma^6 u_{\Lambda_c}(v) =0$ independent of $\tilde \Gamma_i$,
and there are no contributions from $\Pi^{5,6}$ in Fig.~\ref{fig:LambdaC_Pi} 
at order $\frac{1}{m_c}$.

The calculation for $\Lambda_c \to K + \bar N $ with the  operators $\mathcal{\tilde O}^1$ ($\mathcal{\tilde O}^2$) is analogous,
see Eqn.~\eqref{eqn:BNV_low_energy}, which receive contributions from $\Pi^1,\Pi^4$ ($\Pi^2,\Pi^3$). This corresponds to two diagrams in Fig.~\ref{fig:LambdaC_Pi} contributing to each operator.
We obtain 
\begin{equation}
\begin{aligned}
    i \mathcal{\tilde M}^1 &= -\sum_i \int d x d u d \omega \frac{ \alpha_s \pi\alpha_i f_p \Phi_K(x) \psi^{(i)}(u,\omega)}{3 \omega u x } \left( -\frac{4}{x}  \Sigma^1  +   \frac{8}{u}  \Sigma^4  \right)  \sqrt{M_{\Lambda_c}}u_{\Lambda_c}(v) \,,        \\
    i \mathcal{\tilde M}^2 &= -\sum_i \int d x d u d \omega \frac{ \alpha_s \pi\alpha_i f_p \Phi_K(x) \psi^{(i)}(u,\omega)}{3 \omega u x } \left( -\frac{4}{x}  \Sigma^3    +\frac{8}{u}  \Sigma^2   \right)  \sqrt{M_{\Lambda_c}}u_{\Lambda_c}(v)      \,,  
\end{aligned}
\end{equation}
for the $\Lambda_c \to K + \bar N $ matrix elements.
The requisite  traces to project out the form factors in Eqn.~\eqref{eqn:form_factors_traces} are given in Tab.~\ref{tab:sigma_traces}.
The column with the trace projecting onto $\tilde F$ for $i=4$ vanishes since $\slashed{n} \gamma^\mu \slashed{n} \tilde \eta_{\mu\nu}=0$.
All other columns that are entirely zero vanish due to an odd number of Gamma-matrices, and $\psi^{(4)}(u,\omega)$ does not contribute. 
\begin{table}[]
    \centering
    \begin{tabular}{c|c c c c c |c c c c c}
         &\multicolumn{5}{c|}{$\text{Tr}[ P_R \Sigma^I]$} &\multicolumn{5}{c}{$\text{Tr}[ \slashed{\bar n} P_R \Sigma^I]$}   \\ \toprule
      I    &i = &1 &2 &3 &4  &i = &1 &2 &3 &4  \\\midrule  
      1    & &0 &$-4  \omega \delta_{X,R}$    & $8 \omega  \delta_{X,R}$&0& &$-16 x m_c\delta_{X,L}$ &0 &0 &0  \\ 
      2    & &0 &$4 u \omega \delta_{X,R}$  & $8 u \omega  \delta_{X,R}$&0& &0 &0 &0 &0  \\ 
      3    & &0 &$-4\omega \delta_{X,R}$    & $-8 \omega  \delta_{X,R}$ &0&&$-16 x m_c\delta_{X,L}$ &0 &0 &0   \\  
      4    & &0 &$4u\omega \delta_{X,R}$    & $-8u \omega  \delta_{X,R}$ &0& &0 &0 &0 &0    
    \end{tabular}
    \caption{Traces of the $\Sigma^I$ defined in Eqn.~\eqref{eqn:sigma_def} for all terms in the twist expansion $i = 1,2,3,4$ in 
    Eqn.~\eqref{eqn:twist_expansion}. There are no contributions at this order from $\Sigma^{5,6}$, see text.}
    \label{tab:sigma_traces}
\end{table}

\subsection{$\Lambda_c$ Form factors}

The form factors for $\Lambda_c \to \pi + \bar N$ read $\tilde F^{3,R} = 0$, and 
\begin{equation} \label{eq:FFpi}
    \begin{aligned}
          F^{3,L} &= -\tilde F^{3,L}  =  -\int_0^1 d x \int_0^1 d u \int_0^\infty d \omega \frac{ 2 \pi  \alpha_s f_{\Lambda_c} f_\pi}{9   u x \omega  }  \Phi_\pi(x)   \psi^{(1)}(u,\omega)  \,, \\
        F^{3,R} &=  \int_0^1 d x \int_0^1 d u \int_0^\infty d \omega \frac{ (1+ 2x) \pi  \alpha_s f_{\Lambda_c} f_\pi   }{9   u x^2 m_c  } \Phi_\pi(x) \psi^{(2)}(u,\omega) \,.
    \end{aligned}
\end{equation}

The form factors for $\Lambda_c \to K + \bar N$ read, amended by the relations given in Eqn.~(\ref{eqn:form_factor_relations}),
\begin{equation}
\label{eqn:F1L}
\begin{aligned}
    F^{1,L} & =- \int_0^1 d x \int_0^1 d u \int_0^\infty d \omega \frac{  \pi \alpha_s  f_{\Lambda_c} f_K }{ 9 \omega  u x } \Phi_K(x) \psi^{(1)}(u,\omega)   \,, \\
      F^{1,R} &=  -\int_0^1 d x \int_0^1 d u \int_0^\infty d \omega \frac{ (1+2x )  \pi  \alpha_s f_{\Lambda_c} f_K  }{18 m_c  u x^2 } \Phi_K(x)\left(  \psi^{(2)}(u,\omega)  - \psi^{(3)}(u,\omega) \right) \,,  \\
    F^{2,R}&=-\int_0^1 d x \int_0^1 d u \int_0^\infty d \omega \frac{  ( 1+2x )  \pi  \alpha_s f_{\Lambda_c} f_K  }{18  m_c  u x^2 } \Phi_K(x)\left(  \psi^{(2)}(u,\omega)   + \psi^{(3)}(u,\omega) \right) \,. \\
\end{aligned}
\end{equation}

Inspecting Eqns.~(\ref{eq:FFpi}, (\ref{eqn:F1L}) one suspects already  endpoint divergencies from the $dx$ and $du$ integration, although some powers of $1/x$ and $1/u$ which stem from propagator denominators
may cancel with the LCDA's. Eqn.~(\ref{eq:internal}) identifies the diagrams $\Pi^{1,3}$ with the gluon attached to the initial down-quark as the one with the
worst $1/x$-behaviour.

For the $\Lambda_c$ LCDAs, we use input from $\Lambda_b$ LCDAs based on Ref.~\cite{Ball:2008fw},
\begin{equation}
\label{eqn:Psi_parametrization}
    \begin{aligned}
        \psi^{(1)}(u,\omega) &= \Theta\left( 2 s_0 - \omega \right)\frac{15}{2 N}  \omega^2 u ( 1- u) \int_{\omega/2}^{s_0} \mathrm{d} s \exp(- s/\tau) \left(s - \omega/2\right)  \,,\\
        \psi^{(2)}(u,\omega) &= \Theta\left( 2 s_0 - \omega \right)\frac{15}{4 N} \omega\int_{\omega/2}^{s_0} \mathrm{d} s \exp(- s/\tau) \left(s - \omega/2\right)^2  \,,\\
        \psi^{(3)}(u,\omega) &= \Theta\left( 2 s_0 - \omega \right)\frac{15}{4 N} ( 2u  -1 )  \omega \int_{\omega/2}^{s_0} \mathrm{d} s \exp(- s/\tau) \left(s - \omega/2\right)^2  \,,\\
    \end{aligned}
\end{equation}
with  $ 0.4 \leq \tau \leq \SI{0.8}{\GeV}$, $s_0 = \SI{1.2}{\GeV}$ and normalization
\begin{equation}
    N = \int_0^{s_0} ds s^5 \exp(-s/\tau) \,.
\end{equation}
The pseudoscalar LCDAs \cite{Braun:2003rp} are parametrized according to 
\begin{equation}
\label{eqn:Phi_parametrization}
    \Phi_P(x) = 6 x (1- x) \left[ 1 + \sum_{n = 1}^2 a_n^P C_n^{(3/2)}(2 x - 1) \right ]\,,
\end{equation}
where $C^{(l)}_{n}$ are Gegenbauer polynomials with coefficients $a_n^P$, which are summarized with other input values in Tab.~\ref{tab:input_values_LCDA}.
We use $\alpha_s(\mu_{hs}) \simeq 0.4$ at the hard-spectator interaction scale $\mu_{hs}=\sqrt{\Lambda _\text{QCD} m_c}$.
The leading power contribution for $F^{1,L},F^{3,L}$ is finite, as all factors of $1/u$($1/x)$ cancel with the baryon-(meson)-LCDA.  
Numerical values of $F^{1,L}(0),F^{3,L}(0)$ including uncertainties and the central value for $\tau = 0.6$ are given in (\ref{eqn:form_factors_numeric}).

\begin{figure}
    \centering
    \includegraphics[width=0.45\linewidth]{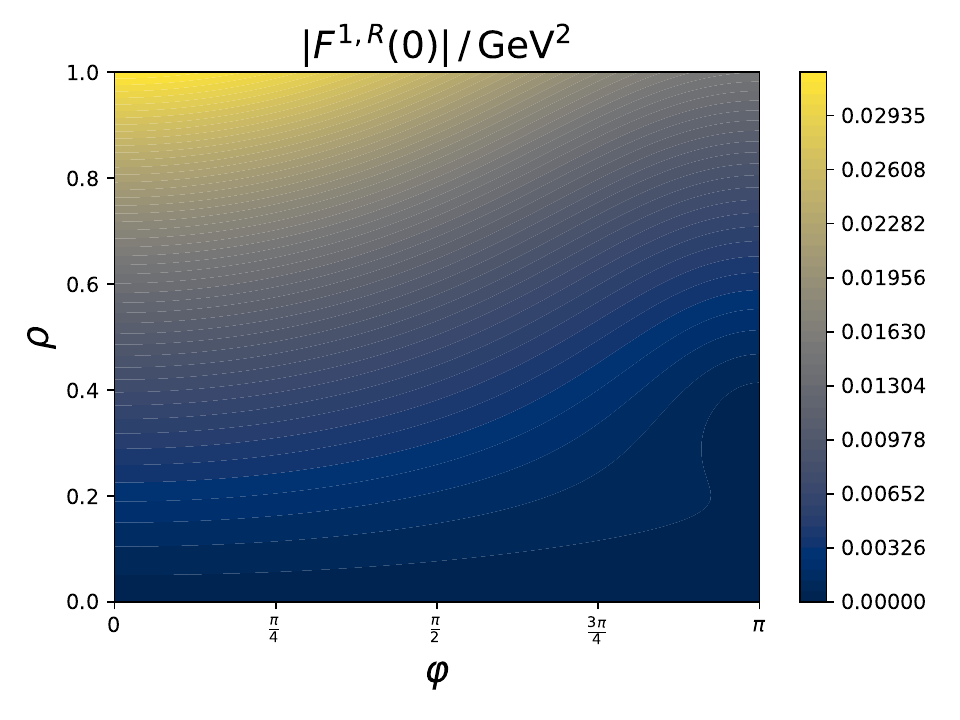}
    \includegraphics[width=0.45\linewidth]{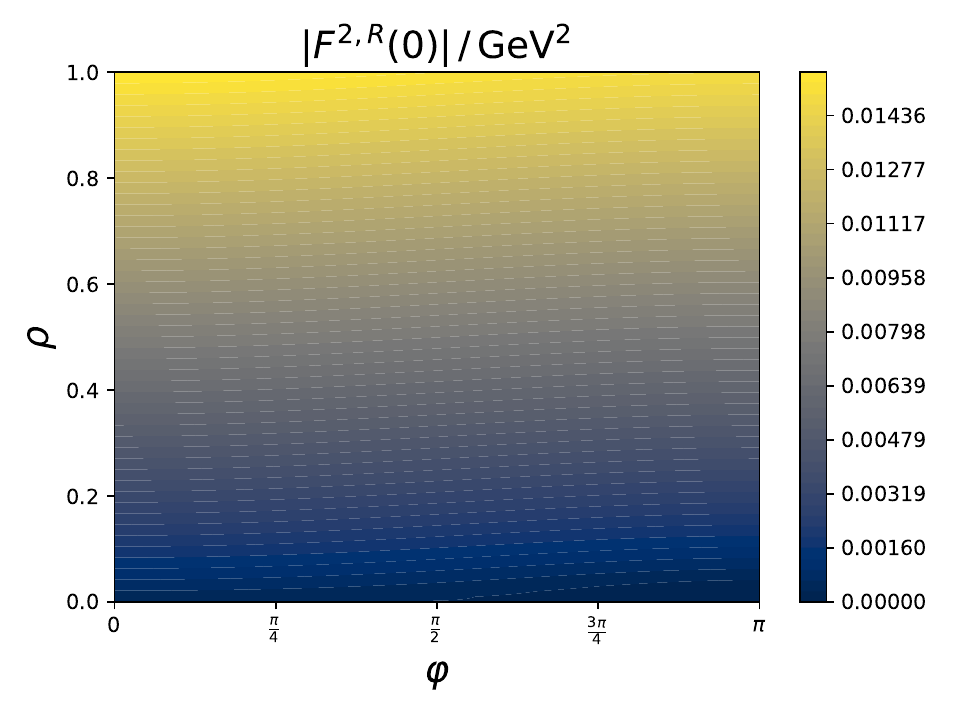}
    \includegraphics[width=0.45\linewidth]{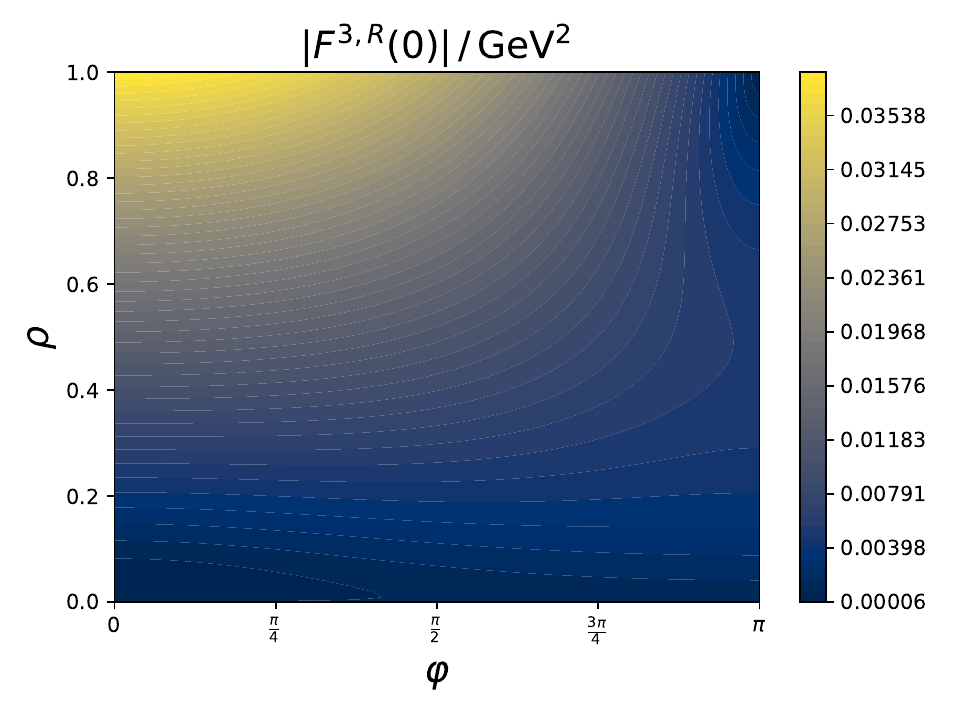}
    \includegraphics[width=0.45\linewidth]{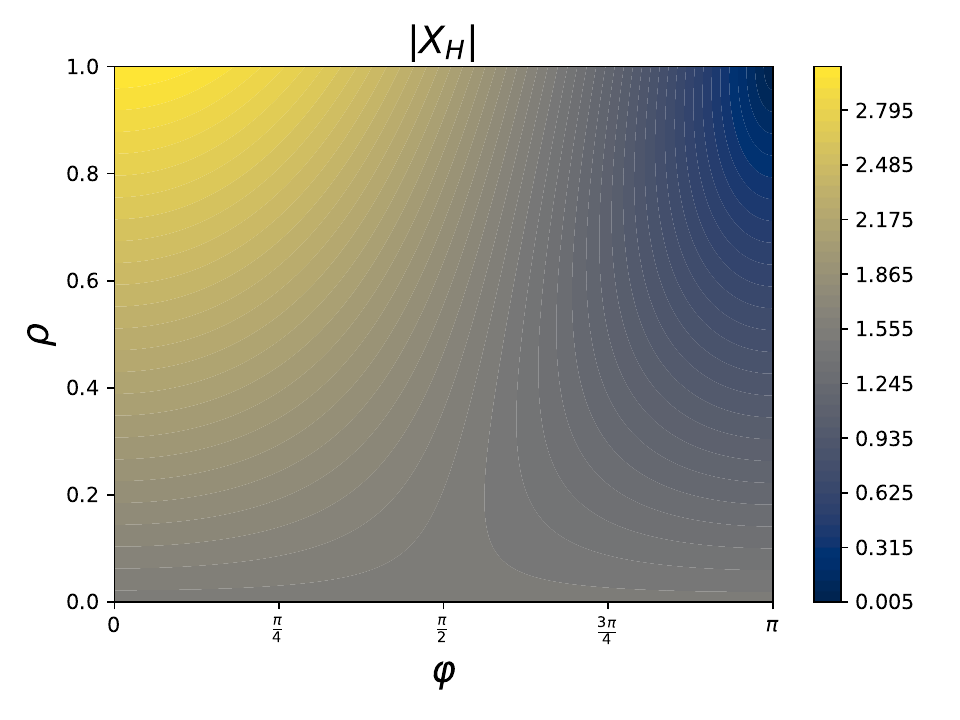}
    \caption{Absolute values of the form factors $F^{1,R},F^{2,R},F^{3,R}$ and $|X_H|$ against the infrared cutoff parameters $\varphi, \rho$
    from (\ref{eq:XH}). 
     Form factors are symmetric under $\varphi \to 2 \pi -\varphi$ and therefore we show ranges from $0$ to $\pi$. 
     }
 \label{fig:form_factors}
\end{figure}

The form factors $F^{I,R}$ arise at twist 3.
For $F^{2,R}$ the $1/u$ divergence cancels, as $\psi^{(2)}(u,\omega)   + \psi^{(3)}(u,\omega) \sim u $, and only the $1/x$ divergence remains.
For $F^{1,R},F^{3,R}$ both endpoint divergences persist.
We regulate the ill-defined and unphysical integrals     $ \int_0^1 d u/ u, \int_0^1 d x/ x$ following Ref.~\cite{Beneke:2001ev},  by replacing them with
 \begin{equation} \label{eq:XH}
    X_H = (1 + \rho \exp ( i \varphi)) \log(m_{\Lambda_c}/\Lambda)\,   ,
 \end{equation}
 which corresponds to an $\mathcal{O}(1)$ uncertainty for the parameters $\rho \lesssim 1$ and arbitrary phase $\varphi$.
 We use $ \Lambda = \SI{0.5}{\GeV}$.
The magnitudes of the form factors $F^{I,R}$ are shown in Fig.~\ref{fig:form_factors}. Ranges are sizeable, from essentially zero to
$O(10^{-2})\, \text{GeV}^2$. 
$F^{2,R}$ has a milder dependence on the cut-off parameters than the others, which is expected due to a partial cancellation of the endpoint divergences.
The suppression of $F^{3,R}$ along $\varphi \sim \pi$ can be understood since  $u$-integration yields $F^{3,R} \propto X_H \int dx \int d \omega \ldots$
and $|X_H| \sim \sqrt{1+\rho^2 +2 \rho \cos \varphi}$ vanishes for $\varphi \to \pi$ and $\rho \to 1$. Similarly for $F^{1,R} \propto \left(X_H - 1 \right)\int dx \int d \omega \ldots  $, 
which leads to suppression around $\rho \to 0.34$ and $\varphi \to \pi$.

\begin{table}[t]
\centering
\begin{tabular}{|ll|}
\hline
$m_{\Lambda_c} = 2.29~\text{GeV}$ & \cite{ParticleDataGroup:2024cfk} \\
$\tau_{\Lambda_c} = \SI{202.6 e-15}{\s}$ & \cite{ParticleDataGroup:2024cfk} \\

     $f_{\Lambda_C} = 0.028\pm 0.002 ~\text{GeV}^3$ & \cite{Groote:1996em}  \\

$f_{\pi} = 0.1304 \pm 0.0002~\text{GeV}$ & \cite{Grossman:2015cak} \\
     $f_K = 0.1562 \pm 0.0007~\text{GeV}$ & \cite{Grossman:2015cak} \\

$a^{\pi}_1 = 0$ & \cite{Grossman:2015cak} \\
     $a^{\pi}_2 = 0.29 \pm 0.08$ & \cite{Grossman:2015cak} \\

$a^{K}_1 = -0.07 \pm 0.04$ & \cite{Grossman:2015cak} \\
     $a^{K}_2 = 0.24 \pm 0.08$ & \cite{Grossman:2015cak} \\

\hline
\end{tabular}
\caption{The input parameters used for the numerical analysis of the $\Lambda_c$ QCDF results.}
\label{tab:input_values_LCDA}
\end{table}

\section{UV completions \label{app:UV}}
\subsection{UV lagrangians}

It is commonly accepted that EFT analyses constraining certain operators are better motivated when one can show that these operators can arise in simple UV models not excluded by other experimental probes. They also help—under additional assumptions—to relate experimental probes that are otherwise unrelated in a general EFT analysis. In this appendix, we reproduce the UV Lagrangians that generate, at tree level, the EFT operators studied in this work, and we list the additional EFT terms they induce at tree level.

Under $SU(3)_{C}\times SU(2)_L \times U(1)_Y\times U(1)_{B}$, our light degrees of freedom are $u_R\sim (3,1,2/3,1/3)$, $d_R\sim (3,1,-1/3,1/3)$, $q_L\sim(3,2,1/6,1/3)$, $N\sim(1,1,0,-1)$. It is straightforward to check that, omitting family, color indices and chirality projection, this generates

\begin{equation}
\mathcal{L}=\frac{C^{qqdN}}{\Lambda^2}\epsilon^{ab}  ( \bar{q}^{C}_{a}  q_{b} ) ( \bar{d}_{\gamma}^{C} N )+ \frac{C^{uddN}}{\Lambda^2} ( \bar{u}^{C} d ) ( \bar{d}^{C} N ) \, .
\end{equation}

Dimension-four interactions that can generate four-fermion operators involving one 
$N$ and three SM quarks must feature a boson that couples both to two SM quarks and to 
$N$ together with another SM quark. The latter implies that the heavy boson must be a color triplet with $B=-2/3$ and also leaves us with three possibilities, again omitting color, family indices and chirality projection\footnote{In principle one can conceive other bosons with analogous interactions coupling with the left-handed modes of $N$, like $\Psi^{\mu,*}\bar{d}^{C}\gamma_{\mu}N$. However, they never generate the other needed vertex given the remaining restrictions.}
\begin{equation}
\begin{aligned}
\mathcal{L}_1&\supset -y^{Nu} \Psi^\dagger\, \bar{u}^{C} N \to \Psi\sim (3,1,2/3,-2/3) \to \mathcal{L}_1\supset -\tilde{y}^{dd} \Psi \, \bar{d}^{C}d \, ,\\
\mathcal{L}_2&\supset -y^{Nd} \Phi^\dagger\, \bar{d}^{C} N \to \Phi\sim (3,1,-1/3,-2/3) \to \mathcal{L}_2\supset -y^{ud} \Phi  \,\bar{u}^{C}d -y^{QQ} \, \epsilon^{ab} \, \Phi   \, \bar{q}_{a }^{C}q_{b}  \,  ,\\
\mathcal{L}_3&\supset -x^{NQ} X^{\mu,\dagger,a}\, \bar{q}^{C}_a \gamma_ {\mu} N \to X^{\mu}_a\sim (3,2,1/6,-2/3) \to \mathcal{L}_3\supset -x^{Qd} \, \, X^{\mu}_{a} \epsilon^{ab}  \, \bar{q}^{C}_b \gamma_{\mu} d  \,  .
\end{aligned}
\end{equation}
Notice that all the terms in these models are identical to the ones obtained in the corresponding models with right-handed neutrinos $\nu_R$ instead of vector-like dark fermions, as the spinor structure of the allowed operators allows to freely trade $N$ (or $N_R$) by $\nu_R$. Consequently, they generate the same tree-level $D=6$ EFT lagrangians and we can then compare with results in the literature (written in a different notation), as we do in Table \ref{tab:UV_mediators}.

\subsection{EFT lagrangians at tree level}
By using the classical equations of motion, it is easy to check that whenever one has a Lagrangian of the form
\begin{align}
\mathcal{L}\supset -M_{S}^2|S|^2+y_{j}S J_{j}+y_{j}^{\dagger}S^{\dagger}J_{j}^{\dagger}+M_{V}^2|V|^2+x_{j}V^{\mu}J_{\mu\,j}+x_{j}^{\dagger}V^{\mu \dagger}J_{\mu\, j}^\dagger \, ,
\end{align}
its EFT form at tree level can be written as
\begin{equation}
\mathcal{L}\supset \frac{y_{j}y_{k}^{\dagger}}{M_{S}^2}J_{j}J_{k}^{\dagger}-\frac{x_{j}x_{k}^{\dagger}}{M_{V}^2}J^{\mu}_{j}J_{\mu\, k}^{\dagger} \, .
\end{equation}

The relevant interaction terms of the UV Lagrangian, keeping for definiteness all the indices except for the spinor ones (in matrix form), are
\begin{equation}
\begin{aligned}
\mathcal{L}\supset &-\, y^{Nu}_{p} \;\Psi{}^{\dagger,\alpha}\, \bar{u}{}^{C}_{R,\alpha,p} N _R\, 
- \, \epsilon_{prs} \, \tilde{y}^{dd}_{p} \, \epsilon^{\alpha\beta\gamma} \;\Psi{}_{\alpha} \, \bar{d}{}^{C}_{R,\beta,r} d{}_{R,\gamma,s}
\\ 
&-y^{Nd}_{p} \; \Phi{}^{\dagger,\alpha}\, \bar{d}{}^{C}_{R,\alpha,p} N_R \, 
-\, y^{ud}_{rs} \, \epsilon^{\alpha\beta\gamma} \; \Phi_{\alpha} \, \bar{u}^{C}_{R,\beta,r} d_{R,\gamma,s} \, 
-\, y^{QQ}_{rs} \, \epsilon^{ab} \, \epsilon^{\alpha\beta\gamma} \; \Phi_{\alpha}   \, \bar{q}_{L,a,\beta,r}^{C}q_{L,b,\gamma,s} 
\\ 
&-\,x^{NQ}_{p} \; X^{\mu,\dagger,a,\alpha}\, \bar{q}^{C}_{L,a,\alpha,p} \gamma_ {\mu} N_R \, 
-\, x^{Qd}_{rs} \, \epsilon^{ab} \, \epsilon^{\alpha\beta\gamma}\; X^{\mu}_{a,\alpha}   \, \bar{q}^{C}_{L,b,\beta,r} \gamma_{\mu} d_{R,\gamma,s}  \, 
+ \mathrm{h.c.}
\end{aligned}
\end{equation}

After a somewhat tedious re-shuffling of spinor, color, family, and weak indices, we find 
the lagrangian $\mathcal{L}_1$ for model 1  given in Eqn.~(\ref{eqn:LQ_model_1}), and

\begin{equation}
\begin{aligned} \label{eq:model2}
M^{2}_{\Phi}\mathcal{L}_2&=\frac{1}{2}y^{ud}_{rs} \, y^{*,ud}_{r's'} \; (\bar{u}_{R.r'}\gamma^{\mu}u_{R,r} \, \bar{d}_{R,s'}\gamma_{\mu}d_{R,s}-\bar{u}_{R,r'}\gamma^{\mu}d_{R,s} \, \bar{d}_{R,s'}\gamma_{\mu}u_{R,r}) 
\\
&+\frac{1}{2} \, y^{QQ}_{rs} \, y^{*,QQ}_{r's'} \;  \epsilon^{ab} \epsilon_{a'b'} \left( \bar{q}_{L,a',r'}\gamma^{\mu} q_{L,a,r} \, \bar{q}_{L,b',s'} \gamma_{\mu}q_{L,b,s} + \bar{q}_{L,a',r'}\gamma^{\mu} q_{L,a,s} \, \bar{q}_{L,b',s'} \gamma_{\mu}q_{L,b,r} \right)  
\\
&+\frac{1}{2}\, y^{*,Nd}_{p}\, y^{Nd}_{p'} \,\bar{d}_{R,p}\gamma^{\mu}d_{R,p'} \, \bar{N}_R\gamma_{\mu}  N_R    
\\
&+\Big[-\frac{1}{2}\, y^{ud}_{rs}\, y^{*,QQ}_{r's'} \,  \,\epsilon_{a'b'}\left( \bar{q}_{L,b',s'}d_{R,s} \, \bar{q}_{L,a',r'} u_{R,r} -\frac{1}{4} \, \bar{q}_{L,b',s'}\sigma^{\mu\nu}d_{R,s} \, \bar{q}_{L,a',r'}\sigma_{\mu\nu}u_{R,r}\right) 
\\
&+\frac{1}{2}y^{ud}_{rs} y^{*,QQ}_{r's'}  \,\epsilon_{a'b'}\Big( \bar{q}_{L,b',\beta,s'}d_{R,\gamma,s}  \bar{q}_{L,a',\gamma ,r'} u_{R,\beta,r} -\frac{1}{4} \bar{q}_{L,b',\beta,s'}\sigma^{\mu\nu}d_{R,\gamma,s}  \bar{q}_{L,a',\gamma,r'}\sigma_{\mu\nu}u_{R,\beta,r}\Big)    
\\
&+y^{ud}_{rs}\, y^{Nd}_{p'} \, \epsilon^{\alpha\beta\gamma} \, \bar{u}_{R,r,\alpha}^{C}  d_{R,s,\beta}\,\bar{N}_R^{C}d_{R,p',\gamma}    
\\
&+     y^{QQ}_{rs}\, y^{Nd}_{p'} \, \epsilon^{ab} \epsilon^{\alpha\beta\gamma} \,\bar{q}_{L,a,r,\beta}^{C} q_{L,b,\gamma,s} \, \,\bar{N}_R^{C}d_{R,p',\alpha} + \mathrm{h.c.} \Big] \quad ,
\end{aligned}
\end{equation}
and 
\begin{equation}
\begin{aligned}
-M^{2}_{V}\mathcal{L}_3&=-x_{rs}^{QD}\, x_{r's'}^{*,QD}\, (\bar{q}_{L,a,r'}\gamma^{\mu}q_{L,a,r}\,\bar{d}_{R,s'}\gamma_{\mu}d_{R,s}+2\bar{q}_{L,a,r'}d_{R,s}\, \bar{d}_{R,s'}q_{L,a,r})  
\\
&-x^{*,NQ}_{p'} \, x^{NQ}_{p} \, \bar{q}_{L,a,p'}\gamma^{\mu}q_{L,a,p}\,\bar{N}_R \gamma_{\mu} N_R 
\\
&+\big(2\, x_{rs}^{QD} \, x^{NQ}_{p} \, \epsilon^{\alpha\beta\gamma}\, \epsilon^{ab} \,  \bar{q}^{C}_{L,b,r,\beta}q_{L,a,p,\alpha}\, \bar{N}_R^{C} d_{R,s,\gamma} +\mathrm{h.c.} \big)\quad .
\end{aligned}
\end{equation}

\bibliographystyle{apsrev4-1}
\bibliography{references}

\end{document}

%% file: feynman/xsec.tex
\begin{tikzpicture}[scale=1.0]

   \begin{scope}[shift={(6,0)}]  \begin{feynman}
  
    \vertex [crossed dot](ww1){};
    \vertex[crossed dot,right =0.5cm of ww1](ww2){};
    \vertex [left=1.cm of ww1] (ml);
    \vertex [right=1.cm of ww2] (mr);
    \vertex [above=0.7cm of ml] (u1);
    \vertex [left=0.3cm of u1] (u) {\(u_{R,i}\)};
    \vertex [below=0.7cm of ml] (u2);
    \vertex [left=0.3cm of u2] (u3) {\( d_{R,j}\)};
    \vertex [above=0.7cm of mr] (l1);
    \vertex [right=0.2cm of l1] (l) {\(\bar d_{R,k}\)};
    \vertex [below=0.7cm of mr] (l2);
    \vertex [right=0.2cm of l2] (l3) {\(\bar N_R\)};
  
    \diagram* {
    (u) -- [thick,fermion] (ww1),
    (u3) -- [thick,fermion] (ww1),
    (l3) -- [thick,fermion] (ww2),
    (l) -- [thick,fermion] (ww2),
      };

  \end{feynman} 
 \node at (0,-1.3) {\Large$\propto C_{ijk}^{uddN} $  };
 \node at (3.5,0) {$=$};
\end{scope} 
\begin{scope}[shift={(0,0)}]  \begin{feynman}
  
    \vertex (ww1);
    \vertex[crossed dot,above =0.05cm of ww1] (ww2){};
    \vertex[crossed dot,below =0.05cm of ww1] (ww3){};
    \vertex [left=1.5cm of ww1] (ml);
    \vertex [right=1.5cm of ww1] (mr);
    \vertex [above=0.7cm of ml] (u1);
    \vertex [left=0.3cm of u1] (u) {\(u_{R,i}\)};
    \vertex [below=0.7cm of ml] (u2);
    \vertex [left=0.3cm of u2] (u3) {\( d_{R,j}\)};
    \vertex [above=0.7cm of mr] (l1);
    \vertex [right=0.2cm of l1] (l) {\(\bar d_{R,k}\)};
    \vertex [below=0.7cm of mr] (l2);
    \vertex [right=0.2cm of l2] (l3) {\(\bar N_R\)};
  
    \diagram* {
    (u) -- [thick,fermion] (ww2),
    (u3) -- [thick,fermion] (ww3),
    (l3) -- [thick,fermion] (ww3),
    (l) -- [thick,fermion] (ww2),
      };

  \end{feynman} 
  \node at (0,-1.3 ) {\Large$\propto C_{ikj}^{uddN} $  };
  \node at (3.25,0) {$+$};
\end{scope} 
\begin{scope}[shift={(12,0)}]  \begin{feynman}
  
    \vertex[crossed dot] (ww1){};
    \vertex [left=1.cm of ww1] (ml);
    \vertex [right=1.cm of ww1] (mr);
    \vertex [above=0.7cm of ml] (u1);
    \vertex [left=0.3cm of u1] (u) {\(u_{R,i}\)};
    \vertex [below=0.7cm of ml] (u2);
    \vertex [left=0.3cm of u2] (u3) {\( d_{R,j}\)};
    \vertex [above=0.7cm of mr] (l1);
    \vertex [right=0.2cm of l1] (l) {\(\bar d_{R,k}\)};
    \vertex [below=0.7cm of mr] (l2);
    \vertex [right=0.2cm of l2] (l3) {\(\bar N_R\)};
  
    \diagram* {
    (u) -- [thick,fermion] (ww1),
    (u3) -- [thick,fermion] (ww1),
    (l3) -- [thick,fermion] (ww1),
    (l) -- [thick,fermion] (ww1),
      };

  \end{feynman} 
\end{scope}


\begin{scope}[shift={(3,-3)}]  \begin{feynman}
  
    \vertex (ww1);
    \vertex[crossed dot,above =0.05cm of ww1] (ww2){};
    \vertex[crossed dot,below =0.05cm of ww1] (ww3){};
    \vertex [left=1.5cm of ww1] (ml);
    \vertex [right=1.5cm of ww1] (mr);
    \vertex [above=0.7cm of ml] (u1);
    \vertex [left=0.3cm of u1] (u) {\(u_{L,i}\)};
    \vertex [below=0.7cm of ml] (u2);
    \vertex [left=0.3cm of u2] (u3) {\( d_{R,j}\)};
    \vertex [above=0.7cm of mr] (l1);
    \vertex [right=0.2cm of l1] (l) {\(\bar d_{L,k}\)};
    \vertex [below=0.7cm of mr] (l2);
    \vertex [right=0.2cm of l2] (l3) {\(\bar N_R\)};
  
    \diagram* {
    (u) -- [thick,fermion] (ww2),
    (u3) -- [thick,fermion] (ww3),
    (l3) -- [thick,fermion] (ww3),
    (l) -- [thick,fermion] (ww2),
      };

  \end{feynman} 
  \node at (0,-1.3 ) {\Large$\propto 2 C_{ikj}^{qqdN} $  };
  \node at (3.25,0) {$=$};
\end{scope} 
\begin{scope}[shift={(9,-3)}]  \begin{feynman}
  
     \vertex[crossed dot] (ww1){};
    \vertex [left=1.cm of ww1] (ml);
    \vertex [right=1.cm of ww1] (mr);
    \vertex [above=0.7cm of ml] (u1);
    \vertex [left=0.3cm of u1] (u) {\(u_{L,i}\)};
    \vertex [below=0.7cm of ml] (u2);
    \vertex [left=0.3cm of u2] (u3) {\( d_{R,j}\)};
    \vertex [above=0.7cm of mr] (l1);
    \vertex [right=0.2cm of l1] (l) {\(\bar d_{k,L}\)};
    \vertex [below=0.7cm of mr] (l2);
    \vertex [right=0.2cm of l2] (l3) {\(\bar N_R\)};
  
    \diagram* {
    (u) -- [thick,fermion] (ww1),
    (u3) -- [thick,fermion] (ww1),
    (l3) -- [thick,fermion] (ww1),
    (l) -- [thick,fermion] (ww1),
      };

  \end{feynman}


\end{scope}

     \begin{scope}[shift={(3,-6)}]  \begin{feynman}
  
   \vertex [crossed dot](ww1){};
    \vertex[crossed dot,right =0.5cm of ww1](ww2){};
    \vertex [left=1.cm of ww1] (ml);
    \vertex [right=1.cm of ww2] (mr);
    \vertex [above=0.7cm of ml] (u1);
    \vertex [left=0.3cm of u1] (u) {\(u_{L,i}\)};
    \vertex [below=0.7cm of ml] (u2);
    \vertex [left=0.3cm of u2] (u3) {\( d_{L,j}\)};
    \vertex [above=0.7cm of mr] (l1);
    \vertex [right=0.2cm of l1] (l) {\(\bar d_{R,k}\)};
    \vertex [below=0.7cm of mr] (l2);
    \vertex [right=0.2cm of l2] (l3) {\(\bar N_R\)};
  
    \diagram* {
    (u) -- [thick,fermion] (ww1),
    (u3) -- [thick,fermion] (ww1),
    (l3) -- [thick,fermion] (ww2),
    (l) -- [thick,fermion] (ww2),
      };

  \end{feynman} 
 \node at (0,-1.3) {\Large$\propto 2 C_{ijk}^{qqdN} $  };
 \node at (3.25,0) {$=$};
\end{scope} 

\begin{scope}[shift={(9,-6)}]  \begin{feynman}
  
      \vertex[crossed dot] (ww1){};
    \vertex [left=1.cm of ww1] (ml);
    \vertex [right=1.cm of ww1] (mr);
    \vertex [above=0.7cm of ml] (u1);
    \vertex [left=0.3cm of u1] (u) {\(u_{L,i}\)};
    \vertex [below=0.7cm of ml] (u2);
    \vertex [left=0.3cm of u2] (u3) {\( d_{L,j}\)};
    \vertex [above=0.7cm of mr] (l1);
    \vertex [right=0.2cm of l1] (l) {\(\bar d_{R,k}\)};
    \vertex [below=0.7cm of mr] (l2);
    \vertex [right=0.2cm of l2] (l3) {\(\bar N_R\)};
  
    \diagram* {
    (u) -- [thick,fermion] (ww1),
    (u3) -- [thick,fermion] (ww1),
    (l3) -- [thick,fermion] (ww1),
    (l) -- [thick,fermion] (ww1),
      };

  \end{feynman} 

\end{scope} 

\end{tikzpicture}
  

%% file: feynman/LambdaC.tex
\begin{tikzpicture}[scale=1.0] 

\begin{scope}[shift={(0,-5)}] \begin{feynman}
  
    \vertex [crossed dot] (ww1){};
    \vertex [below= 0.3cm of ww1,crossed dot] (ww2){};
    \vertex [left=1.cm of ww1] (ml);
    \vertex [right=1.cm of ww1] (mr);
    \vertex [above=1.0cm of ml] (u1);
    \vertex [left=0.3cm of u1] (u) {\(d\)};
    \vertex [below=1cm of ml] (u2);
    \vertex [left=0.3cm of u2] (u3) {\( c\)};
    \vertex [below = 1cm of u2] (u4);
    \vertex [left=0.3cm of u4] (u5) {\( u\)};
    \vertex [above=1.0cm of mr] (l1);
    \vertex [right=0.3cm of l1] (l) {\(\bar N\)};
    \vertex [below=1cm of mr] (l2);
    \vertex [right=0.3cm of l2] (l3) {\(\bar d (\bar s )\)};
       \vertex [below = 1cm of l2] (l4);
    \vertex [right=0.7cm of l4] (l5) {\( u\)};

        \vertex at ($(u5)!0.5!(l5)$) (gpoint2);
        \vertex[below = 0.05cm of gpoint2](y){\(y\)};
     \vertex at ($(ww2)!0.4!(l3)$) (gpoint1);
     \vertex[above = 0.05cm of gpoint1](z){\(z\)};
     
    \diagram* {
    (u) -- [thick,fermion] (ww1),
    (u3) -- [thick,fermion] (ww2),
    (l3) -- [thick,fermion] (ww2),
    (l) -- [thick,fermion] (ww1),
    (u5) -- [thick,fermion] (l5),
        (gpoint1) -- [gluon, thick] (gpoint2),
      };
  \end{feynman}  \draw [decorate,decoration={brace,amplitude=5pt,mirror},thick]
    (u.north west) -- (u5.south west)
    node [midway, left=4pt] {\(\Lambda_c\)};
    \draw [decorate,decoration={brace,amplitude=5pt},thick]
    (l3.north east) -- (l5.south east)
    node [midway, right=4pt] {\(\pi^+(K^+)\)}; 
    \node at (-2.5,1.5) {\large $\Pi^2$ };
    \end{scope}

  \begin{scope}[shift={(8,-5)}] \begin{feynman}
  
    \vertex  (ww3);
    \vertex [left= 0.01cm of ww3,crossed dot] (ww1){};
    \vertex [right= 0.01cm of ww3,crossed dot] (ww2){};
    \vertex [left=1.cm of ww3] (ml);
    \vertex [right=1.cm of ww3] (mr);
    \vertex [above=1cm of ml] (u1);
    \vertex [left=0.3cm of u1] (u) {\(d\)};
    \vertex [below=1cm of ml] (u2);
    \vertex [left=0.3cm of u2] (u3) {\( c\)};
    \vertex [below = 1cm of u2] (u4);
    \vertex [left=0.3cm of u4] (u5) {\( u\)};
    \vertex [above=1cm of mr] (l1);
    \vertex [right=0.3cm of l1] (l) {\(\bar N\)};
    \vertex [below=1cm of mr] (l2);
    \vertex [right=0.3cm of l2] (l3) {\(\bar d(\bar s)\)};
       \vertex [below = 1cm of l2] (l4);
    \vertex [right=0.7cm of l4] (l5) {\( u\)};

        \vertex at ($(u5)!0.5!(l5)$) (gpoint2);
        \vertex[below = 0.05cm of gpoint2](y){\(y\)};
     \vertex at ($(ww2)!0.4!(l3)$) (gpoint1);
     \vertex[above = 0.05cm of gpoint1](z){\(z\)};
    \diagram* {
    (u) -- [thick,fermion] (ww1),
    (u3) -- [thick,fermion] (ww1),
    (l3) -- [thick,fermion] (ww2),
    (l) -- [thick,fermion] (ww2),
    (u5) -- [thick,fermion] (l5),
    (gpoint1) -- [gluon, thick] (gpoint2),
      };
  \end{feynman} 
   \draw [decorate,decoration={brace,amplitude=5pt,mirror},thick]
    (u.north west) -- (u5.south west)
    node [midway, left=4pt] {\(\Lambda_c\)};
    \draw [decorate,decoration={brace,amplitude=5pt},thick]
    (l3.north east) -- (l5.south east)
    node [midway, right=4pt] {\(\pi^+(K^+)\)};
   \node at (-2.5,1.5) {\large $\Pi^4$ };
    \end{scope}
\begin{scope}[shift={(8,0)}] \begin{feynman}
  
    \vertex [crossed dot] (ww1){};
    \vertex [below= 0.3cm of ww1,crossed dot] (ww2){};
    \vertex [left=1.cm of ww1] (ml);
    \vertex [right=1.cm of ww1] (mr);
    \vertex [above=1.0cm of ml] (u1);
    \vertex [left=0.3cm of u1] (u) {\(d\)};
    \vertex [below=1cm of ml] (u2);
    \vertex [left=0.3cm of u2] (u3) {\( c\)};
    \vertex [below = 1cm of u2] (u4);
    \vertex [left=0.3cm of u4] (u5) {\( u\)};
    \vertex [above=1.0cm of mr] (l1);
    \vertex [right=0.3cm of l1] (l) {\(\bar N\)};
    \vertex [below=1cm of mr] (l2);
    \vertex [right=0.3cm of l2] (l3) {\(\bar d(\bar s)\)};
       \vertex [below = 1cm of l2] (l4);
    \vertex [right=0.7cm of l4] (l5) {\( u\)};

        \vertex at ($(u5)!0.5!(l5)$) (gpoint2);
        \vertex[below = 0.05cm of gpoint2](y){\(y\)};
     \vertex at ($(ww1)!0.4!(u)$) (gpoint1);
     \vertex[above = 0.05cm of gpoint1](z){\(z\)};
     
    \diagram* {
    (u) -- [thick,fermion] (ww1),
    (u3) -- [thick,fermion] (ww2),
    (l3) -- [thick,fermion] (ww2),
    (l) -- [thick,fermion] (ww1),
    (u5) -- [thick,fermion] (l5),
        (gpoint1) -- [gluon, thick] (gpoint2),
      };
  \end{feynman}  \draw [decorate,decoration={brace,amplitude=5pt,mirror},thick]
    (u.north west) -- (u5.south west)
    node [midway, left=4pt] {\(\Lambda_c\)};
    \draw [decorate,decoration={brace,amplitude=5pt},thick]
    (l3.north east) -- (l5.south east)
    node [midway, right=4pt] {\(\pi^+(K^+)\)}; 
    \node at (-2.5,1.5) {\large $\Pi^3$ };
    \end{scope}

  \begin{scope}[shift={(0,0)}] \begin{feynman}
  
    \vertex  (ww3);
    \vertex [left= 0.01cm of ww3,crossed dot] (ww1){};
    \vertex [right= 0.01cm of ww3,crossed dot] (ww2){};
    \vertex [left=1.cm of ww3] (ml);
    \vertex [right=1.cm of ww3] (mr);
    \vertex [above=1cm of ml] (u1);
    \vertex [left=0.3cm of u1] (u) {\(d\)};
    \vertex [below=1cm of ml] (u2);
    \vertex [left=0.3cm of u2] (u3) {\( c\)};
    \vertex [below = 1cm of u2] (u4);
    \vertex [left=0.3cm of u4] (u5) {\( u\)};
    \vertex [above=1cm of mr] (l1);
    \vertex [right=0.3cm of l1] (l) {\(\bar N\)};
    \vertex [below=1cm of mr] (l2);
    \vertex [right=0.3cm of l2] (l3) {\(\bar d(\bar s)\)};
       \vertex [below = 1cm of l2] (l4);
    \vertex [right=0.7cm of l4] (l5) {\( u\)};

   \vertex at ($(u5)!0.5!(l5)$) (gpoint2);
     \vertex[below = 0.05cm of gpoint2](y){\(y\)};
       \vertex at ($(ww1)!0.4!(u)$) (gpoint1);
     \vertex[above = 0.05cm of gpoint1](z){\(z\)};
    \diagram* {
    (u) -- [thick,fermion] (ww1),
    (u3) -- [thick,fermion] (ww1),
    (l3) -- [thick,fermion] (ww2),
    (l) -- [thick,fermion] (ww2),
    (u5) -- [thick,fermion] (l5),
    (gpoint1) -- [gluon, thick] (gpoint2),
      };
  \end{feynman} 
   \draw [decorate,decoration={brace,amplitude=5pt,mirror},thick]
    (u.north west) -- (u5.south west)
    node [midway, left=4pt] {\(\Lambda_c\)};
    \draw [decorate,decoration={brace,amplitude=5pt},thick]
    (l3.north east) -- (l5.south east)
    node [midway, right=4pt] {\(\pi^+(K^+)\)};
   \node at (-2.5,1.5) {\large $\Pi^1$ };
    \end{scope}

\begin{scope}[shift={(0,-10)}] \begin{feynman}
  
    \vertex [crossed dot] (ww1){};
    \vertex [below= 0.3cm of ww1,crossed dot] (ww2){};
    \vertex [left=1.cm of ww1] (ml);
    \vertex [right=1.cm of ww1] (mr);
    \vertex [above=1.0cm of ml] (u1);
    \vertex [left=0.3cm of u1] (u) {\(d\)};
    \vertex [below=1cm of ml] (u2);
    \vertex [left=0.3cm of u2] (u3) {\( c\)};
    \vertex [below = 1cm of u2] (u4);
    \vertex [left=0.3cm of u4] (u5) {\( u\)};
    \vertex [above=1.0cm of mr] (l1);
    \vertex [right=0.3cm of l1] (l) {\(\bar N\)};
    \vertex [below=1cm of mr] (l2);
    \vertex [right=0.3cm of l2] (l3) {\(\bar d (\bar s )\)};
       \vertex [below = 1cm of l2] (l4);
    \vertex [right=0.7cm of l4] (l5) {\( u\)};

        \vertex at ($(u5)!0.5!(l5)$) (gpoint2);
        \vertex[below = 0.05cm of gpoint2](y){\(y\)};
     \vertex at ($(ww2)!0.4!(u3)$) (gpoint1);
     \vertex[above = 0.05cm of gpoint1](z){\(z\)};
     
    \diagram* {
    (u) -- [thick,fermion] (ww1),
    (u3) -- [thick,fermion] (ww2),
    (l3) -- [thick,fermion] (ww2),
    (l) -- [thick,fermion] (ww1),
    (u5) -- [thick,fermion] (l5),
        (gpoint1) -- [gluon, thick] (gpoint2),
      };
  \end{feynman}  \draw [decorate,decoration={brace,amplitude=5pt,mirror},thick]
    (u.north west) -- (u5.south west)
    node [midway, left=4pt] {\(\Lambda_c\)};
    \draw [decorate,decoration={brace,amplitude=5pt},thick]
    (l3.north east) -- (l5.south east)
    node [midway, right=4pt] {\(\pi^+(K^+)\)}; 
    \node at (-2.5,1.5) {\large $\Pi^5$ };
    \end{scope}
  \begin{scope}[shift={(8,-10)}] \begin{feynman}
  
    \vertex  (ww3);
    \vertex [left= 0.01cm of ww3,crossed dot] (ww1){};
    \vertex [right= 0.01cm of ww3,crossed dot] (ww2){};
    \vertex [left=1.cm of ww3] (ml);
    \vertex [right=1.cm of ww3] (mr);
    \vertex [above=1cm of ml] (u1);
    \vertex [left=0.3cm of u1] (u) {\(d\)};
    \vertex [below=1cm of ml] (u2);
    \vertex [left=0.3cm of u2] (u3) {\( c\)};
    \vertex [below = 1cm of u2] (u4);
    \vertex [left=0.3cm of u4] (u5) {\( u\)};
    \vertex [above=1cm of mr] (l1);
    \vertex [right=0.3cm of l1] (l) {\(\bar N\)};
    \vertex [below=1cm of mr] (l2);
    \vertex [right=0.3cm of l2] (l3) {\(\bar d(\bar s)\)};
       \vertex [below = 1cm of l2] (l4);
    \vertex [right=0.7cm of l4] (l5) {\( u\)};

        \vertex at ($(u5)!0.5!(l5)$) (gpoint2);
        \vertex[below = 0.05cm of gpoint2](y){\(y\)};
     \vertex at ($(ww1)!0.4!(u3)$) (gpoint1);
     \vertex[above = 0.05cm of gpoint1](z){\(z\)};
    \diagram* {
    (u) -- [thick,fermion] (ww1),
    (u3) -- [thick,fermion] (ww1),
    (l3) -- [thick,fermion] (ww2),
    (l) -- [thick,fermion] (ww2),
    (u5) -- [thick,fermion] (l5),
    (gpoint1) -- [gluon, thick] (gpoint2),
      };
  \end{feynman} 
   \draw [decorate,decoration={brace,amplitude=5pt,mirror},thick]
    (u.north west) -- (u5.south west)
    node [midway, left=4pt] {\(\Lambda_c\)};
    \draw [decorate,decoration={brace,amplitude=5pt},thick]
    (l3.north east) -- (l5.south east)
    node [midway, right=4pt] {\(\pi^+(K^+)\)};
   \node at (-2.5,1.5) {\large $\Pi^6$ };
    \end{scope}
  \end{tikzpicture}